\documentclass[aps,pre,reprint,superscriptaddress,showpacs,notitlepage,a4]{revtex4-2}
\usepackage{amssymb,amsmath}
\usepackage{booktabs}
\usepackage{soul}
\usepackage{graphicx}
\usepackage{tabularx}
\usepackage{multirow}
\usepackage{comment}
\usepackage{mathtools}
\usepackage{cases}
\usepackage[export]{adjustbox}
\usepackage{overpic}
\usepackage[colorlinks=true, linkcolor=black, urlcolor=blue, citecolor=blue]{hyperref}
\usepackage{mathscinet}
\usepackage{bm} 
\usepackage{float}
\usepackage{dcolumn}
\usepackage{bm}
\usepackage{bigints}
\usepackage{enumitem}
\usepackage{xcolor}
\usepackage[normalem]{ulem}

\newcommand{\tcpu}{t_\mathrm{CPU}}
\newcommand{\esat}{\varepsilon_\mathrm{sat}}
\newcommand{\dx}{\Delta x}
\newcommand{\dt}{\Delta t}
\newcommand\norm[1]{\lVert#1\rVert}
\def\P{{\sf Prob}}

\begin{document}

\title{Numerical methods for quasi-stationary distributions}

\author{Sara Oliver-Bonafoux}
\affiliation{Instituto de Física Interdisciplinar y Sistemas Complejos IFISC (CSIC-UIB), Campus UIB, 07122 Palma de Mallorca, Spain.}
\author{Javier Aguilar}
\affiliation{Instituto de Física Interdisciplinar y Sistemas Complejos IFISC (CSIC-UIB), Campus UIB, 07122 Palma de Mallorca, Spain.}
\affiliation{Departamento de Electromagnetismo y Física de la Materia and Instituto Carlos I de Física Teórica
y Computacional. Universidad de Granada. E-18071, Granada, Spain}
\affiliation{Laboratory of Interdisciplinary Physics, Department of Physics and Astronomy
“G. Galilei”, University of Padova, Padova, Italy}
\author{Tobias Galla}
\affiliation{Instituto de Física Interdisciplinar y Sistemas Complejos IFISC (CSIC-UIB), Campus UIB, 07122 Palma de Mallorca, Spain.}
\author{Raúl Toral}
\affiliation{Instituto de Física Interdisciplinar y Sistemas Complejos IFISC (CSIC-UIB), Campus UIB, 07122 Palma de Mallorca, Spain.}

\date{\today}

\begin{abstract}
In stochastic processes with absorbing states, the \emph{quasi-stationary distribution} provides valuable insights into the long-term behaviour prior to absorption. In this work, we revisit two well-established numerical methods for its computation. The first is an iterative algorithm for solving the non-linear equation that defines the quasi-stationary distribution. We generalise this technique to accommodate general Markov stochastic processes, either with discrete or continuous state space, and with multiple absorbing states. The second is a Monte Carlo method with resetting, for which we propose a single-trajectory approach that uses the trajectory’s own history to perform resets after absorption. In addition to these methodological contributions, we provide a detailed analysis of implementation aspects for both methods. We also compare their accuracy and efficiency across a range of examples. The results indicate that the iterative algorithm is generally the preferred choice for problems with simple boundaries, while the Monte Carlo approach is more suitable for problems with complex boundaries, where the implementation of the iterative algorithm is a challenging task.
\end{abstract}
 
\maketitle

\section{Introduction}
 
The stochastic decay towards absorbing states is used to model a plethora of phenomena, such as species extinction~\cite{azaele2016statistical,volkov2003neutral}, the fade-out of infectious diseases~\cite{keeling,pastor2001epidemic}, first passage times of Brownian particles~\cite{krapivsky2010kinetic,redner2001guide}, and the emergence of consensus in opinion dynamics~\cite{castellano2009statistical,axelrod1997dissemination}. The stationary distribution of processes with absorbing states concentrates the entire probability mass in these states~\cite{hinrichsen2000non,marro2005nonequilibrium}. However, the transient dynamics leading to absorption can vary across different processes, and it can be challenging to characterise these transient states analytically or numerically. For example, in systems with metastable states, the time to reach absorption can be exceedingly long, and it is sometimes not feasible to observe absorption using standard simulation methods~\cite{naasell2011extinction}. If, on the other hand, the process is biased towards absorption, the decay is usually rapid, requiring very fine temporal discretisations for proper sampling~\cite{Magalang2023}. Although Wentzel--Kramers--Brillouin (WKB) approaches provide analytical methods for these transient states (see, e.g.,~\cite{assaf2010extinction}), their validity is restricted to the limit of small noise.

The distribution of the process conditioned on not having been absorbed is referred to as the \emph{quasi-stationary distribution}. It characterises fluctuations both for processes with slow and with fast transients~\cite{Nasell1996,Vezzani2023,van2013quasi,naasell2011extinction}. Several quantities of interest can be evaluated from quasi-stationary distributions, including average survival times ~\cite{assaf2010extinction,collet2013quasi} and exit probabilities~\cite{collet2013quasi}. It has also been shown that quasi-stationary distributions can be used to sample likelihoods for inference problems~\cite{pollock2020quasi} and unlikely paths~\cite{aguilar2022sampling,lelievre2016partial}.

This collection of applications explains the interest in methods to obtain quasi-stationary distributions. As we will describe in more detail below, the quasi-stationary distribution fulfills a non-linear differential equation that can only be solved analytically for a few toy models~\cite{van2013quasi,naasell2011extinction,collet2013quasi,cavender1978quasi}. In absence of such an exact solution, a common approach is to introduce an auxiliary process whose stationary distribution approximates the quasi-stationary distribution of the process of interest~\cite{van2013quasi,Nasell1996,Vezzani2023}, or neglecting the non-linear term in the equation describing the quasi-stationary distribution~\cite{assaf2010extinction,assaf2017wkb}. However, similar to WKB techniques, these approximations are only valid in the small-noise limit when absorptions are rare events. When these conditions are not fulfilled, significant discrepancies between the approximate and exact distributions have been observed~\cite{Nasell1996,naasell2011extinction}.

The preferred strategy to solve the full non-linear equation for quasi-stationary distributions is thus the use of numerical methods. In~\cite{Nasell1996,naasell2011extinction,dickman2002quasi,van2013quasi}, iterative schemes and shooting methods are discussed to find the quasi-stationary distribution of a discrete-state model. The non-linear equation for the quasi-stationary distribution can be interpreted as an eigenvalue equation~\cite{collet2013quasi} (with the eigenvalue to be determined self-consistently), and thus numerical tools to compute eigenvalues and eigenvectors can be used to obtain the quasi-stationary distribution in discrete-state processes~\cite{Meleard}.

As far as we are aware, general numerical methods for solving quasi-stationary distributions for arbitrary processes are still lacking. In this work, we expand the ideas of~\cite{Nasell1996,de2005simulate} and propose an iterative algorithm and a simulation method to obtain the quasi-stationary distribution for multi-dimensional Markov processes with discrete or continuous state spaces, and with one or multiple absorbing states. We compare the two methods on the basis of selected examples, and discuss which method would be the preferred numerical strategy for these examples. Ultimately, we conclude that the iterative method is the preferred option for problems with simple boundaries, as it is efficient and easy to implement. In contrast, Monte Carlo simulations are more suitable for problems with complicated boundaries, where the implementation of the iterative algorithm is cumbersome.

The paper is organised as follows: In Sec.~\ref{sec:definition}, we define the quasi-stationary distribution and present the main equations for processes with discrete and continuous state spaces. In Sec.~\ref{sec:NumericalMethods}, we extend and improve previously existing numerical methods to obtain quasi-stationary distributions. These are an iterative algorithm (Sec.~\ref{sec:Iterative}) and a Monte Carlo resetting method (Sec.~\ref{sec:MonteCarlo}). In Sec.~\ref{sec:Examples}, we apply the numerical methods to several stochastic models with absorbing states. In Sec.~\ref{sec:DiscussionMethods}, we evaluate the efficiency of the approaches and provide additional technical details on the numerical implementations. Conclusions and a discussion of the results can be found in Sec.~\ref{sec:Conclusions}. Further details and analytical calculations are presented in the Appendices.

\section{Definition and basic equations}\label{sec:definition}

\subsection{Discrete states}

We consider a general Markov process $n(t)$ in continuous time $t$ and with discrete states $n\in\Omega=\{0,1,2,\dots\}$. The probability distribution of the process, $P(n,t)=\P\left[n(t)=n\right]$, satisfies the master equation
\begin{equation}\label{eq:master}
\partial_t P(n,t)=\sum_{n'}[W(n'\to n)P(n',t)-W(n\to n')P(n,t)],
\end{equation}
with transition rates defined as
\begin{equation}
W(n\to n')=\lim_{\Delta t\to0}\frac{\P[n(t+\Delta t)=n'|n(t)=n]}{\Delta t}. 
\end{equation}
In compact vector form, Eq.~\eqref{eq:master} is written as
\begin{equation}\label{eq:mastervector}
\begin{split}
\partial_t \bm{P}(t)&={\mathbb L}\bm{P}(t),
\end{split}
\end{equation}
where $\bm{P}(t)$ is a vector with components $P(n,t)$, and ${\mathbb L}$ is a matrix with elements
\begin{equation}\label{eq:transition_matrix}
\mathbb L_{n,m} = W(m\to n)-\delta_{n,m}\sum_{n'} W(n\to n').
\end{equation}
The stationary distribution, $P^\text{st}(n)=\lim_{t\to\infty} P(n,t)$, is found by setting the left-hand-side of Eq.~\eqref{eq:master} to zero.

An absorbing state $n^*$ is a state such that \makebox{$W(n^*\to n)=0$} for all $n\neq n^*$. If such a state is reached, the dynamics stops. The stationary distribution of a stochastic process with a set $\mathcal{S}=\{n^*_i\}_{i=1,\dots,N_\mathrm{A}}$ of absorbing states consists of a linear combination of Kronecker delta functions
\begin{equation}\label{eq:st_distribution}
 P^\text{st}(n)=\sum_{i=1}^{N_\mathrm{A}} p_i \, \delta_{n,n_i^*},
\end{equation}
where $p_i\in[0,1]$ is the absorption probability associated to state $n_i^*$, and $\sum_{i=1}^{N_A}p_i=1$. These probabilities can depend on the initial distribution of the process, \makebox{$\bm P(t = 0)$}. The absorption time is defined as the first time at which the process hits one of the absorbing states,
\begin{equation}
 \tau_{\rm abs} = \min_t\left[{n(t)\in \mathcal{S}}\right].
\end{equation}
We write $\tau$ for the average of the stochastic quantity $\tau_{\rm abs}$. 

The stationary distribution provides no information about the dynamics leading to absorption, or the statistics of fluctuations prior to absorption. To characterise these transient properties, one instead considers the distribution of the process $n(t)$ conditioned on not having reached absorption at time $t$,
\begin{equation}\label{eq:conditioned_distrib}
 Q(n,t)=\P\left[n(t)=n|n(t)\notin \mathcal{S}\right],
\end{equation}
which is naturally defined only for states that are not absorbing. For later convenience it is useful to set $Q(n^*, t) = 0, \, \forall n^* \in \mathcal{S}$. The distribution $Q(n,t)$ (for $n\notin \mathcal{S}$) fulfills an evolution equation similar to the master equation of the unconditioned process \cite{Nasell1996, collet2013quasi}
\begin{equation}\label{eq:evolution_QSD}
\begin{split}
\partial_t \bm{Q}(t)&={\mathbb L} \bm{Q}(t) +J_\mathcal{S} \, \bm{Q}(t),
\end{split}
\end{equation}
where $\bm{Q}(t)$ is a vector with components $Q(n,t)$, and $J_\mathcal{S}$ is the probability flux of $\bm{Q}(t)$ into the set of absorbing states,
\begin{equation}\label{eq:flux_absorbing}
 J_\mathcal{S}=\sum_{n^*\in \mathcal{S}}\sum_{n\in\Omega} W(n\to n^*) \, Q(n,t).
\end{equation}

We can think of Eq.~\eqref{eq:evolution_QSD} as describing a process related to the original stochastic dynamics, but in which trajectories that reach an absorbing state are reset to a non-absorbing state. More precisely, the last term in Eq.~\eqref{eq:evolution_QSD} indicates that the probability flux into the absorbing states is redistributed among the set of non-absorbing states in proportion to the probability mass in those states. This term is non-linear in $\bm{Q}(t)$ due to the dependence of $J_\mathcal{S}(t)$ on $\bm{Q}(t)$. The states that are most probable under $Q(n,t)$ benefit most from the redistribution of probability. 

The stationary solution of Eq.~\eqref{eq:evolution_QSD} is the quasi-stationary distribution. 
It satisfies the non-linear equation
\begin{equation}\label{eq:QSD}
{\mathbb L}\bm{Q} = -J_\mathcal{S} \bm{Q}.
\end{equation}
Eq.~\eqref{eq:QSD} may admit multiple solutions depending on the initial distribution of the process. Here, we focus on initial distributions concentrated entirely in a single state, in which case the quasi-stationary distribution is referred to as the Yaglom limit~\cite{collet2013quasi,meleard2012quasi}.

If the process is started from a quasi-stationary distribution, the mean absorption time is equal to~\cite{collet2013quasi}
\begin{equation}\label{eq:MTE}
 \tau = \frac{1}{J_\mathcal{S}}.
\end{equation}

\subsection{Continuous states}\label{sec:continuous_states}

We now consider a one-dimensional stochastic process $x(t)$ in continuous space and time. The probability density of the process, \makebox{$P(x,t)dx=\P\left[x(t)\in[x,x+dx]\right]$}, follows the Fokker--Planck equation 
\begin{equation}
\partial_t P(x,t)=\mathbb L P(x,t),
\end{equation}
where the operator $\mathbb L$ is given by
\begin{equation}\label{eq:FP_operator}
\mathbb L P(x,t)= -\partial_x\left[ A(x)P(x,t)\right]+\frac{1}{2} \partial^2_x\left[ B(x)P(x,t)\right]. 
\end{equation}

The quantities $A(x)$ and $B(x)$ are the drift and diffusion functions associated to the It\^{o} process
\begin{equation}\label{eq:Ito}
 \dot{x}(t) = A(x(t)) +\sqrt{B(x(t))}\xi(t),
\end{equation}
where $\xi(t)$ is a Gaussian white noise with mean $\langle \xi(t)\rangle=0$ and correlations $\langle \xi(t)\xi(t')\rangle=\delta(t-t')$. 

\subsubsection{Distinction between natural and artificial absorbing states}

As in the discrete case, an absorbing state is a point~$x^*$ at which the dynamics stops. For systems with a continuous state space, we distinguish between two types of absorbing states, following, for example, \cite{VanKam}.

The first type consists of states $x^*$ at which both the drift and diffusion functions vanish, i.e., \makebox{$A(x^*) = B(x^*) = 0$}. Consequently, when the process reaches such a state, the dynamics halts according to Eq.~\eqref{eq:Ito}. This type of absorbing states are referred to as \emph{natural absorbing states}~\cite{Lipowski2001, VanKam}. The absorbing nature of these states is intrinsically tied to the dynamics of the process. Natural absorbing states appear only in systems with multiplicative noise and often arise in the continuous-state approximation of discrete population models.

The second type of absorbing states refers to states that are not intrinsically absorbing in the stochastic process. In other words, either $A(x)$ or $B(x)$ (or both) remain non-zero at such states. However, one can `artificially' impose that realisations that reach a given state~$x^*$ are absorbed and removed. Mathematically, this is implemented through absorbing boundary conditions, i.e., by demanding that \makebox{$P(x^*, t) = 0$}~\cite{VanKam, jacobs2010stochastic, redner2001guide, gardiner1985handbook}. States of this type are referred to as \emph{artificial absorbing states}~\cite{VanKam}. Artificial absorbing boundary conditions are commonly used to study first-passage events~\cite{gardiner1985handbook,redner2001guide,jacobs2010stochastic}.

\subsubsection{Treatment of absorbing states}\label{sec:continuous_absorbingstates}

Writing again $\mathcal{S}$ for the set of absorbing states, the probability density of the process conditioned on not having been absorbed is given by
\begin{equation}\label{eq:def_q(x,t)}
 Q(x,t) \, dx=\P\left[x(t)\in [x,x+dx]|x(t)\not \in \mathcal{S}\right].
\end{equation}
The time evolution and the stationary distribution of this conditioned probability are given by the equivalent of Eqs.~\eqref{eq:evolution_QSD} and~\eqref{eq:QSD}, respectively, using the continuous Fokker--Planck operator defined in Eq.~\eqref{eq:FP_operator}. The expression for the flux $J_\mathcal{S}$ depends on the configuration of absorbing states of the process. 

If there is a single (natural or artificial) absorbing state~$x_1^*$ and the dynamics starts at $x>x_1^*$, the flux into the absorbing state is 
\begin{equation}\label{eq:flux_continuousx1}
 J_\mathcal{S}= -J(x^*_1),
\end{equation}
where
\begin{equation}\label{eq:flux_definition}
 J(x)= A(x)Q(x)- \frac{1}{2}\partial_x \left[B(x) Q(x)\right].
\end{equation}
We can also introduce a reflecting boundary at \makebox{$x=L>x_1^*$} such that the dynamics is restricted to the interval $(x_1^*,L)$. This condition corresponds to a vanishing probability flux at the boundary, i.e., $J(L)=0$.

If there are two absorbing states $x_1^* < x_2^*$ (again natural or artificial) and the dynamics starts at a point $x\in(x_1^*,x_2^*)$, the flux into the set of absorbing states is 
\begin{equation}\label{eq:flux_continuousx1x2}
 J_\mathcal{S}= J(x^*_2) - J(x^*_1).
\end{equation}

If there are more than two absorbing states \makebox{$x_1^*<x_2^*<x_3^*<\dots$} and the dynamics starts at \makebox{$x\in(x_i^*, \, x_{i+1}^*)$}, the process is effectively reduced to a system with two absorbing states: $x_i^*$ and $x_{i+1}^*$. 

The conditioned process must follow the same dynamics and satisfy the same boundary conditions as the original process. Therefore, the quasi-stationary distribution for any artificial absorbing state $x^*$ must also satisfy the boundary condition $Q(x^*, t) = 0$ for all $t$.

Determining the quasi-stationary distribution analytically is a challenging task in both the discrete and continuous settings. Consequently, numerical methods are the preferred tool. In the next section, we describe and develop such approaches.

\section{Numerical methods}\label{sec:NumericalMethods}

Several numerical methods have been proposed to estimate the quasi-stationary distribution of Markov stochastic processes in continuous time, corresponding to the stationary solution of Eq.~\eqref{eq:evolution_QSD} or its equivalent for continuous state spaces. While the time integration of this equation naturally converges to the quasi-stationary distribution, it requires selecting an appropriate integration time step. 

A more direct strategy consists in solving the time-independent Eq.~\eqref{eq:QSD}. A useful approach is based on iterative algorithms, in which an initial guess for the quasi-stationary distribution is refined through successive updates until convergence is achieved. However, to the best of our knowledge, existing implementations (e.g., \cite{Nasell1996, dickman2002quasi, dickman2002numerical}) are restricted to specific classes of stochastic processes, such as {\em one-step} discrete processes (that is, processes involving only events of the form \makebox{$n\to n\pm 1$}) with a single absorbing state. Here, we extend previous approaches to accommodate more general Markov processes.

Quasi-stationary distributions can also be estimated via Monte Carlo simulation techniques. A common approach consists in incorporating a resetting mechanism for absorbed trajectories, thus enabling the simulation of the quasi-stationary state. A widely used implementation involves simulating multiple realisations of the process and estimating the quasi-stationary distribution from the empirical distribution of final states (see, e.g.,~\cite{Vezzani2023,meleard2012quasi}). Here, we propose the simulation of a single trajectory. The quasi-stationary distribution is obtained from the distribution of residence times accumulated during the simulation.

\subsection{Iterative algorithm}\label{sec:Iterative}

Building on \cite{Nasell1996, dickman2002quasi, dickman2002numerical}, we present an iterative algorithm applicable to general Markov processes, with discrete or continuous state spaces, and accommodating any number of absorbing states.

\subsubsection{Discrete states}\label{sec:IterativeDiscrete}

We first consider a stochastic process $n(t)$ with discrete states $n$ in a finite set $\Omega$. The process evolves in continuous time according to transition rates \makebox{$W(n \to n')$}. As before, we denote by $\mathcal{S}$ the set of absorbing states. We also write~$\mathcal{R}$ for the set of non-absorbing states (\makebox{$\Omega=\mathcal{S}\cup \mathcal{R}$}, \makebox{$\mathcal{R}\cap \mathcal{S}=\emptyset$}).

The starting point is the relation
\begin{align}\label{eq:Q_iterative0}
Q(n)&= \frac{\sum\limits_{n'\in\Omega} W(n' \to n) \, Q(n')}{\sum\limits_{n'\in\Omega} W(n \to n') - \sum\limits_{n^*\in \mathcal{S}} \sum\limits_{n'\in\Omega} W(n'\to n^*) \, Q(n')} \nonumber \\
&\equiv f(n,\bm{Q}),
\end{align}
which is obtained by combining Eqs.~\eqref{eq:transition_matrix}, \eqref{eq:flux_absorbing} and~\eqref{eq:QSD}. 

Eq.~(\ref{eq:Q_iterative0}) is the basis for a numerical method to estimate the quasi-stationary distribution. Starting from an initial guess $\bm{Q}^{(0)}$, we use an iterative scheme with over-relaxation~\cite{dickman2002numerical}
\begin{equation}\label{eq:Iterative_Scheme}
Q(n)^{(k+1)} = s \, Q(n)^{(k)} + (1 - s) \, f\left(n,\bm{Q}^{(k)}\right), \, n\in \mathcal{R}
\end{equation}
where $k = 0, 1, 2,\dots$ indicates the iteration step, and $0 \le s < 1$ is the relaxation factor. We note that the distribution defined by the right-hand side of Eq.~\eqref{eq:Q_iterative0} is not necessarily normalised. Consequently, we carry out a normalisation after each step of the iteration.

We continue the iteration in Eq.~(\ref{eq:Iterative_Scheme}) until there is no further improvement in convergence. This is indicated by the stabilisation of the norm of the difference \makebox{$\norm{\bm{Q}^{(k+1)}-\bm{Q}^{(k)}}$} (with $\norm{\bm{a}}^2=\sum_{n\in \mathcal{R}}a(n)^2$) at a constant value. We write $\bm{\hat{Q}}$ for the numerical distribution obtained at the end of the iteration process.

In some cases, the norm $\norm{\bm{Q}^{(k+1)}-\bm{Q}^{(k)}}$ becomes zero within machine precision. When this does not occur, the iteration typically enters a period-$2$ cycle. As discussed in Sec.~\ref{sec:Iterative_RelaxationInitialCondition}, in certain pathological cases the norm stabilises at a macroscopic value, indicating that the probabilities $Q^{(k)}(n)$ at consecutive iterations genuinely differ. In contrast, for suitable choices of the over-relaxation parameter $s$ and initial distribution $\bm{Q}^{(0)}$, the differences between consecutive iterates in these 2-cycles are very small, typically less than $10^{-13}$. In such cases, the resulting distribution can still be regarded as a reliable estimate of the quasi-stationary distribution.

If high accuracy in the tails of the distribution is required, it is necessary to additionally monitor convergence specifically in that region, since small tail probabilities may have little impact on the overall norm. However, we did not encounter such a case in the simulations reported in this work. 

Our tests of the numerical algorithm include examples in which the analytical quasi-stationary distribution $\bm Q^{\rm exact}$ is known. In those cases, we will report the error in the numerical estimate $\bm{\hat{Q}}$ as 
\begin{equation}\label{eq:TotalErrorDiscrete}
 \varepsilon = ||\bm{Q}^{\rm exact}-\bm{\hat{Q}}||.
\end{equation}

In Sec.~\ref{sec:Iterative_RelaxationInitialCondition}, we investigate the convergence behaviour of the iterative algorithm as a function of the over-relaxation factor $s$ and the initial guess $\bm{Q}^{(0)}$. 

The mean time to extinction can be estimated using Eq.~\eqref{eq:MTE} together with the flux into the set of absorbing states $\hat{J}_\mathcal{S}$ derived from the final distribution $\hat{\bm{Q}}$.

\subsubsection{Continuous states}\label{sec:IterativeContinuous}

We now focus on a Markov stochastic process $x(t)$ in a one-dimensional continuous state space \makebox{$x \in \Omega = [0, L]$} that evolves in continuous time according to the It\^{o} process in Eq.~\eqref{eq:Ito}. We assume that the state $x=0$ is absorbing, while the boundary $x=L$ can be either reflecting or absorbing. 

The quasi-stationary distribution fulfills
\begin{equation}\label{eq:FP_equation}
J_\mathcal{S}Q(x) =\partial_x [A(x)Q(x)] - \frac{1}{2} \partial^2_x [B(x)Q(x)],
\end{equation}
as can be seen by combining Eqs.~\eqref{eq:QSD} and~\eqref{eq:FP_operator}.
To proceed, we discretise the interval $[0, L]$ into subintervals of length $\dx$, and write $x_i = i \dx$ for $i = 0, 1, \dots, N$ with $N = L / \dx$. The use of centred differences to approximate the derivatives leads to
\begin{align}\label{eq:Qcontinuous_iterative0}
J_\mathcal{S} Q_i &=\frac{A_{i+1} Q_{i+1} - A_{i-1} Q_{i-1}}{2 \dx} - \nonumber \\
&- \frac{1}{2} \frac{B_{i+1} Q_{i+1} - 2 B_i Q_i + B_{i-1} Q_{i-1}}{(\dx)^2}+O[(\dx)^2].
\end{align}
We have used the notation $Q_i=Q(x_i)$, and similar for $A_i, B_i$. From this equation, we directly obtain the relation
\begin{align}\label{eq:Qcontinuous_iterative1}
Q_i &= \frac{Q_{i+1} (B_{i+1} - \dx A_{i+1}) + Q_{i-1} (B_{i-1} + \dx A_{i-1})}{2 [B_i - (\dx)^2 J_\mathcal{S}]} \nonumber \\
&\equiv f_i(\bm{Q}),
\end{align}
which is valid for $i=1,\dots,N-1$. The approximation in Eq.~(\ref{eq:Qcontinuous_iterative0}) means that corrections in Eq.~(\ref{eq:Qcontinuous_iterative1}) are of order $(\dx)^4$ or higher. We have introduced the vector notation $\bm{Q}=(Q_0,\dots,Q_N)$. In analogy to Eq.~\eqref{eq:Iterative_Scheme}, we start from an initial guess $\bm{Q}^{(0)}$ and use the following iterative scheme \begin{equation}\label{eq:Iterative_Scheme_Continuous}
Q_i^{(k+1)} = s \, Q_i^{(k)} + (1 - s) \, f_i\left(\bm{Q}^{(k)}\right), \, i=1,\dots,N-1,
\end{equation}
followed by a normalisation of the numerical probability distribution at each step of the algorithm. The expression for the flux $J_\mathcal{S}$, as well as the iteration scheme at the boundaries of the state space, depends on the nature of the endpoints $x=0$ and $x=L$. In the examples, $x=0$ is always an absorbing state, either natural or artificial. We distinguish between the following cases:
\begin{enumerate}

\item If $x=0$ is a natural absorbing point, the value $Q_0$ does not enter in Eq.~\eqref{eq:Qcontinuous_iterative1}, as it is multiplied by either $A_0$ or $B_0$, both of which are zero.

\item If $x=0$ is an artificial absorbing point, we set $Q_0^{(k)}=0$ throughout the iteration (following the discussion in Sec.~\ref{sec:continuous_absorbingstates}). 

\item If $x=L$ is a reflecting boundary, $x=0$ is the only absorbing state. The probability flux $J_\mathcal{S}$ is therefore given by $-J(0)$ [see Eq.~\eqref{eq:flux_continuousx1}]. Approximating the derivative in the definition of flux using forward differences, we obtain
\begin{equation}\label{eq:flux_0absorbing}
J_\mathcal{S} = \frac{B_1 Q_1}{2 \dx},
\end{equation}
where we have used that either $Q_0 = 0$ (artificial absorbing state) or $A_0 = B_0 = 0$ (natural absorbing state), and neglected corrections of order $\dx$. 

The reflecting boundary condition at \makebox{$x = L$} corresponds to setting $J(L) = 0$. By discretising the derivative appearing in this condition using backward differences, we have (up to corrections of order~$\dx$) 
\begin{equation}\label{eq:QN_iterative}
f_N(\bm{Q}) = \frac{B_{N-1}}{B_N - 2 \dx A_N} Q_{N - 1},
\end{equation}
to be used in the iterative algorithm for the calculation of $Q_N^{(k)}$.

\item If $x = L$ is an absorbing point, the flux $J_\mathcal{S}$ into the set of absorbing states is given by \makebox{$J(L) -J(0)$} [see Eq.~\eqref{eq:flux_continuousx1x2}]. By approximating the derivatives in the flux at $x = 0$ and $x = L$ using forward and backward differences, respectively, the flux $J_\mathcal{S}$ can be written as
\begin{equation}\label{eq:flux_0Labsorbing}
 J_\mathcal{S} = \frac{B_{N-1} Q_{N-1} + B_1 Q_1}{2 \dx},
 \end{equation}
where we have used that for any absorbing state $x^*$ either $Q(x^*) = 0$ or $A(x^*) = B(x^*) = 0$.
If \makebox{$x=L$} is an artificial absorbing point, we set \makebox{$Q_N^{(k)}=0$} throughout the calculation. Conversely, if $x=L$ is a natural absorbing point, the value $Q_N$ does not enter in Eq.~\eqref{eq:Qcontinuous_iterative1}.
\end{enumerate}

As in the discrete case, we assess convergence by monitoring the quantity $\norm{\bm{Q}^{(k+1)}-\bm{Q}^{(k)}}$, with \makebox{$\norm{\bm{a}}^2=\sum_{i=1}^{N}a_i^2$}. We denote by $\hat{\bm{Q}}$ the distribution obtained at the end of the iteration process. Once convergence is attained, if $x = 0$ (resp. $x = L$) is a natural absorbing state, we estimate $Q_0$ (resp. $Q_N$) by linear interpolation, i.e., setting $\hat{Q}_0 = 2\hat{Q}_1 - \hat{Q}_2$ (resp. $\hat{Q}_N = 2\hat{Q}_{N-1} - \hat{Q}_{N-2}$).

In examples for which the exact distribution $\bm{Q}^{\rm exact}$ is known, we calculate the error as 
\begin{equation}\label{eq:TotalErrorContinuous}
 \varepsilon = \Delta x \,\norm{\bm{Q}^{\rm exact}-\hat{\bm{Q}}}.
\end{equation}

We note that the discretisation step $\Delta x$ needs to be chosen sufficiently small to ensure that the expression on the right-hand side of Eq.~(\ref{eq:Qcontinuous_iterative1}) remains non-negative.

In Sec.~\ref{sec:Iterative_dx}, we examine how the computation time and the error in the quasi-stationary distribution depend on discretisation step $\Delta x$ for selected problems with known analytical solution. 

\subsection{Monte Carlo resetting method}\label{sec:MonteCarlo}

Monte Carlo methods can be used to estimate the quasi-stationary distribution by sampling trajectories of the process conditioned on non-extinction. The simplest approach involves sampling a large number of independent realisations and discarding those that reach an absorbing state. After a sufficiently long time, the empirical distribution of the surviving trajectories provides an estimate of the quasi-stationary distribution. This method has obvious limitations. In order to obtain an accurate estimate, simulations must be run for long times. However, as time progresses, the set of surviving samples steadily decreases, reducing statistical accuracy. Therefore, generating sufficiently many surviving trajectories at sufficiently long times implies a high computational cost.

An alternative approach involves a resetting mechanism for trajectories that reach absorption. The non-linear term in Eq.~(\ref{eq:evolution_QSD}), which governs the evolution of the process conditioned on non-extinction, represents an influx of probability into each non-absorbing state $n$ given by $J_\mathcal{S}Q(n,t)$. This term indicates that a proportion $Q(n,t)$ of the total probability flux into the set of absorbing states, $J_\mathcal{S}$, is redistributed into state~$n$. This can be interpreted as resetting any trajectory that reaches absorption into one of the non-absorbing states, with $Q(n,t)$ representing the probability of choosing state $n$ for this resetting. The main challenge is that the true distribution $\bm{Q}(t)$ is unknown, as this is precisely the quantity we aim to determine. Therefore, one must rely on approximations or estimates for $\bm{Q}(t)$. The specific method used to construct this estimator defines the specific Monte Carlo algorithm.

One approach consists of replacing $\bm{Q}$ with a uniform distribution for the purposes of the resetting. This approximation appears to perform well only in systems where reaching the absorbing state is rare~\cite{Vezzani2023}. A more commonly used technique involves simulating multiple realisations of the process, and then estimating~$\bm{Q}$ via the empirical distribution of states~\cite{meleard2012quasi}. We refer to this approach as the \emph{multiple-trajectory method}. Specifically, a suitable simulation algorithm is used to generate~$M$ trajectories of the original process. Except for the resetting, these trajectories operate independently, and are generated in parallel. Whenever a trajectory reaches an absorbing state at a time $t+\dt$, it is reset to the state at time $t$ of a randomly selected trajectory from the set of $M$ trajectories. 

In this work, we adopt an alternative approach, first proposed in~\cite{blanchet2013empirical,de2005simulate}, which we refer to as the \emph{single-trajectory method}. This technique consists of simulating a single realisation of the process. Whenever the run reaches an absorbing state, it is reset according to the empirical distribution of residence times for this realisation. 

In most examples presented later in the paper, the single-trajectory approach proves more efficient than the multiple-trajectory technique, as less computation time is required to reach a given target error (see Appendix~\ref{Appendix:MC_MSComparison}). We now provide explicit rules for the implementation of this method, presenting separately the cases of discrete and continuous variables.

\subsubsection{Processes with discrete variables}\label{sec:MonteCarloDiscrete}

We consider a Markov process $n(t)$ as described in Sec.~\ref{sec:IterativeDiscrete}. The Gillespie algorithm~\cite{gillespie1976general, gillespie1977exact} can be used to generate a single trajectory of the process, starting at time $t_0 = 0$ in a non-absorbing state $n_0$. At each simulation step $k = 0, 1, 2, \dots$, a random time increment $\Delta t_k$ is drawn according to 
$\Delta t_k = -\ln (u_k)/ W(n_k)$, where $u_k$ is a uniform random number in the interval $(0, 1]$, and \makebox{$W(n_k) = \sum_{n'} W(n_k \to n')$} is the total escape rate from the current state $n_k$. The system jumps to $n_{k+1}=n$ with probability $W(n_k\to n)/W(n_k)$. Time is incremented to $t_{k+1}=t_k+\Delta t_k$. 

If the state $n_{k+1}$ is non-absorbing, the simulation proceeds. If, however, $n_{k+1}$ is an absorbing state, the trajectory is reset. To do this, a state is drawn from the empirical distribution $\bm{Q}^{(k)}$, which is obtained as
\begin{equation}\label{QitMC}
Q(n)^{(k)} = \dfrac{T(n,k)}{T(k)},
\end{equation}
where $T(n,k)$ is the total time the process has spent in state $n$ up to step $k$, and \makebox{$T(k)$} is the total duration of the process up to step $k$.

This procedure continues until the simulation reaches a pre-defined end time $T$. Alternatively, one can set a total number $K$ of Gillespie steps, such that, on average, \makebox{$T=K\langle\Delta t\rangle$}, where $\langle \Delta t\rangle$ is the average duration of each step. Provided $K$ (equivalently, $T$) are large, both alternatives lead to the same results. We expect that the estimate $\bm{Q}^{(k)}$ converges to a limiting distribution $\bm{\hat{Q}}$ for large $k$. We assess convergence by monitoring the norm $\|\bm{Q}^{(k)}\|=\left[\sum_n (Q^{(k)}(n))^2\right]^{1/2}$. 

This Monte Carlo approach, like the alternative Monte Carlo techniques described earlier, introduces bias because it does not simulate the exact dynamics of the process conditioned on non-extinction. Specifically, in the resetting step the algorithm does not have access to the true distribution $\bm{Q}^{\rm exact}(t)$, but needs to rely on an estimate. As a result, the total error $\varepsilon=\norm{\bm{\hat{Q}}-\bm{Q}^{\rm exact}}$ in the numerical estimate of the quasi-stationary distribution has two contributions: statistical error, which arises from the inherent randomness of the simulation combined with a finite sample size, and bias error. That is:
\begin{equation}\label{eq:ErrorStatisticalPlusErrorBias}
 \varepsilon = \sqrt{\varepsilon_\mathrm{statistical}^2 + \varepsilon_\mathrm{bias}^2}.
\end{equation}
The statistical error scales with the total number of simulation steps as $\varepsilon_\mathrm{statistical} \propto K^{-1/2}$.

From a simulation run, the mean time to extinction can be estimated as
\begin{equation}\label{eq:MTE_MonteCarlo}
\hat{\tau} = \frac{\sum_{i=1}^{N_\mathrm{abs}} \tau_i}{N_\mathrm{abs}},
\end{equation}
where $N_\mathrm{abs}$ represents the total number of absorptions in the run, and $\tau_i$ denotes the elapsed time between absorptions.

\subsubsection{Processes with continuous variables}\label{sec:MonteCarloContinuous}

Let us first focus on a Markov process with a single continuous variable $x(t)$, as described in Sec.~\ref{sec:IterativeContinuous}. In particular, we consider Eq.~\eqref{eq:Ito} with $x=0$ an absorbing state. While specialised simulation approaches are available to generate trajectories of processes with multiplicative noise~\cite{dornic2005integration, moro2004numerical,toral2014stochastic}, the Euler--Maruyama scheme satisfies the precision requirements of our work and we therefore proceed using this integration method. The integration is initialised at time $t_0 = 0$ at a non-absorbing point within the state space. The simulation proceeds in discrete time steps of constant duration $\Delta t$. 

To approximate the quasi-stationary distribution $Q(x, t)$, we discretise the state space into intervals of length $\dx$, \makebox{$I_i = \left[(i-1) \dx, \, i \dx\right]$} for \makebox{$i = 1, \dots, N$}, with $N = L / \dx$, and count the number of times $C(i,k)$ the process has visited the interval $I_i$ up to that time. Numbering time steps by $k$, this means
\begin{equation}\label{eq:single_estimator_continuous}
 Q^{(k)}_i = \frac{C(i,k)}{k \dx}.
\end{equation}

If the process is absorbed at step $k + 1$, that is, if $x_{k+1} \leq 0$ (or if $x_{k+1} \geq L$, in case the upper barrier $x = L$ is also absorbing), we draw an interval $I_i$ from the distribution \makebox{$\bm{Q}^{(k)}$} and then place the process at a random position uniformly distributed in that interval. If the boundary $x = L$ is not absorbing and $x_{k+1} \geq L$, we apply reflecting boundary conditions, and replace $x_{k+1}$ by $2L - x_{k+1}$, to prevent transitions out of the state space. 

As in the discrete case, the convergence criterion in the continuous setting is based on the stabilisation of the quantity $\|\bm{Q}^{(k)}\|$ at a constant value (where \makebox{$\norm{\bm{a}}^2=\sum_{i=1}^{N}a_i^2$}). We denote the final estimate of the quasi-stationary distribution by $\hat{\bm{Q}}$. 

Both for discrete and continuous states, this Monte Carlo resetting method involves two sources of error: statistical error and bias error, as described by Eq.~\eqref{eq:ErrorStatisticalPlusErrorBias}. For continuous states, the bias arises not only from the use of an estimate of the quasi-stationary distribution to perform the resets, but also from the discretisation of time and state space. For processes with known analytical solution, the total error (due to both statistical error and bias error) can be computed using Eq.~\eqref{eq:TotalErrorContinuous}. In Sec.~\ref{sec:MonteCarlo_dxdt}, we examine how the total error depends on the discretisation steps $\dx$ and $\dt$ for several problems with known analytical solution. 

The method can be straightforwardly generalised to multi-dimensional continuous stochastic processes, \makebox{$\bm x(t) = [x^{(1)}(t), \dots, x^{(d)}(t)]$}, where each component $x^{(\ell)}(t)$ evolves according to an It\^{o} process as in Eq.~\eqref{eq:Ito} with drift and diffusion functions $A^{(\ell)}(\bm x)$ and $B^{(\ell)}(\bm x)$, respectively. Each variable $x^{(\ell)}(t)$ has a state space \makebox{$\Omega_\ell = [x^{(\ell)}_{m}, x^{(\ell)}_{M}]$}, which we discretise using intervals of length $\Delta x^{(\ell)}$. Boundary conditions are specified based on the particular problem under study.

\section{Applications}\label{sec:Examples}

In this section, we test the iterative algorithm and the Monte Carlo method across several examples. We consider eight different systems, each with a single stochastic variable (four with discrete state space and four with continuous state space), as well as two processes with multiple degrees of freedom. When analytical solutions for the quasi-stationary distribution are available, we test the numerical results against the analytical predictions. A summary of the characteristics of the processes considered in this work can be found in Appendix~\ref{Appendix:SummaryProcesses}.

In all problems involving a single stochastic variable, we set the relaxation factor to $s = 0.1$ in the iterative algorithm, and perform $K = 10^{11}$ steps in the Monte Carlo resetting method. We use a Kronecker delta distribution as an initial condition for both the iterative algorithm and the simulations. We denote these initial states as $n_0$ or $x_0$ for discrete or continuous state spaces, respectively. The choice of this initial state is based on the characteristics of the process. If the system is biased towards an absorbing state, we place the initial condition near that point. If there is a metastable state, we place the initial condition near that state. The role of the initial condition is mostly irrelevant with some exceptions, as will be discussed in Sec.~\ref{sec:convergence_issues}.

\subsection{Processes with a single discrete variable}\label{sec:ExamplesDiscrete}

The first three examples come from stochastic population dynamics, and each involves a single variable $n=0,1,2, \dots$ representing population size. All three models involve birth and death events. Without loss of generality, we set the parameter describing per capita death rates to one, and write $R$ for the coefficient characterising the birth rate. Effectively, $R$ represents the ratio of birth to death rates in all three models. The state $n = 0$ describes the extinction of the population, and is absorbing.

The first model is the linear branching process~\cite{karlin2014first}, with transition rates given by
\begin{equation}\label{eq:Rates_Branching}
 W(n\to n+1) = R n, \quad W(n\to n-1) = n. 
\end{equation}
This process only exhibits a quasi-stationary regime if $R < 1$.

The second model is an immigration-death process, in which particles are introduced at constant immigration rate (independent of the current population size) from an external reservoir. The transition rates for this model are
\begin{align}\label{eq:Rates_Mix}
W(n\to n+1) &= \begin{cases}
R, & \text{for } n \geq 1, \\
0, & \text{for } n = 0,
\end{cases} \nonumber \\
W(n\to n-1) &= n, \quad \text{for } n \geq 0.
\end{align}

The quasi-stationary distribution of these two models can be determined analytically (see Appendices ~\ref{Appendix:QSD_Discrete_Branching} and~\ref{Appendix:QSD_Discrete_Mix}). 

The third model is the branching-annihilation-decay process introduced in~\cite{assaf2010extinction}. In addition to linear birth and death, this model also includes pairwise annihilation, that is, a two-step process. The transition rates are
\begin{align}\label{eq:Rates_BAD}
 W(n\to n+1) &= R n, \nonumber \\
 W(n\to n-1) &= n, \nonumber \\
 W(n\to n-2) &= \frac{R n (n-1)}{2 V},
\end{align}
 where $V$ is the volume in which particles are contained, or, equivalently, a parameter determining the characteristic population size in the long run \cite{VanKam}. 

As a fourth example, we consider the biased voter model of opinion formation~\cite{czaplicka2022biased}, with transition rates
\begin{align}\label{eq:Rates_VM}
 W(n \to n + 1) &= R \frac{n (N-n)}{N}, \nonumber \\
 W(n \to n - 1) &= \frac{n (N-n)}{N},
\end{align}
where $N$ is the population size, $R$ quantifies the bias towards the preferred opinion, and $0 \le n \le N$ is the number of individuals holding this opinion. This model has two absorbing states, $n = 0$ and $n = N$, describing consensus on either of the two opinion states. The model can also be understood as an evolutionary dynamics of the Moran type, with frequency independent selection (see, e.g.,~\cite{traulsen2009stochastic}).

We note that the examples in Eqs.~(\ref{eq:Rates_Branching}), (\ref{eq:Rates_Mix}) and (\ref{eq:Rates_VM}) are one-step processes, with transitions of the form \makebox{$n\to n\pm 1$}. In contrast, the branching-annihilation-decay process in Eq.~(\ref{eq:Rates_BAD}) involves events of the type $n\to n-2$ and is therefore a multi-step process. We also highlight that the examples in Eqs.~(\ref{eq:Rates_Branching}), (\ref{eq:Rates_Mix}) and (\ref{eq:Rates_BAD}) each have a single absorbing state ($n=0$), whereas the voter model in Eq.~(\ref{eq:Rates_VM}) has two absorbing states. 

\begin{figure}[h]
	\centering	\includegraphics[width=0.49\linewidth]{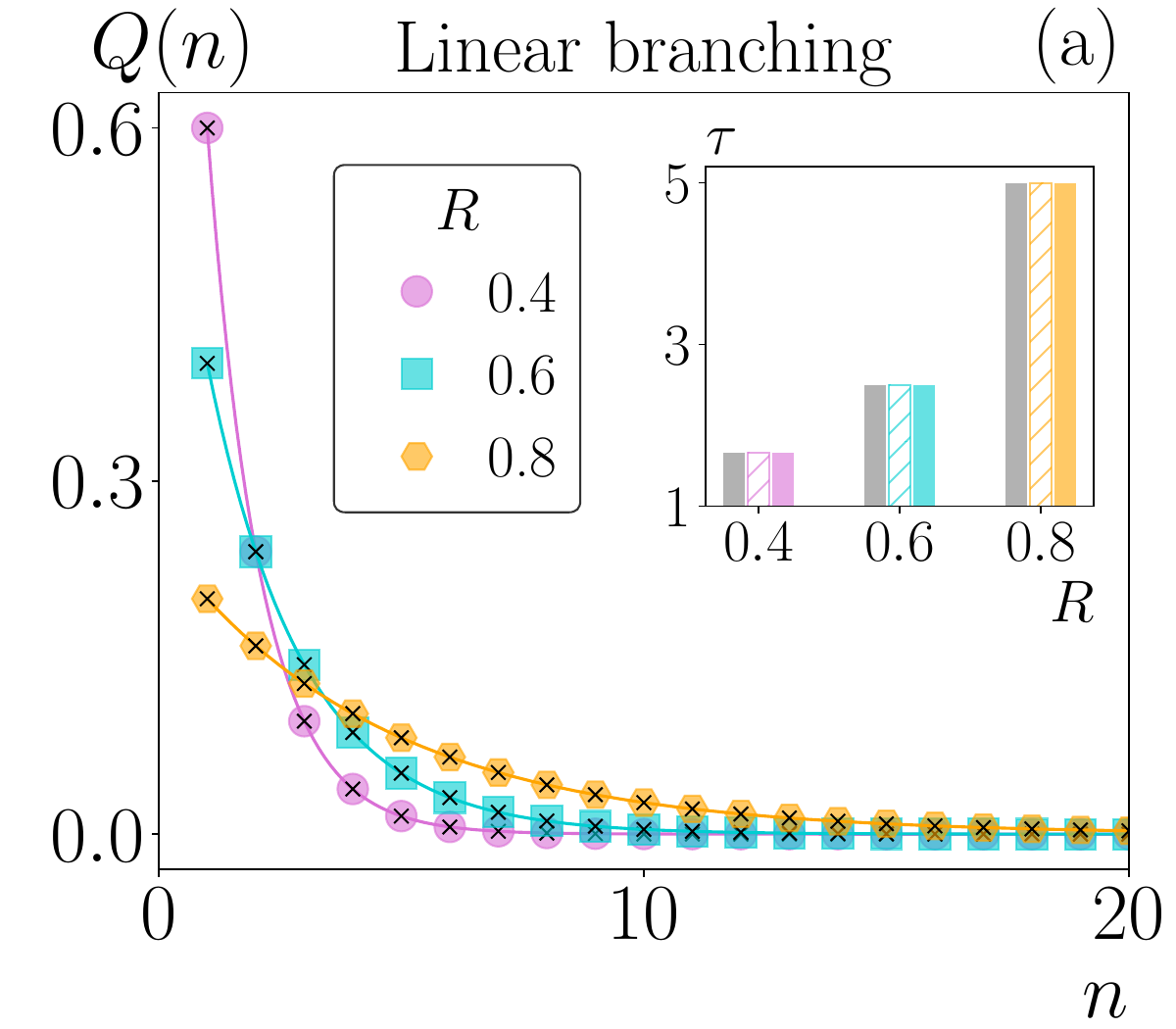}
 \includegraphics[width=0.49\linewidth]{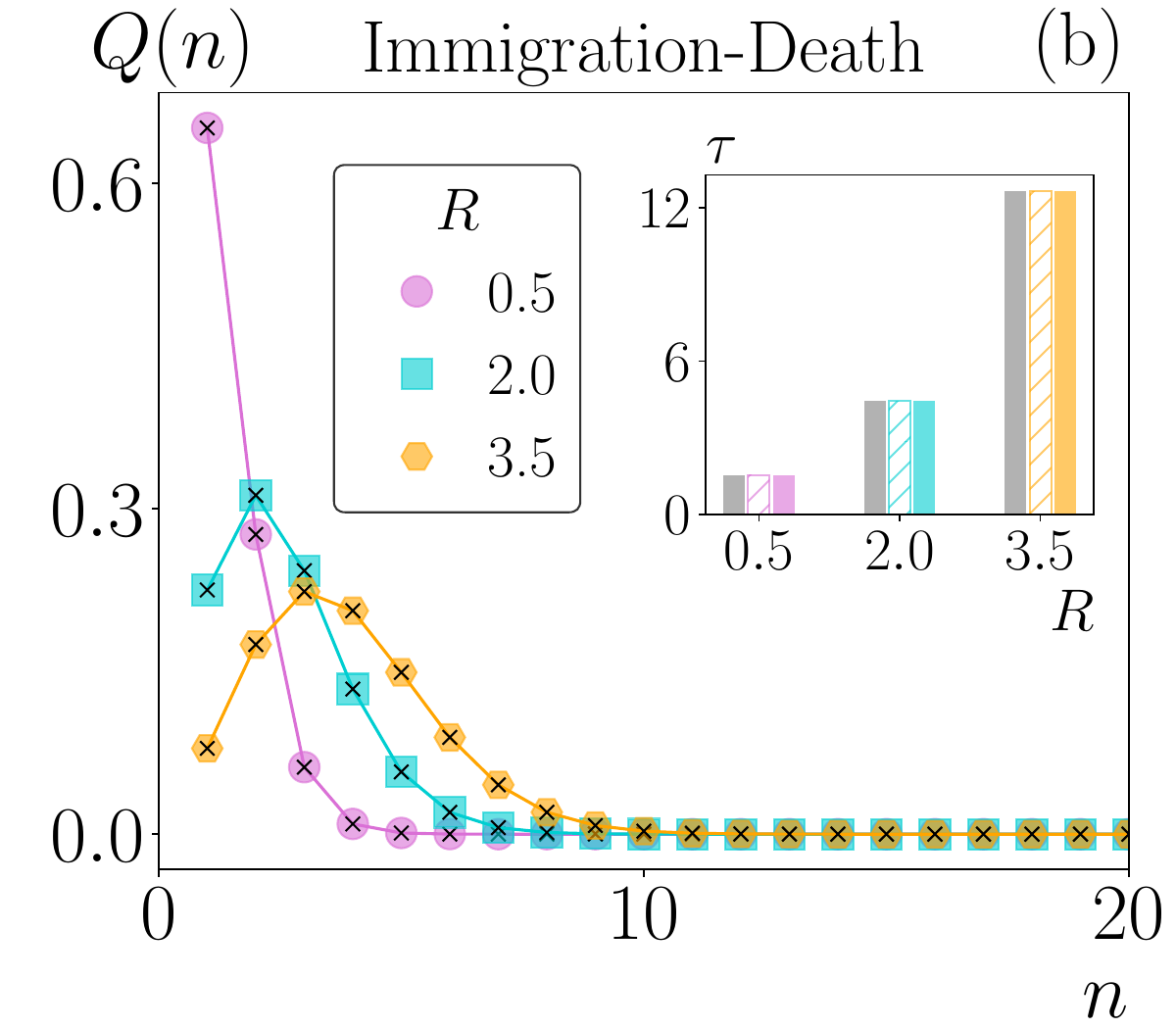}
 \includegraphics[width=0.49\linewidth]{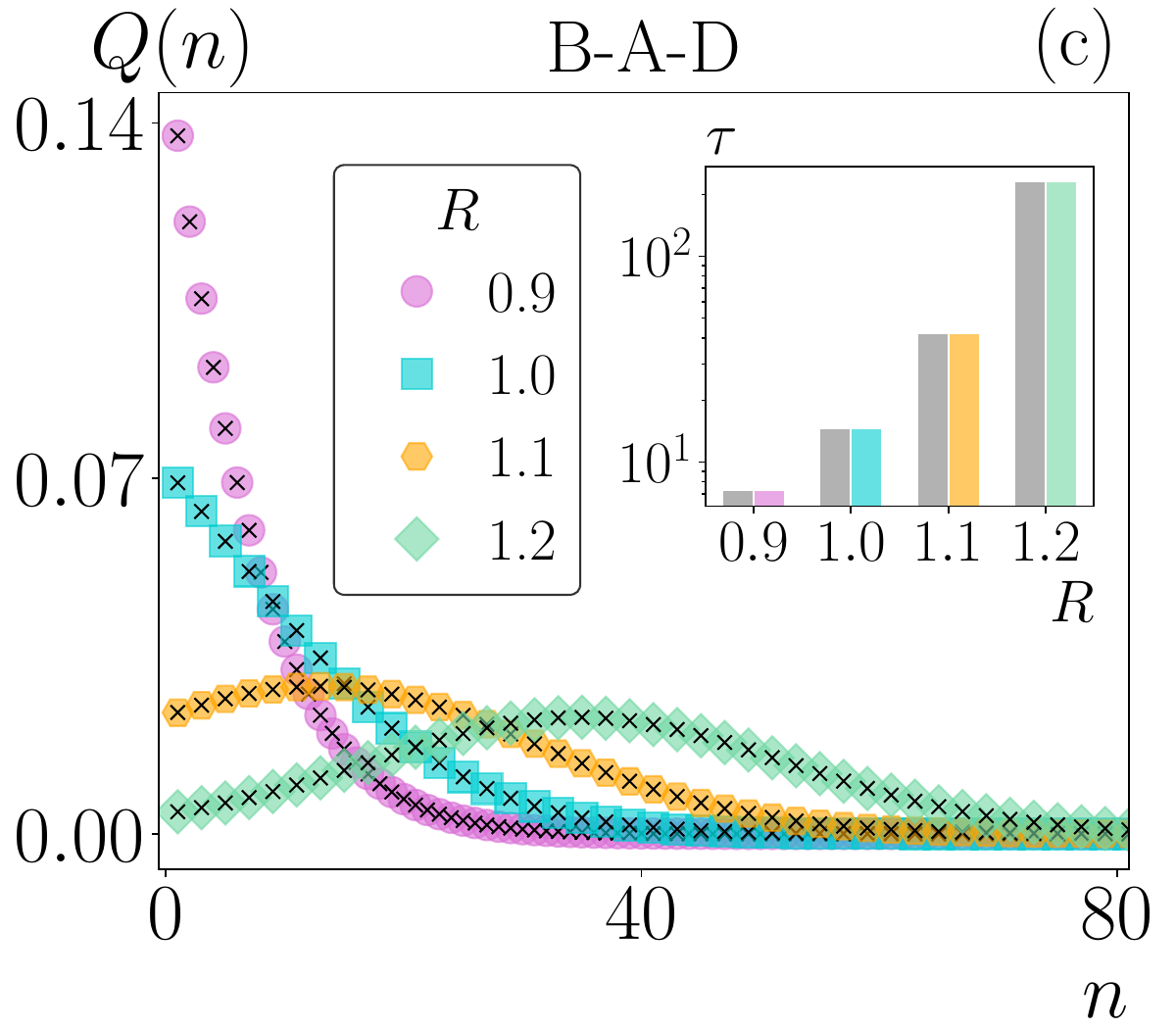}
 \includegraphics[width=0.49\linewidth]{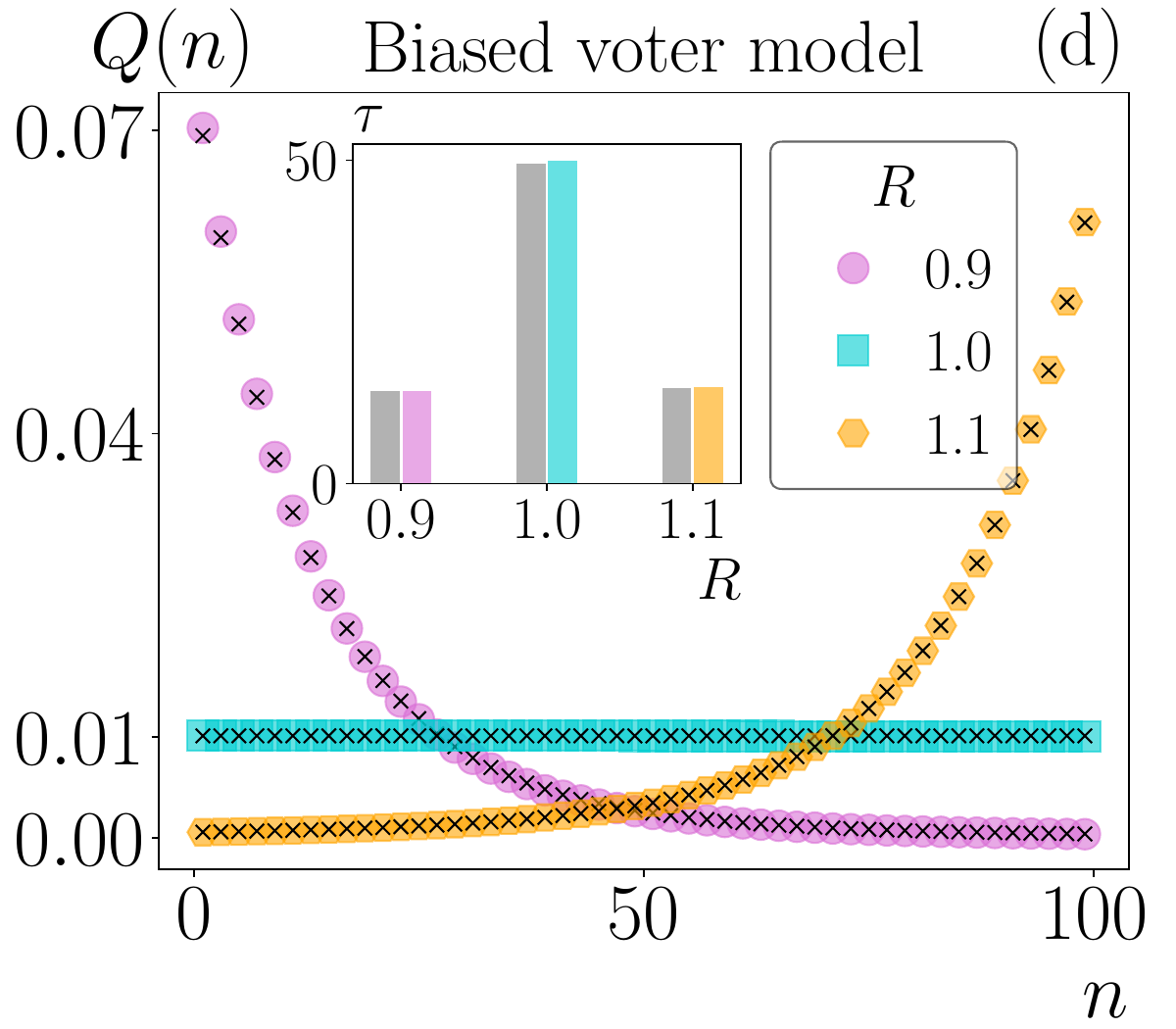}
	\caption{\textbf{Quasi-stationary distribution and mean time to extinction for the discrete processes in Sec.~\ref{sec:ExamplesDiscrete}:} (a) the linear branching process, (b) the immigration-death process, (c) the branching-annihilation-decay (B-A-D) process with $V = 250$, and (d) the biased voter model with \makebox{$N = 100$}. Main panels: Quasi-stationary distribution for different values of the parameter $R$ (see text). Black crosses come from the iterative algorithm, while coloured symbols come from the single-trajectory Monte Carlo method. Additionally, in panels (a)-(b), solid lines represent the analytical solution. For better visualisation, in cases (c) for $R = 1.0, 1.1, 1.2$ and (d), the numerical distribution is plotted only for odd values of~$n$. In both numerical methods, the initial condition has been set to: (a) $n_0 = 10$, (b) $n_0 = R$ (metastable state) (c) $n_0 = 10$ if $R \leq 1$ and $n_0 = 1 + V(1 - 1/R)$ (metastable state) if $R > 1$, and (d) $n_0 = 10, 50, 90$ for $h = 0.9, 1.0, 1.1$, respectively. Additionally, in cases (a)-(c) we have set an upper bound for the number of particles at $N = 150$ in the iterative algorithm. Insets: Mean absorption time, $\tau$, for different choices of $R$. Grey bars come from the iterative algorithm, filled coloured bars come from the Monte Carlo method and, in cases (a) and (b), bars with diagonal lines represent the analytical solution.}
	\label{fig:QSD_Discrete}
\end{figure}

Fig.~\ref{fig:QSD_Discrete} shows the quasi-stationary distribution and mean absorption time for these processes, calculated for several values of the characteristic parameters of the models. The results obtained with both the iterative algorithm and the Monte Carlo resetting method are found to be in agreement with analytical solution when the latter is available [panels (a) and (b)]. These examples illustrate that both numerical approaches allow us to obtain the quasi-stationary distribution for multi-step processes, as well as for processes with multiple absorbing states. 

\subsection{Processes with a single continuous variable}\label{sec:ExamplesContinuous}

We first consider two stochastic processes with an artificial absorbing state at $x = 0$. 

The first process is a biased random walk on the positive real line bounded by a reflecting barrier at $x = L$. The drift and diffusion functions governing this process are~\cite{krapivsky2010kinetic}
\begin{equation}\label{eq:BRW_DriftDiffusion}
 A(x) = f, \quad B(x) = 2D,
\end{equation}
with $f\in\mathbb{R}$ and $D>0$.

The second process is an Ornstein--Uhlenbeck process, with drift and diffusion functions~\cite{VanKam}
\begin{equation}\label{eq:Ornstein_DriftDiffusion}
 A(x) = -fx, \quad B(x) = 2D, 
\end{equation}
with $f > 0$ and $D>0$.

The third example is the linear branching process with a natural absorbing barrier at $x = 0$. The drift and diffusion functions for this process are~\cite{Meleard}
\begin{equation}\label{eq:Branching_DriftDiffusion}
 A(x) = -fx, \quad B(x) = 2Dx,
\end{equation}
with $f > 0$ and $D>0$.

The quasi-stationary distribution of all three processes can be derived analytically (see Appendices~\ref{Appendix:QSD_Continuous_BRW}-\ref{Appendix:QSD_Continuous_Branching}). Note that the last two processes only exhibit a quasi-stationary regime if $f > 0$. 

The fourth example is the continuous formulation of the agent-based biased voter model described in Eq.~\eqref{eq:Rates_VM}. In this problem, $x \in [0, 1]$ is a stochastic continuous variable accounting for the density of individuals holding the preferred opinion. The drift and diffusion functions are given by 
\begin{align}\label{eq:VM_DriftDiffusion}
 A(x) &= (R-1) x (1-x), \nonumber \\
 B(x) &= \frac{(R+1) x (1-x)}{N}.
\end{align}

In the numerical methods, we set the discretisation parameters as follows: $\dx = 5 \cdot 10^{-4}$ in the iterative algorithm; and $\dx = 10^{-3}$ and $\dt = 10^{-4}$ in the Monte Carlo method. The choice of these particular values will be justified in Sec.~\ref{sec:discretisationParameters}. Note that the spatial step size~$\dx$ in the Monte Carlo simulations is larger than in the iterative algorithm. This choice is justified by the fact that, while decreasing $\dx$ in the iterative algorithm leads to an improvement on accuracy, in the Monte Carlo method such a reduction has a negligible impact on accuracy but leads to a significant increase in computation time.

\begin{figure}[ht!]
	\centering
	\includegraphics[width=0.49\linewidth]{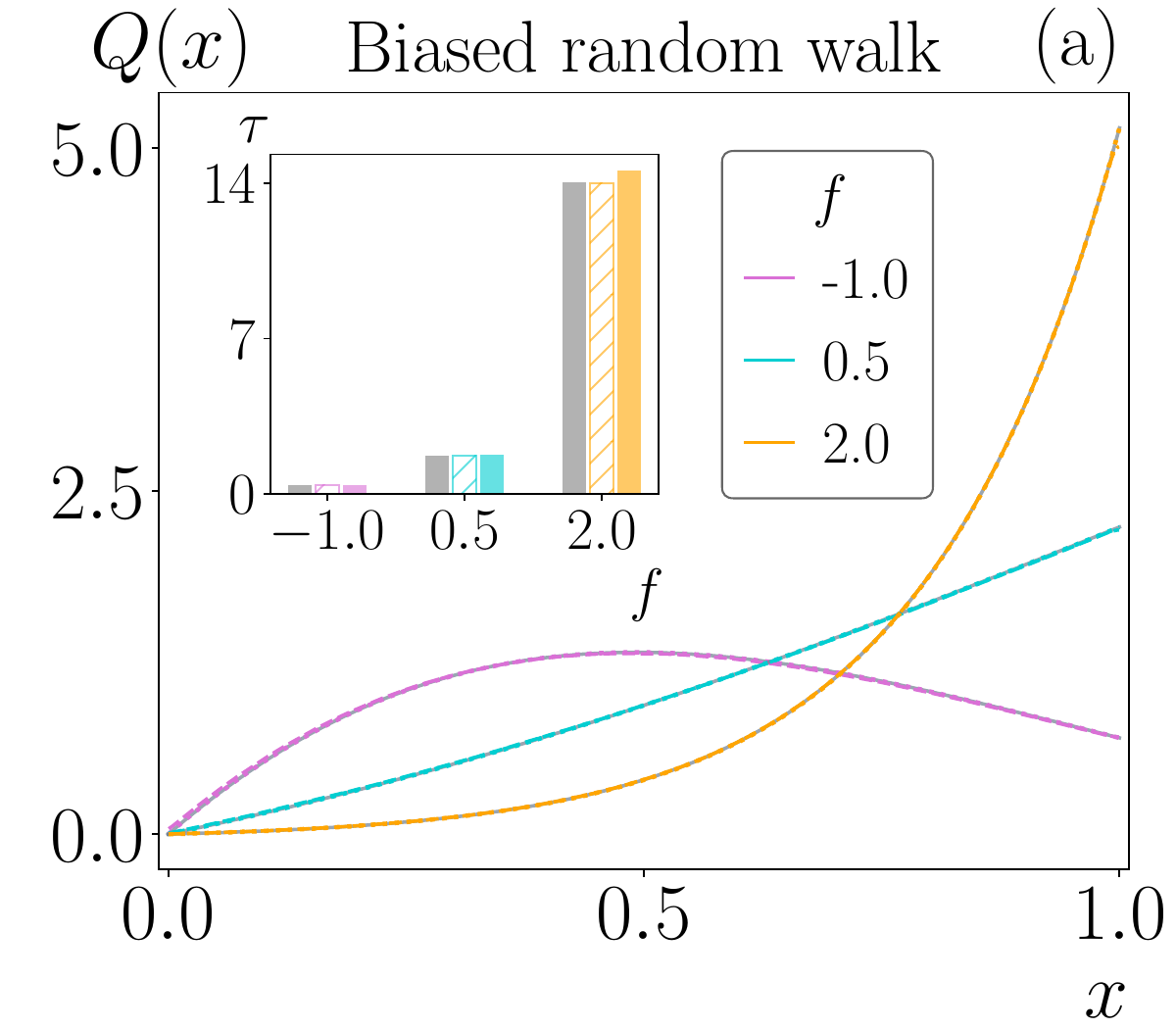}
 \includegraphics[width=0.49\linewidth]{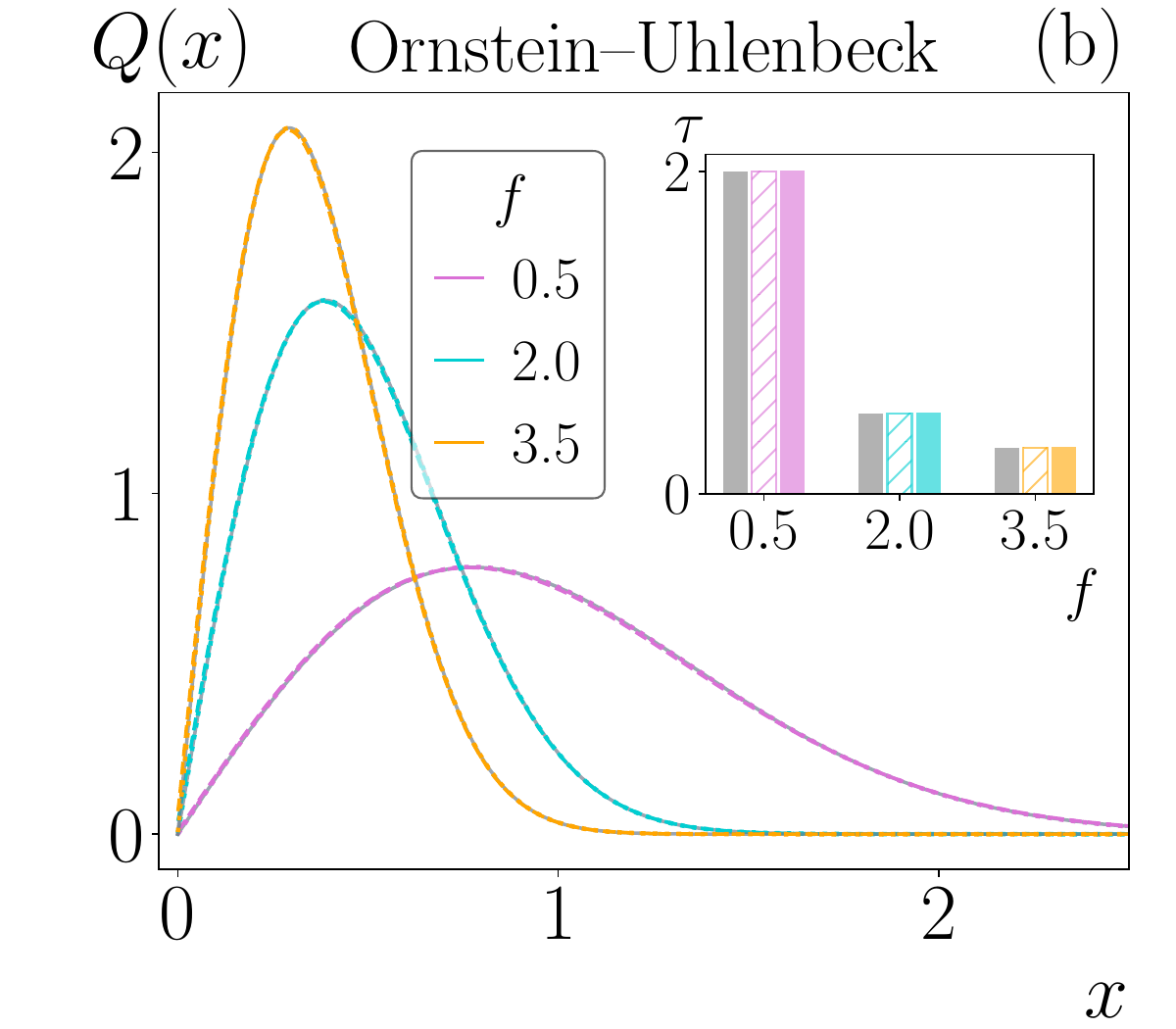}
 \includegraphics[width=0.49\linewidth]{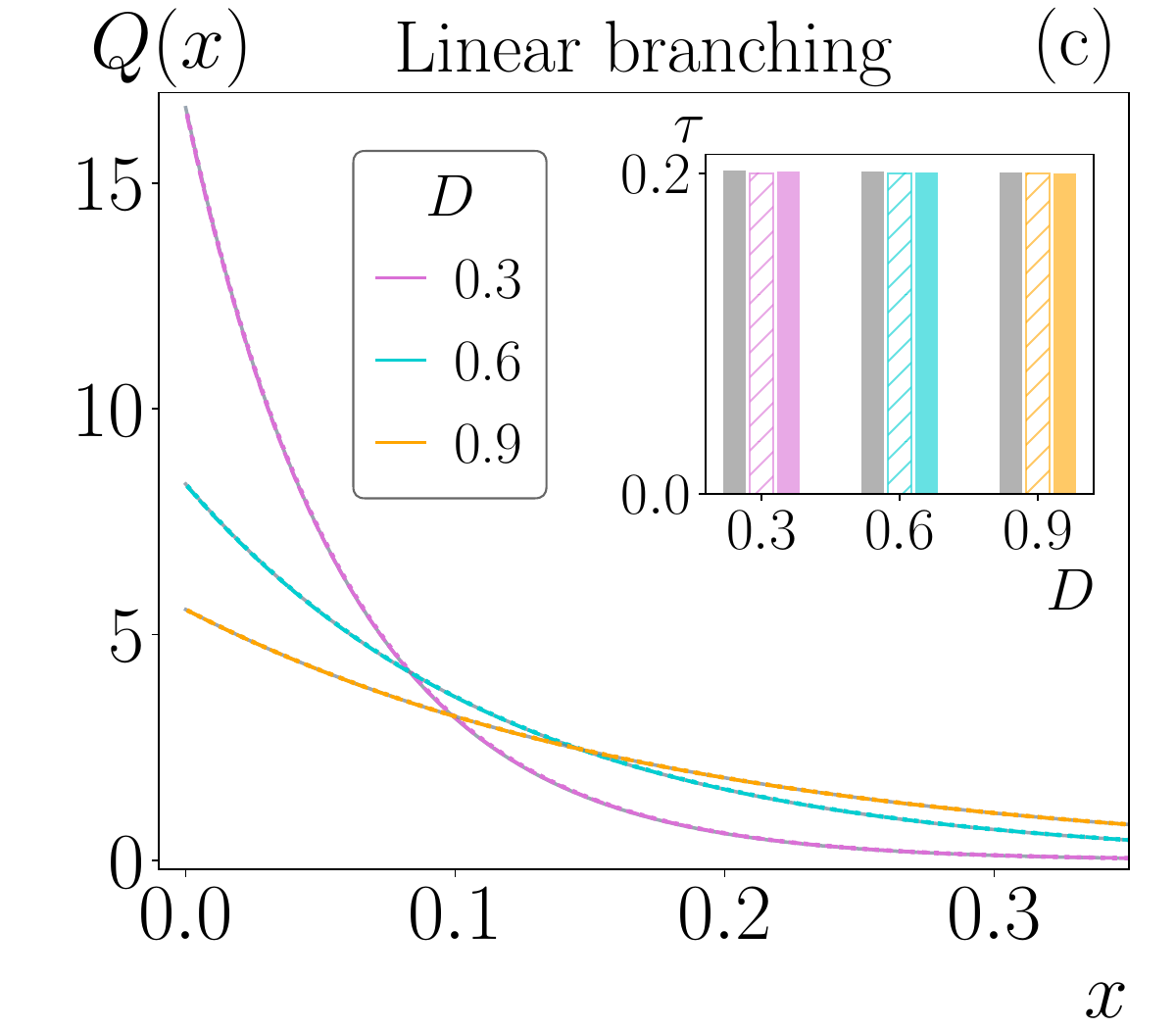}
 \includegraphics[width=0.49\linewidth]{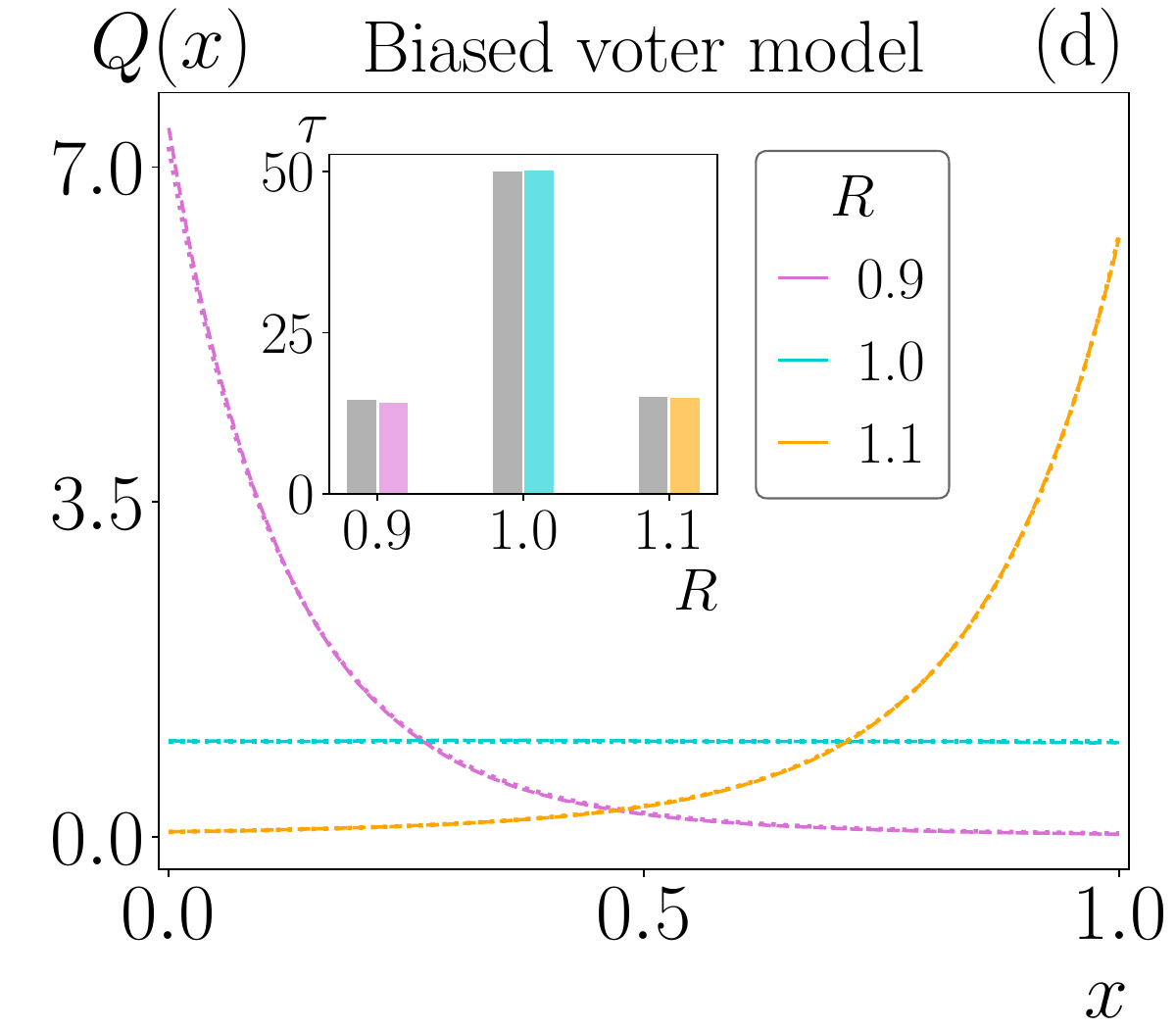}
	\caption{\textbf{Quasi-stationary density function and mean absorption time for the continuous processes in Sec.~\ref{sec:ExamplesContinuous}:} (a) the biased random walk with $D = 0.4$ and $L = 1$, (b) the Ornstein--Uhlenbeck process with $D = 0.3$, (c) the linear branching process with $f = 5$, and (d) the continuous biased voter model with $N = 100$. Main panels: Quasi-stationary density function for different values of a characteristic parameter of the model (see text). Dotted coloured lines correspond to the iterative algorithm, while dashed coloured lines correspond to the Monte Carlo method. Moreover, in panels (a)-(c), solid grey lines represent the analytical solution. In most cases, the different lines are visually indistinguishable. In the numerical methods, we have used as initial condition: (a) $x_0 = 0.1$ for $f = -1$, and $x_0 = 0.9$ for $f = 0.5, 2.0$, (b)-(c) $x_0 = 0.1$, and (d) $x_0 = 0.1, 0.5, 0.9$ for $h = 0.9, 1.0, 1.1$, respectively. Additionally, in cases (b)-(c) we have introduced a reflecting upper boundary in the iterative algorithm at $L = 3$ and $L = 2$, respectively. Insets: Mean absorption time, $\tau$, for different choices of the characteristic parameter. Grey bars come from the iterative algorithm, filled coloured bars come from the Monte Carlo method and, in cases (a)-(c), bars with diagonal lines represent the analytical solution.}
	\label{fig:QSD_Continuous}
\end{figure}

In Fig.~\ref{fig:QSD_Continuous}, we show the quasi-stationary density function for these examples, along with the corresponding mean times to extinction. As in the discrete case, the numerical results agree very well with the theoretical solutions, when the latter are available. This set of examples demonstrates that the numerical approaches are effective to estimate the quasi-stationary distribution and the absorption time for general continuous processes.

\subsection{Processes with multiple degrees of freedom}\label{sec:Examples2D}

We first consider the SIRS model of epidemic spread~\cite{NasellEpidemics2D}. In this model, individuals transition between susceptible (S), infected (I), and recovered (R) states via the reactions \makebox{$\text{S} + \text{I} \xrightarrow{\beta} 2\text{I}$}, $\text{I} \xrightarrow{\gamma} \text{R}$ and $\text{R} \xrightarrow{\mu} \text{S}$. The parameters $\beta$ and $\gamma$ represent the infection and recovery rates, respectively, and $\mu$ characterises the rate at which recovered individuals lose immunity and become susceptible again. We restrict our analysis to the case $\mu=\gamma$. The basic reproduction ratio is defined as $R = \beta / \gamma$, and we fix the timescale by setting $\gamma = 1$ (and hence also $\mu=1$). The state of the population is described by the pair $(m, n)$, where $m$ and $n$ denote the number of susceptible and infected individuals, respectively. The number of recovered individuals is then $N-n-m$, where $N$ is the total population size. The transition rates are
\begin{align}\label{eq:Rates_SIRS}
W[(m, n) \to (m-1, n+1)] &= R \frac{n m}{N}, \nonumber \\
W[(m, n) \to (m, n-1)] &= n, \nonumber \\
W[(m, n) \to (m+1, n)] &= N - n - m.
\end{align}
The dynamics halts completely only when the system reaches the state $(m=N,n=0)$, in which all individuals are susceptible. However, in any state of the form $(m, 0)$---with no infected individuals---no further infection events can occur, meaning the disease has gone extinct. Thus, while $(N,0)$ is the only truly absorbing state, all states $(m, 0)$ are absorbing with respect to the infected population and represent the end of epidemic activity.

We next consider a two-dimensional diffusion process in the $(x,y)$ plane with evolution equations
\begin{equation}\label{eq:2d_evolutionrandomwalk}
 \begin{cases}
 \dot{x} = -\dfrac{x}{r^2}D + \sqrt{2D} \xi_x(t), \\
 \\
 \dot{y} = -\dfrac{y}{r^2} D +\sqrt{2D} \xi_y(t),
 \end{cases}
\end{equation}
where $D > 0$ and $r = \sqrt{x^2 + y^2}$. We assume that the radial variable is confined to \makebox{$r \in [r_0, r_1]$}, with $r = r_0$ an artificial absorbing barrier and $r = r_1$ a reflecting barrier. The circular symmetry allows the quasi-stationary distribution to be obtained analytically (see Appendix~\ref{Appendix:QSD_2D_Brownian}).

\begin{figure}[h]
	\centering
	\includegraphics[width=0.49\linewidth]{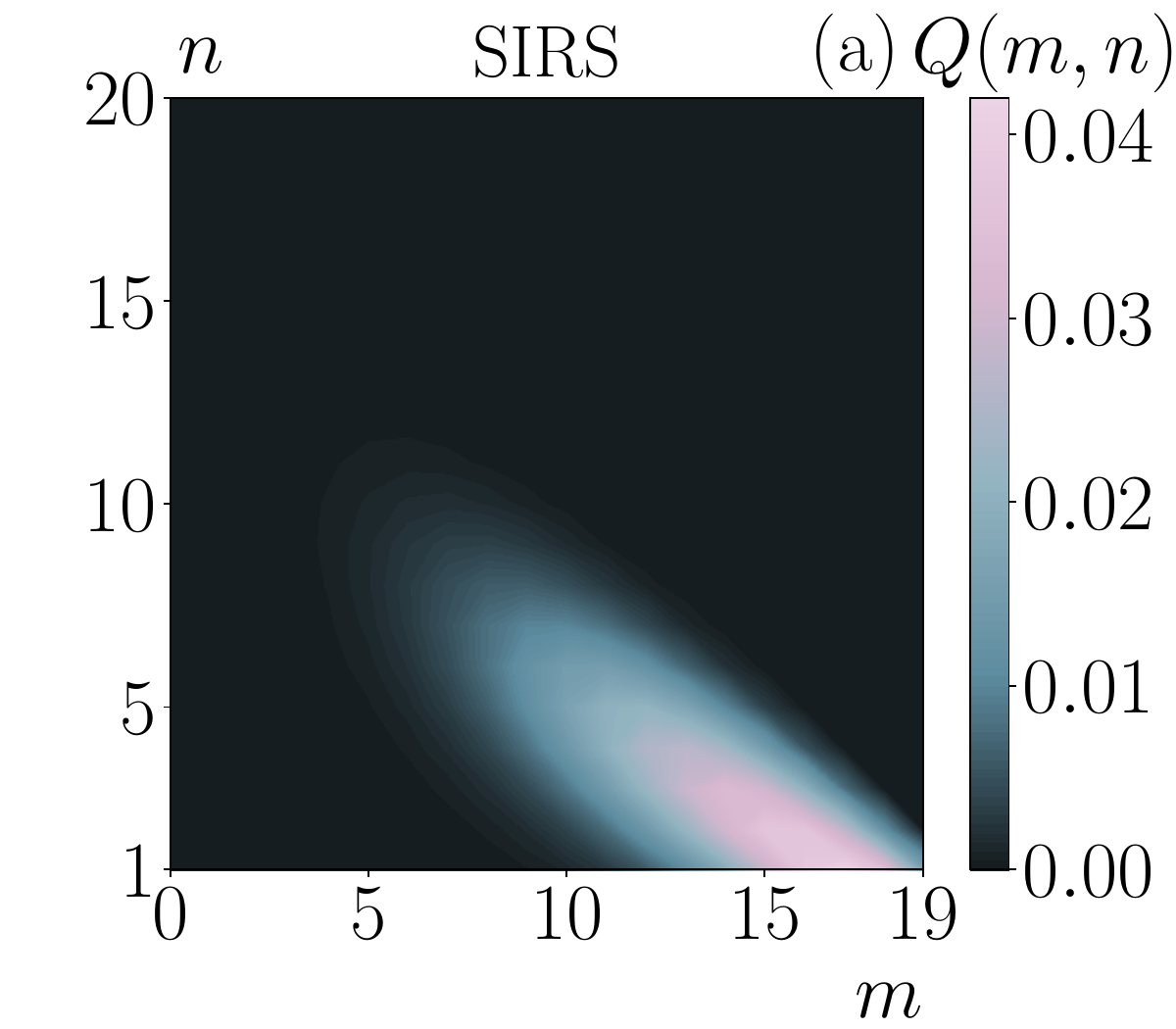}
 \includegraphics[width=0.49\linewidth]{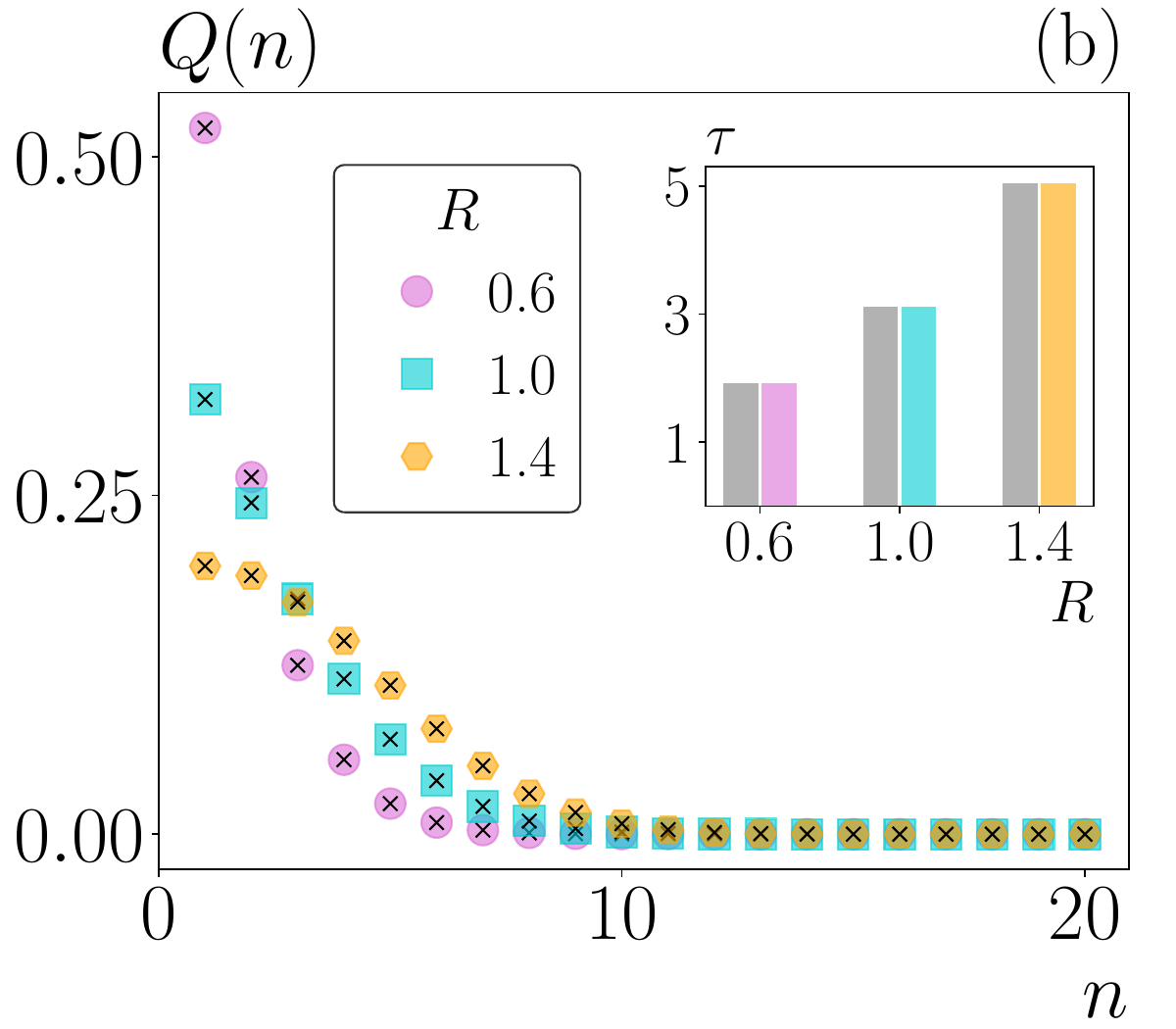}
 \includegraphics[width=0.49\linewidth]{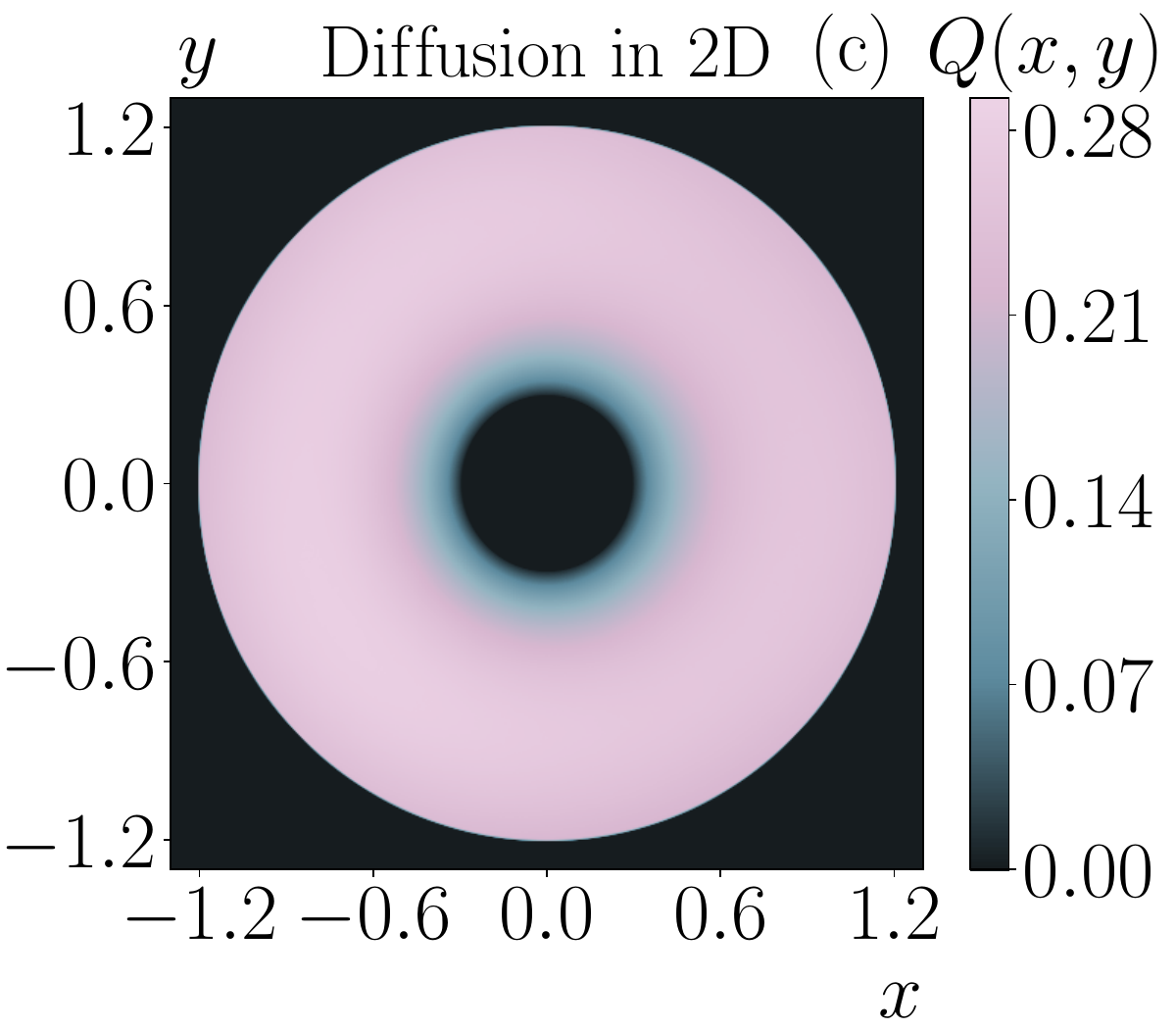}
 \includegraphics[width=0.49\linewidth]{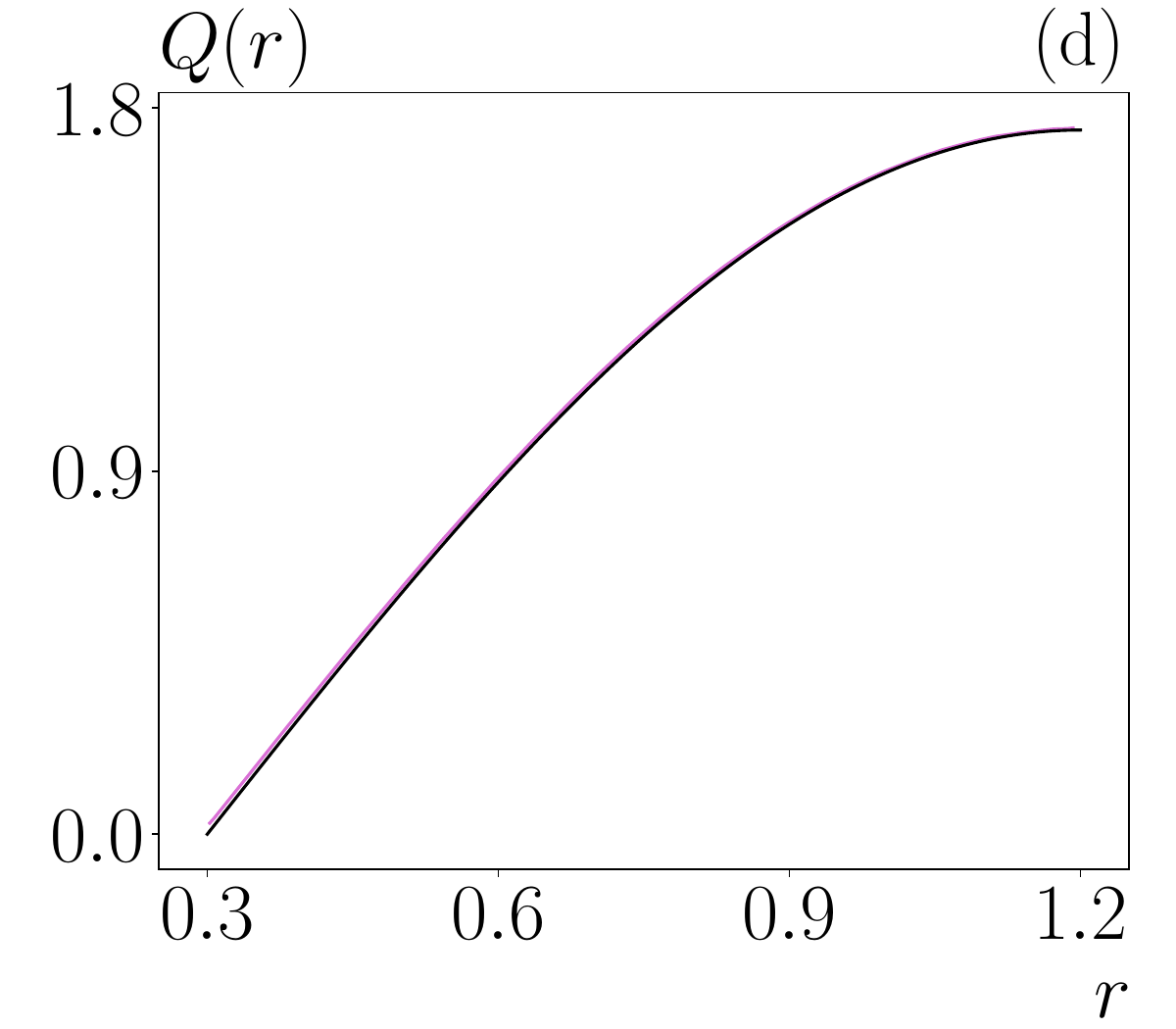}
	\caption{\textbf{Results for the two-dimensional processes in Sec.~\ref{sec:Examples2D}.} (a),(c) Quasi-stationary distribution obtained with the Monte Carlo method for: (a) the SIRS model (with $N = 20$ and $R = 1.4$) for \makebox{$K = 10^9$} steps and initial condition $(m_0, n_0) = \left(\frac{N}{R}, N \frac{R - 1}{2R} \right)$ (metastable state), and (c) the two-dimensional diffusion process with circular symmetry (with $D = 0.5$, $r_0 = 0.3$ and $r_1 = 1.2$) for $K = 10^{11}$ steps, discretisation parameters $\dx = \Delta y = 5 \cdot 10^{-3}$ and $\dt = 10^{-4}$, and initial condition $(x_0, y_0)$ randomly sampled within the annular domain. (b) Marginal quasi-stationary distribution of the number of infected individuals, $Q(n) = \sum_{m=1}^{N-1} Q(m,n)$ for $n>1$, for the SIRS model (with $N = 20$ and varying~$R$) computed using the iterative algorithm with a relaxation factor $s = 0.1$ (black crosses) and the Monte Carlo method (coloured symbols). The initial condition has been set to $(m_0, n_0) = (N-10, 10)$ for $R = 0.6, 1.0$ and identically as in (a) for $R = 1.4$. Inset: Mean time to extinction, defined as the average time the system takes to reach any absorbing state of the form $(m, 0)$ (i.e., with no infected individuals), as a function of $R$. (d) Radial distribution for the two-dimensional diffusion process with circular symmetry (with parameters as in (c)), obtained in the Monte Carlo simulation with $\Delta r = 10^{-3}$ (pink) and compared to the analytical solution (black).}
	\label{fig:QSD_2D}
\end{figure}

Panels (a) and (c) of Fig.~\ref{fig:QSD_2D} display the quasi-stationary distributions of the SIRS model and the two-dimensional diffusion process with circular symmetry, respectively, obtained via Monte Carlo simulations. In order to facilitate comparison of the numerical result for the diffusion process with the corresponding analytical prediction---and since direct comparison in the $(x,y)$ plane through contour plots can be visually challenging---in the simulation we also compute the quasi-stationary distribution of the radial variable. As shown in Fig.~\ref{fig:QSD_2D}(d), the numerical result closely matches the theoretical expression.

Unlike the Monte Carlo method, the iterative algorithm is difficult to apply to multi-dimensional processes with non-trivial boundary shapes. This is the case for the annular domain in the example of the two-dimensional diffusion process. In the SIRS model, in contrast, the simple boundaries allow for a feasible implementation (see Appendix~\ref{Appendix:Iterative_SIRS}). In Fig.~\ref{fig:QSD_2D}(b), we show the marginal quasi-stationary distribution of the number of infected individuals and the mean time to extinction of the SIRS model obtained with both the Monte Carlo method and the iterative algorithm. The results obtained from both methods are in agreement.

\section{Discussion of the methods}\label{sec:DiscussionMethods}

In this section, we compare the two numerical methods (i.e., the iterative approach and the single-trajectory Monte Carlo method), and discuss aspects of their implementation. In Sec.~\ref{sec:Comparison}, we analyse the efficiency and accuracy of the two numerical approaches for selected examples. In Sec.~\ref{sec:Iterative_RelaxationInitialCondition}, we investigate how the initial distribution and relaxation factor affect the convergence of the iterative algorithm. Finally, in Sec.~\ref{sec:discretisationParameters}, we examine the role of the discretisation parameters on the performance of the methods in problems with continuous state space. 

In Appendix~\ref{Appendix:SIScriticality}, we use the SIS model to investigate the effect of varying the system size on memory usage and computation time at criticality.

\subsection{Comparison of the methods}\label{sec:Comparison}

\subsubsection{Efficiency and accuracy}\label{sec:EfficiencyAccuracy}

\begin{figure}[h]
	\centering	
 \includegraphics[width=0.8\linewidth]{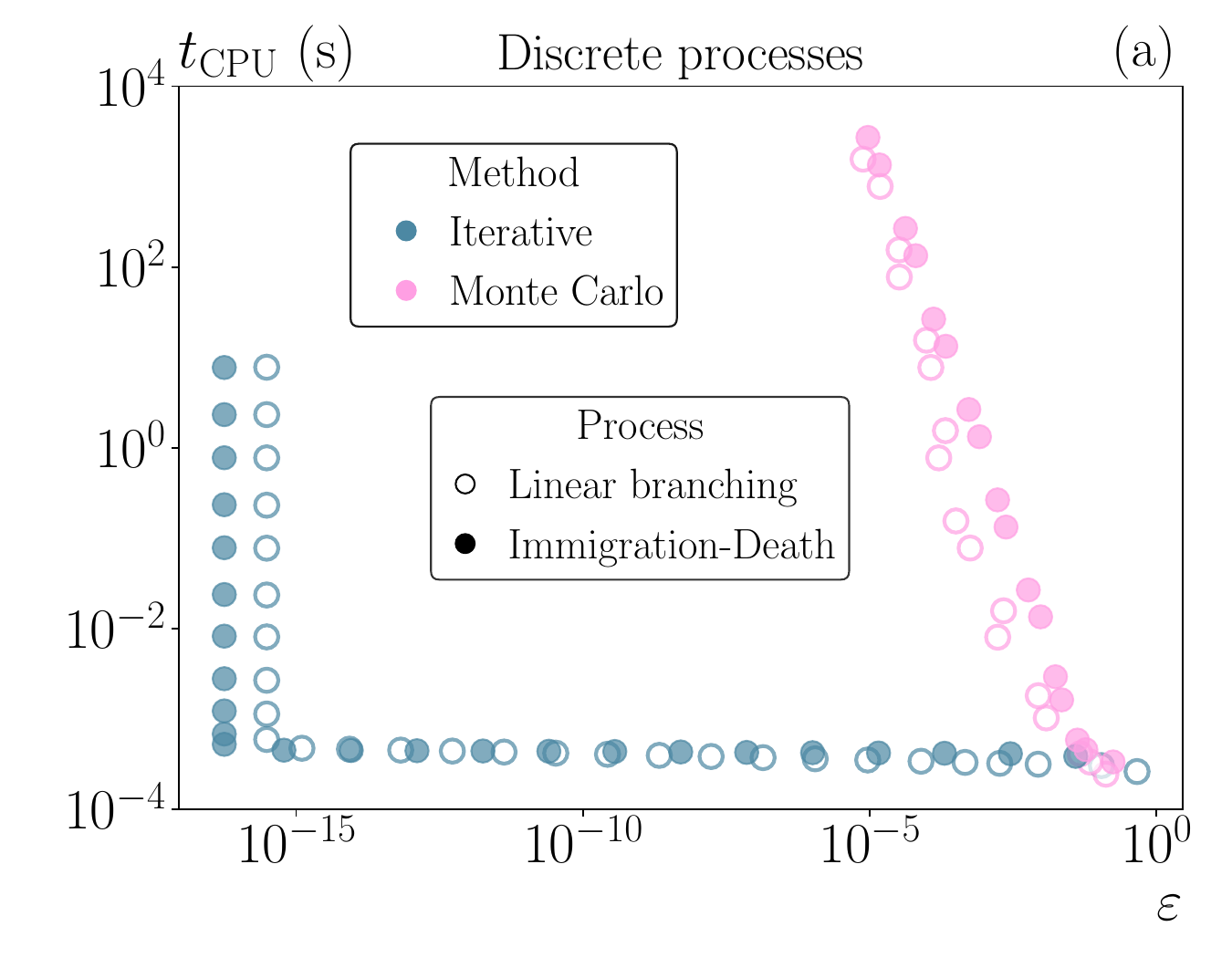}
 \includegraphics[width=0.8\linewidth]{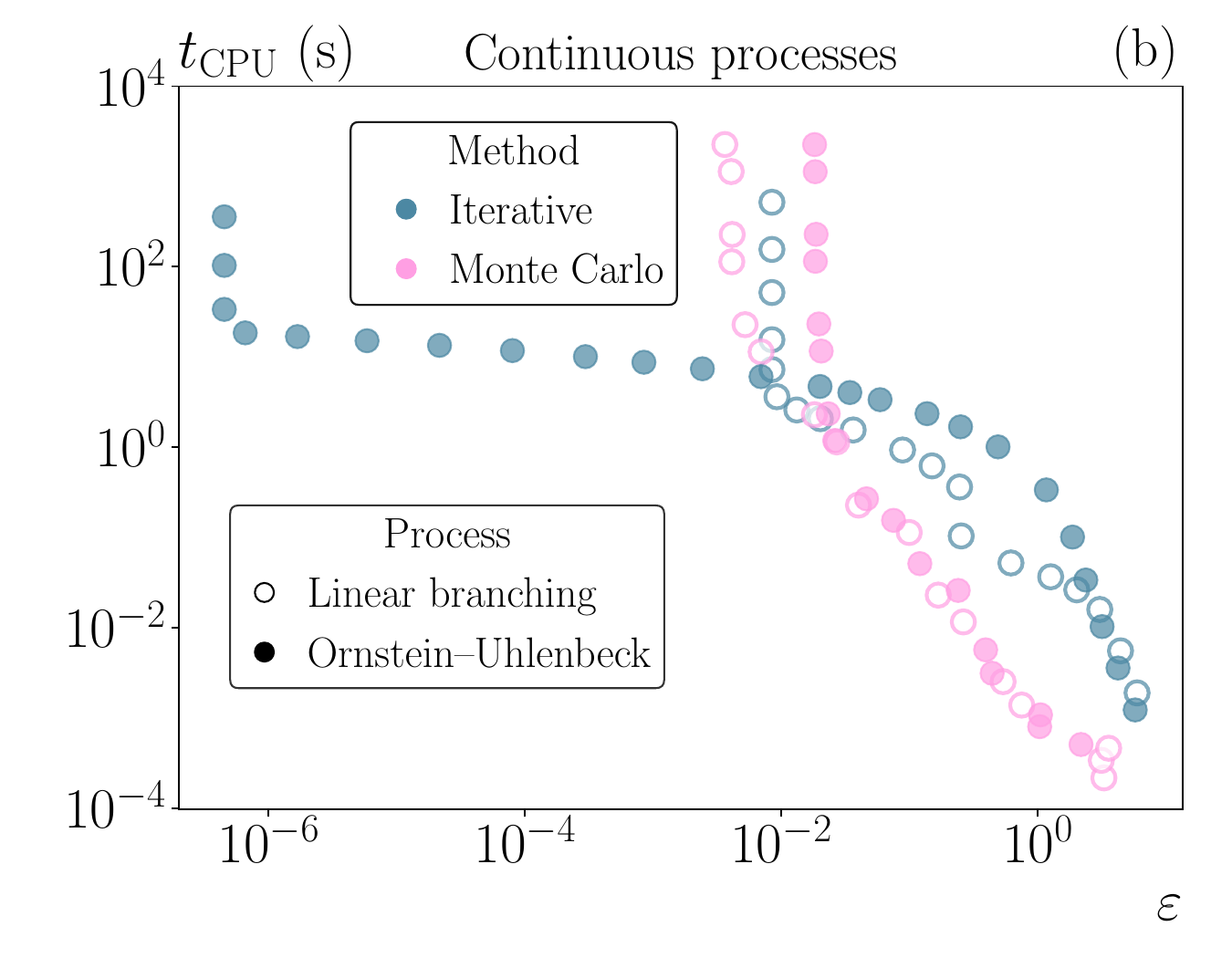}
	\caption{\textbf{Comparison of efficiency and accuracy of the numerical methods.} Computation time (measured in seconds) as a function of the total error in the numerical estimate of the quasi-stationary distribution. Blue and pink symbols come from the iterative algorithm and the Monte Carlo method, respectively. The initial conditions have been set as in Sec.~\ref{sec:Examples}. (a) Data for the discrete linear branching process (empty symbols) and the immigration-death process (filled symbols) described in Sec.~\ref{sec:ExamplesDiscrete}. In both cases we have set $R = 0.6$, used Eq.~\eqref{eq:TotalErrorDiscrete} to compute the errors, and set an upper bound at $N = 150$ in the iterative algorithm. (b) Results for the continuous linear branching process (empty symbols) for $D = 0.3$ and $f = 5$, and the Ornstein--Uhlenbeck process (filled symbols) for $D = 0.3$ and $f = 3.5$ (see Sec.~\ref{sec:ExamplesContinuous}). Errors have been calculated using Eq.~\eqref{eq:TotalErrorContinuous}. In the iterative algorithm, we have introduced a reflecting boundary condition at $L = 2$ for the linear branching process, and at $L = 3$ for the Ornstein--Uhlenbeck process. All simulations in this paper have been run on AMD Epyc processors (models 7402, 7282 and 9754).}
\label{fig:tcpuvserror}
\end{figure}

To compare the efficiency and accuracy of the iterative scheme and the Monte Carlo method, we focus on problems with known analytical quasi-stationary distribution. In Fig.~\ref{fig:tcpuvserror}, we plot the computation time required to achieve a given target error $\varepsilon$ in the quasi-stationary distribution.

Fig.~\ref{fig:tcpuvserror}(a) illustrates that, for the two models with discrete states and available analytical solutions (the linear branching process and the immigration-death process), the iterative algorithm outperforms the Monte Carlo method in both efficiency and accuracy, as it leads to smaller errors in significantly shorter computation times. In the Monte Carlo simulations, where we have performed up to $K = 10^{11}$ steps, the error achieved is on the order of $10^{-6}$ for both processes, as indicated by the upper end of the pink curves. In contrast, the error for the iterative algorithm reached saturation even as the allowed computing time was increased. This is indicated by the vertical parts of the curves for the iterative method (blue). The minimum error achievable with the iterative method is on the order of $10^{-16}$ for the linear branching process and $10^{-17}$ for the immigration-death process. In both cases, this level of accuracy has been achieved with fewer than 2000 iteration steps. These steps have been performed in a computation time approximately $10^5$ times shorter than that required for the Monte Carlo simulations.

We now turn to processes with continuous state space. In Fig.~\ref{fig:tcpuvserror}(b), we present the outcomes for the linear branching process and for the Ornstein--Uhlenbeck process. The latter also captures the behaviour observed for the continuous biased random walk, although we do not show the corresponding data to avoid an overcrowded figure. The Monte Carlo method, where $K = 10^{11}$ simulation steps have been performed, outperforms the iterative algorithm for large target errors ($\varepsilon \gtrsim 10^{-2}$), as seen by the fact that the data shown in pink is below that shown in blue. Conversely, when higher precision (smaller $\varepsilon$) is required, we find that the iterative algorithm is more efficient for the continuous biased random walk and the Ornstein--Uhlenbeck process. The linear branching process is the only case in which the iterative algorithm yields lower long-term accuracy than the Monte Carlo method. 

\subsubsection{Calculation of small probabilities}

An advantage of the iterative algorithm over the Monte Carlo method is that it allows calculating the probability of extremely unlikely states. Such states are rarely visited within achievable simulation times, making it impractical to estimate their probabilities via Monte Carlo methods. Consequently, the iterative algorithm is particularly well-suited for computing the mean time to extinction in regimes where absorption is a rare event, and for sampling rare trajectories~\cite{aguilar2022sampling}. 

\begin{figure}[h]
	\centering	\includegraphics[width=0.8\linewidth]{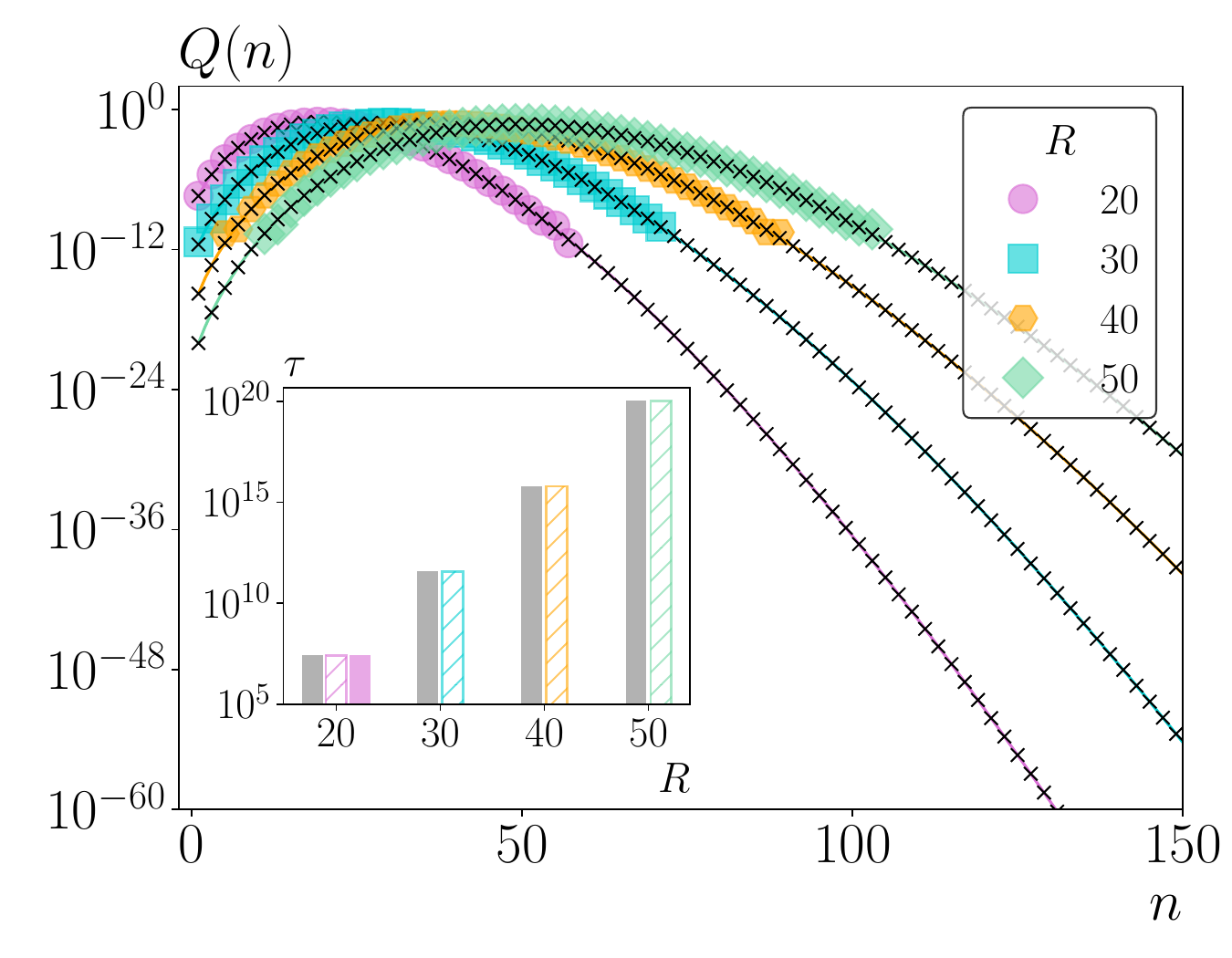}
	\caption{\textbf{Sampling of small probabilities.} Quasi-stationary distribution and mean time to extinction for the immigration-death process introduced in Eq.~\eqref{eq:Rates_Mix} for large values of the reproduction ratio~$R$. Main panel: Quasi-stationary distribution, obtained with the iterative algorithm (black crosses), the Monte Carlo method (coloured symbols), and the analytical solution (solid lines). For better visualisation, the numerical distribution is plotted only for odd values of $n$. In the numerical methods, we have set the initial condition at $n_0 = R$. Moreover, we have set $N = 500$ as the upper bound for the population size in the iterative algorithm. Inset: Mean time to extinction,~$\tau$, as a function of $R$. Results from the iterative algorithm are shown as grey bars, results from the Monte Carlo method as filled coloured bars, and the analytical solution as bars with diagonal lines. As noted in the text, results for the mean time to extinction can only be obtained from the Monte Carlo method within manageable computing time for $R=20$.}
	\label{fig:QSD_Discrete_MixLargeR}
\end{figure}

In Fig.~\ref{fig:QSD_Discrete_MixLargeR}, we show the quasi-stationary distribution and the mean time to extinction for the immigration-death process (see Sec.~\ref{sec:ExamplesDiscrete}) for values of the reproduction ratio larger than those considered in Fig.~\ref{fig:QSD_Discrete}. While the Monte Carlo method allows computing probabilities only down to approximately $10^{-12}$, the iterative algorithm reproduces analytical results down to probabilities as small as $10^{-60}$. Furthermore, the mean time to extinction could be estimated using the Monte Carlo method only for $R = 20$, as no absorption events occurred in the simulations for the other values of $R$. In contrast, all the average absorption times obtained with the iterative algorithm are in excellent agreement with the theoretical predictions.

\subsubsection{Convergence}\label{sec:convergence_issues}

When implemented with a suitable relaxation factor and initial distribution (see Sec.~\ref{sec:Iterative_RelaxationInitialCondition} for a detailed analysis), the iterative algorithm consistently exhibits fast convergence across all processes we considered. In contrast, the Monte Carlo method can suffer from intrinsically slow convergence in certain systems. To illustrate this phenomenon, we introduce a discrete biased random walk with an absorbing state at the origin, with transition rates
\begin{align}\label{eq:Rates_BRW}
W(n \to n + 1) &= \begin{cases}
R, & \text{for } n \geq 1,\\
0, & \text{for } n = 0,
\end{cases} \nonumber \\
W(n \to n - 1) &= \begin{cases}
1, & \text{for } n \geq 1,\\
0, & \text{for } n = 0.
\end{cases} 
\end{align}
As in previous examples, we have fixed the timescale by setting the probability of jumping to the left to one (for $n \geq 1$). The process exhibits a quasi-stationary regime only for $R < 1$, in which case the quasi-stationary distribution can be solved analytically (see Appendix~\ref{Appendix:QSD_Discrete_BRW}).

\begin{figure}[h]
	\centering
	\includegraphics[width=0.49\linewidth]{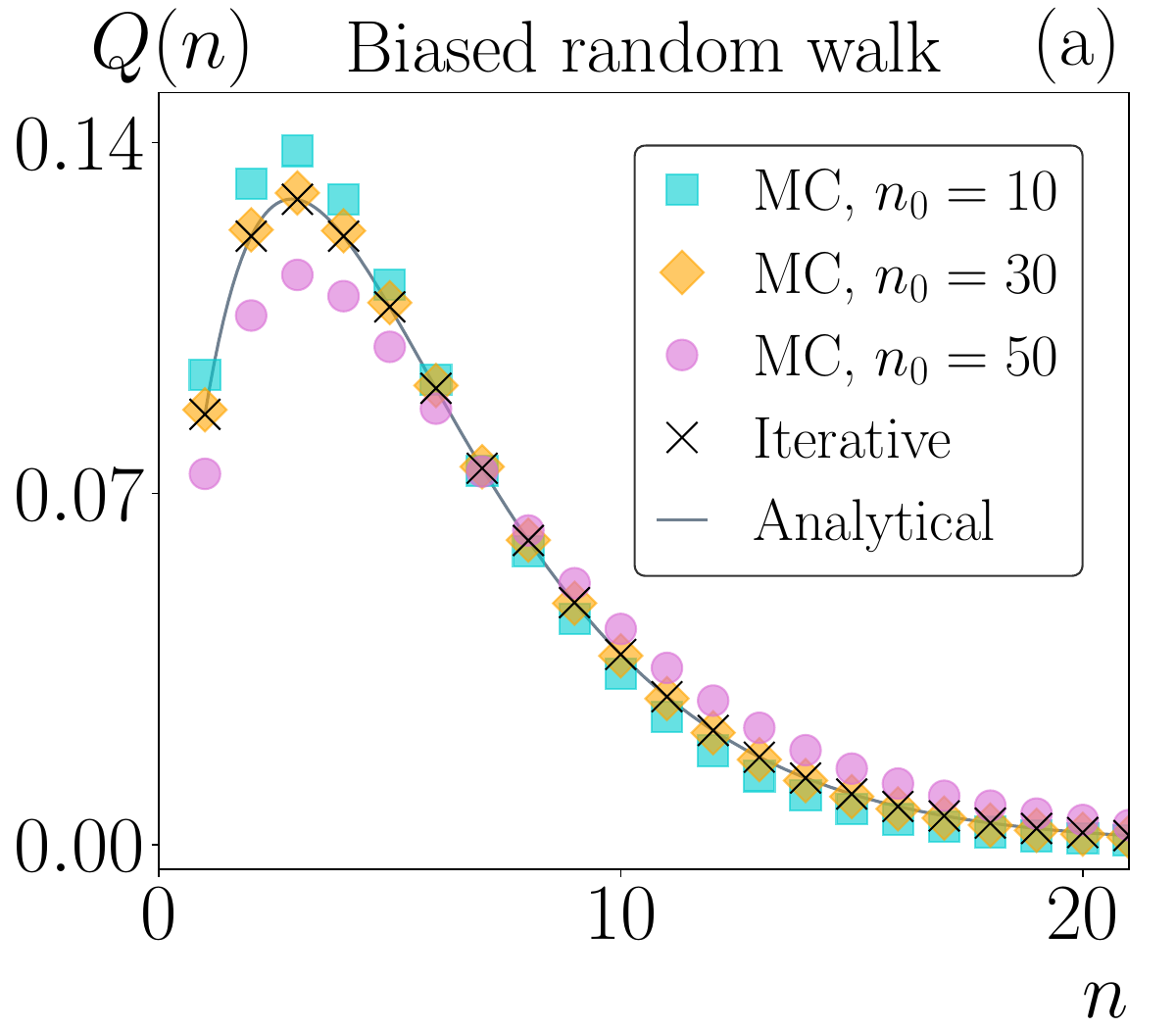}
 \includegraphics[width=0.49\linewidth]{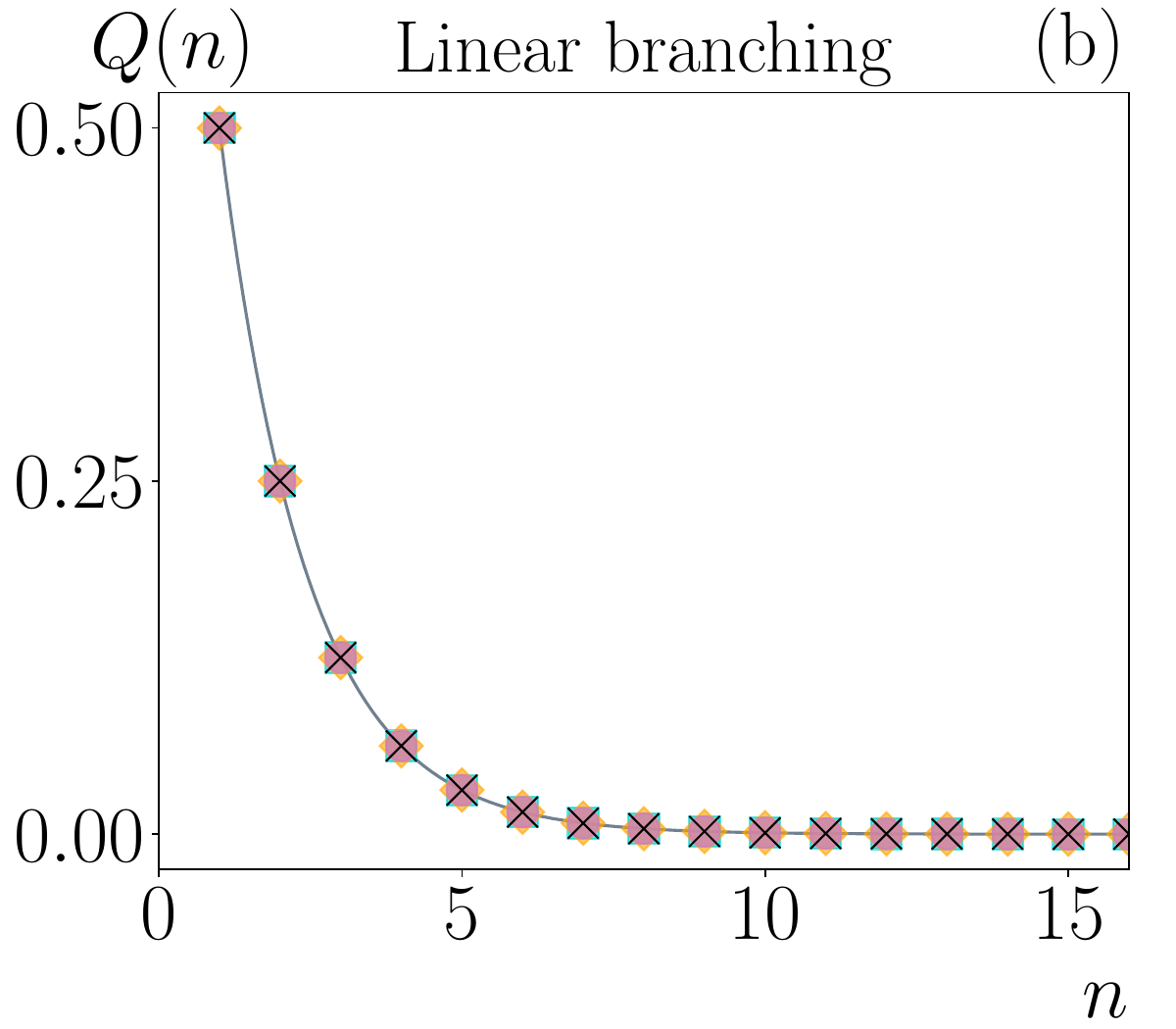}
 \includegraphics[width=0.49\linewidth]{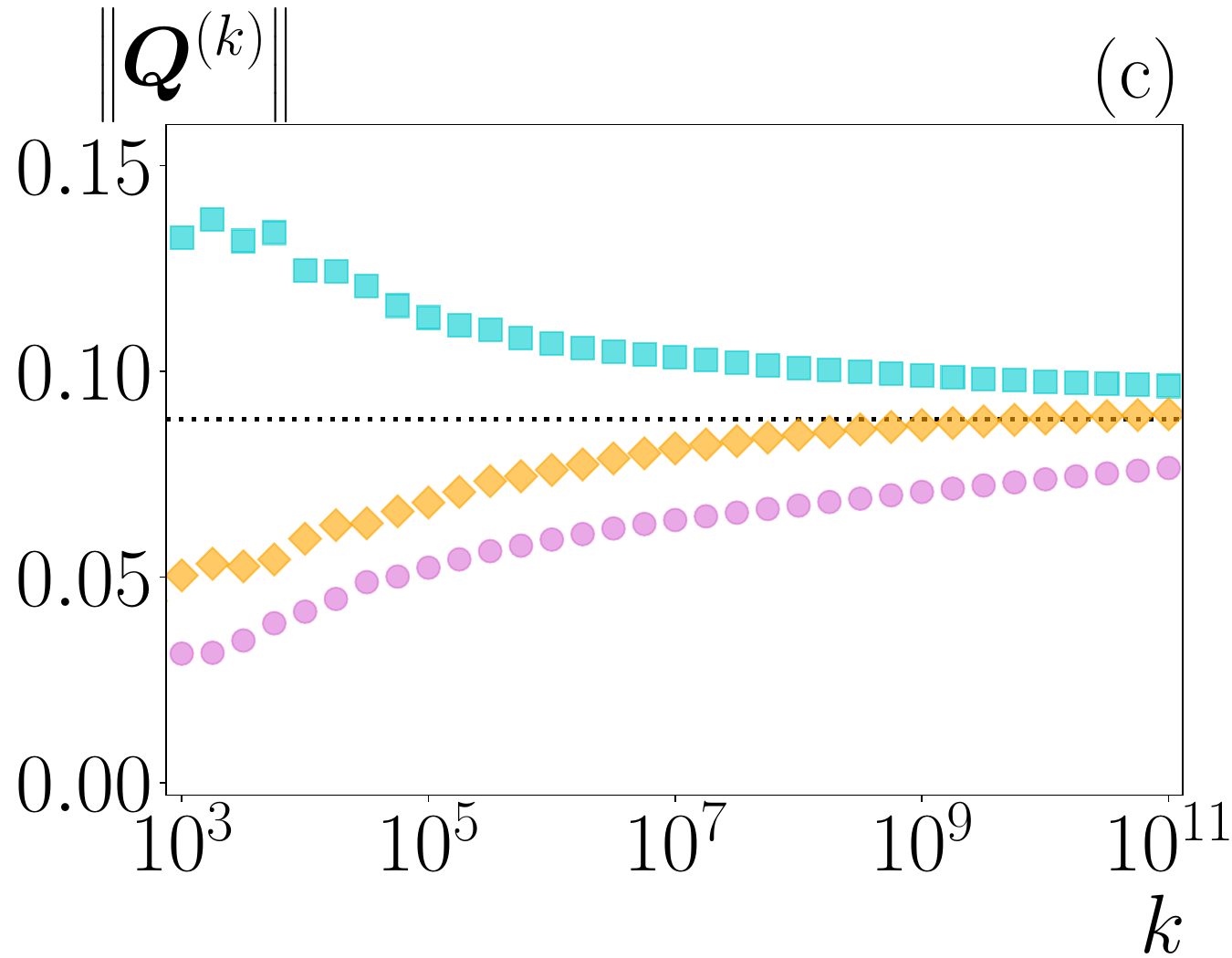}
 \includegraphics[width=0.49\linewidth]{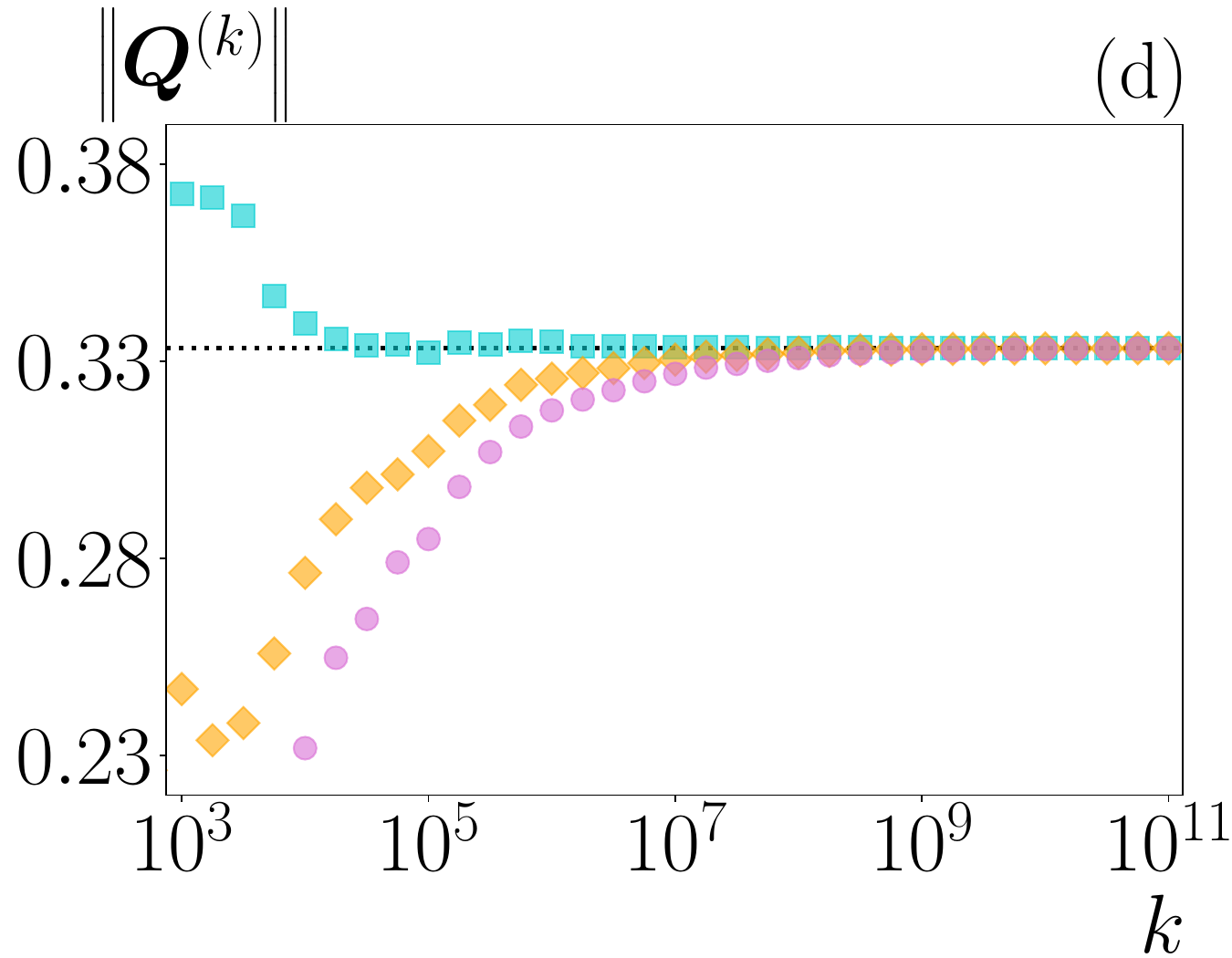}
	\caption{\textbf{Convergence issues in the Monte Carlo method.} (a),(b) Quasi-stationary distribution of the discrete biased random walk in Eq.~\eqref{eq:Rates_BRW} and the discrete linear branching process in Eq.~\eqref{eq:Rates_Branching}, respectively. Coloured symbols represent the outcomes from the Monte Carlo (MC) method with $K = 10^{11}$ simulation steps and three different initial conditions, black crosses correspond to the iterative algorithm with a relaxation factor $s = 0.1$, and solid lines represent the analytical solution. For both processes, we have set $R = 0.5$ and introduced an upper limit at $N = 500$ in the iterative algorithm. (c),(d) Convergence measure of the Monte Carlo method, $\|\bm{Q}^{(k)}\|$, as a function of the simulation step $k$ for the biased random walk and the linear branching process, respectively. Results are presented for the same initial conditions $n_0$ considered in panels (a),(b). Horizontal dotted lines indicate the theoretical value $\|\bm{Q}^{\rm exact}\|$.}
	\label{fig:MC_CIDependence}
\end{figure}

In Fig.~\ref{fig:MC_CIDependence}(a) and (b), we show the quasi-stationary distribution of the biased random walk and of the discrete linear branching process, respectively. We present the results obtained using the Monte Carlo method with three different initial conditions $n_0$, along with those from the iterative algorithm and the analytical solution. 
For both processes, using the initial condition $Q(n)^{(0)} = \delta_{n,n_0}$ in the iterative scheme yields identical outcomes for all three choices of $n_0$, and results for a single representative initial condition are shown. In the case of the linear branching process, which reflects the typical behaviour observed across all examples in Sec.~\ref{sec:Examples}, the results from the Monte Carlo method are also insensitive to the initial condition and closely match both the iterative and analytical solutions. For the biased random walk, in contrast, Monte Carlo results exhibit significant dependence on the initial condition $n_0$, indicating that the bias error intrinsic in this approach remains significant at the end of the simulation. The iterative algorithm, in contrast, yields results in excellent agreement with the analytical solution for all three initial conditions. 

We now analyse the evolution of the convergence measure $\|\bm{Q}^{(k)}\|$ of the Monte Carlo method as a function of $k$ for the three different initial conditions $n_0$. In Fig.~\ref{fig:MC_CIDependence}(c) and (d), we show the results for the biased random walk and the linear branching process, respectively. The results confirm that, by the end of the simulation, the Monte Carlo method has lost memory of the initial condition in the case of the linear branching process [panel (d)], but has not yet converged for the biased random walk [panel~(c)].

\subsection{Effect of relaxation factor and initial distribution on the iterative algorithm}\label{sec:Iterative_RelaxationInitialCondition}

We now examine the impact of the relaxation factor~$s$ and the initial distribution $\bm{Q}^{(0)}$ on the convergence behaviour of the iterative algorithm. We consider two distinct initial distributions: a uniform distribution, and a Kronecker delta distribution.

\begin{figure}[h!]
	\centering
\includegraphics[width=0.8\linewidth]{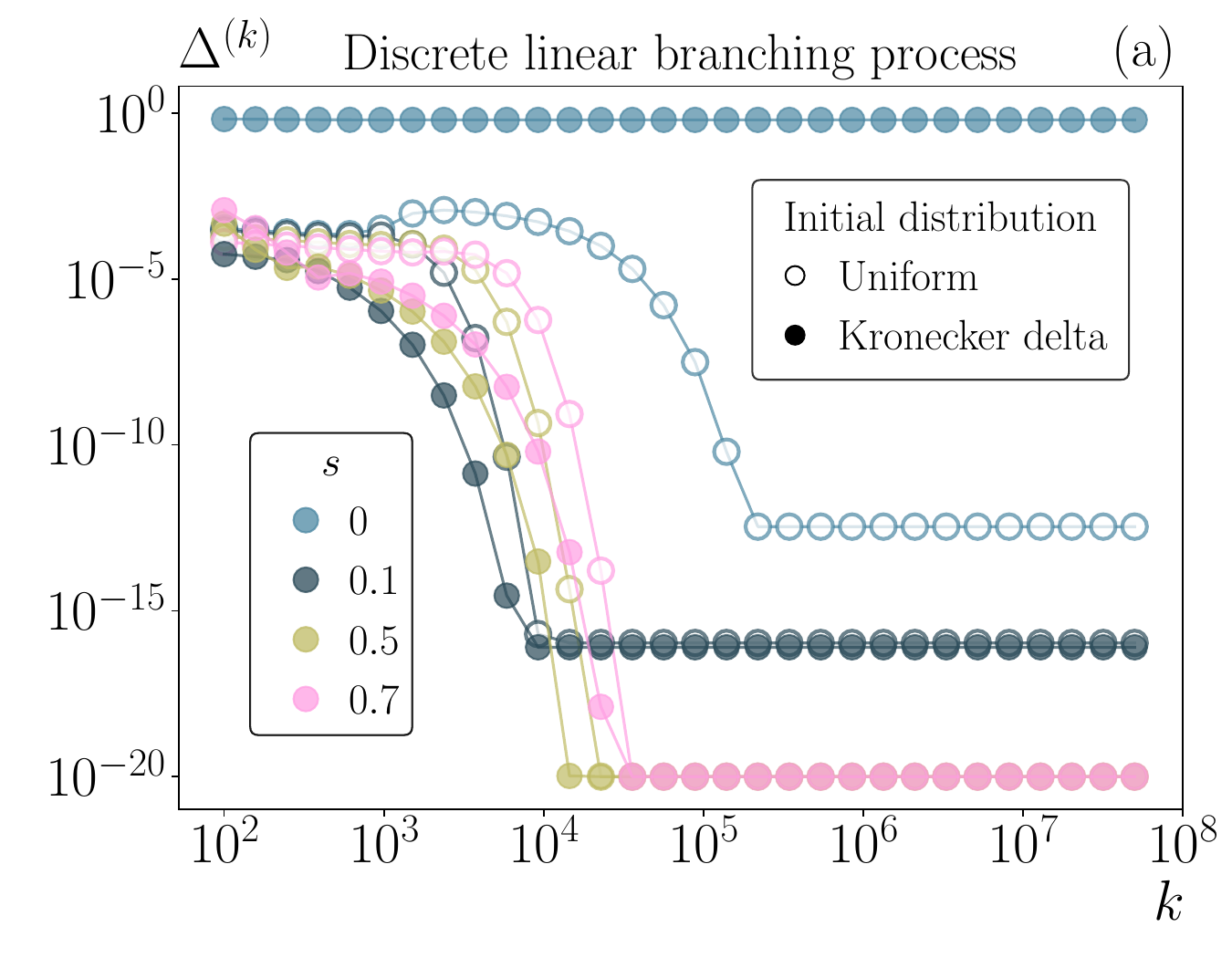}
\includegraphics[width=0.8\linewidth]{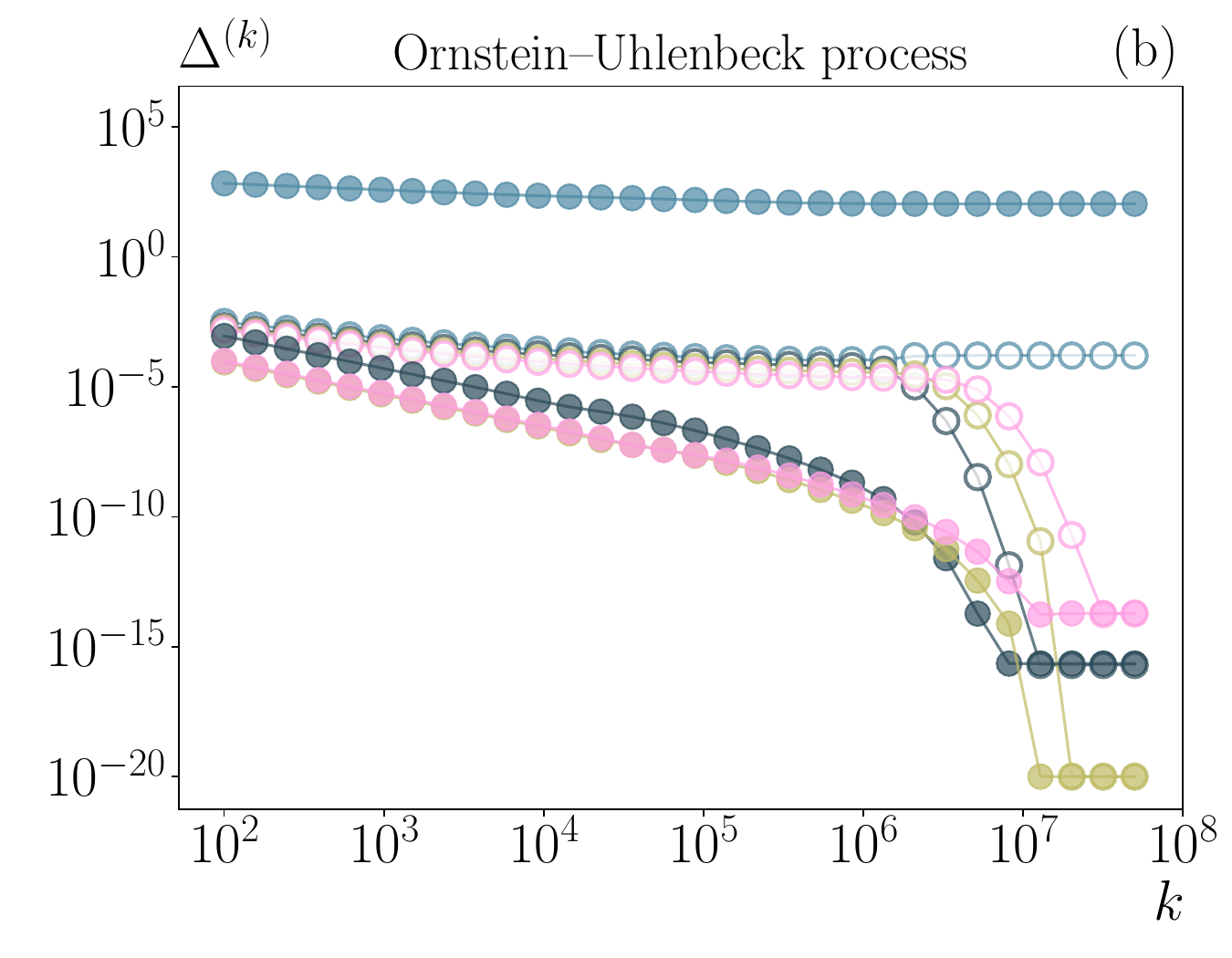}
	\caption{\textbf{Effect of $s$ and initial distribution in the iterative algorithm.} Convergence measure of the iterative algorithm, $\Delta^{(k)} \equiv \norm{\bm{Q}^{(k)} - \bm{Q}^{(k-1)}}$, as a function of the iteration step~$k$. We consider two initial distributions (a uniform distribution and a Kronecker delta distribution), and plot the results for four relaxation factors ($s = 0, 0.1, 0.5, 0.7$). The results correspond to: (a) the discrete linear branching process in Eq.~\eqref{eq:Rates_Branching} with a reproduction ratio $R = 0.8$ and an artificial upper limit at $N = 150$, and (b) the Ornstein--Uhlenbeck process in Eq.~\eqref{eq:Ornstein_DriftDiffusion} with $D = 0.3$ and $f = 3.5$, for which we have introduced a reflecting upper boundary at $L = 3$. In the latter case, the spatial step size has been set to $\dx = 5 \cdot 10^{-4}$. The initial Kronecker delta distribution has been centred at: (a) $n_0 = 10$ and (b) $x_0 = 0.1$. To avoid issues with zero values on the logarithmic scale, all values below the machine precision have been set to $10^{-20}$. For better visualisation, data points are connected with lines. }
\label{fig:Iterative_diff_versus_k}
\end{figure}

In Fig.~\ref{fig:Iterative_diff_versus_k}, we show the evolution of the convergence measure of the algorithm, denoted here by \makebox{$\Delta^{(k)} \equiv \norm{\bm{Q}^{(k+1)} - \bm{Q}^{(k)}}$}, as a function of the iteration step $k$, for the discrete linear branching process [panel (a)] and the Ornstein--Uhlenbeck process [panel (b)]. 

The data in Fig.~\ref{fig:Iterative_diff_versus_k} demonstrates that it is essential to incorporate over-relaxation, as setting $s = 0$ generally results in slow convergence and usually causes the norm~$\Delta^{(k)}$ to stabilise at non-negligible values (blue symbols in the figure). In particular, the use of $s = 0$ and a Kronecker delta distribution as initial condition can lead to pathological distributions, as discussed in Appendix~\ref{Appendix:IterativeQSDOscillations}. For all the values $s > 0$ considered in the analysis, in contrast, the norm $\Delta^{(k)}$ converges to a constant value either smaller than the machine precision, or typically smaller than $10^{-13}$. In this latter case, the algorithm enters a period-$2$ cycle in which the probabilities $Q^{(k)}(n)$ (or $Q_i^{(k)}$, if the state space is continuous) at consecutive iterations alternate between two values that differ only slightly. Despite these small oscillations, the algorithm still provides a reliable numerical estimate of the quasi-stationary distribution.

While some degree of over-relaxation improves performance, excessively large values of $s$ also lead to slow convergence, as can be seen in Fig.~\ref{fig:Iterative_diff_versus_k}. Among the relaxation factors tested ($s=0,0.1,0.2,\dots 0.9$), $s = 0.1$ has consistently yielded the fastest convergence across all processes examined in this work. Efficiency can be further enhanced using a Kronecker delta distribution, as illustrated by the two representative cases shown in the figure. While uniform and Kronecker delta distributions represent extreme choices for the initial distribution, we found that using intermediate options, such as Gaussian distributions, did not yield any significant improvement in convergence over the Kronecker delta initialization.

\subsection{Impact of discretisation parameters in continuous processes}\label{sec:discretisationParameters}

In this section, we examine how discretisation parameters influence the performance of the two numerical methods in the continuous setting. For this study, we consider processes with available analytical quasi-stationary distribution, for which we can calculate the total error in the numerical estimates. In the numerical methods, we set the same initial condition $x_0$ as in Sec.~\ref{sec:ExamplesContinuous} for all processes.

\subsubsection{Impact of $\dx$ in the iterative algorithm}\label{sec:Iterative_dx}

First, we examine the effect of the spatial step size $\dx$ in the performance of the iterative algorithm described in Sec.~\ref{sec:IterativeContinuous}. For this analysis, we set the relaxation factor to $s = 0.1$.

\begin{figure}[h]
	\centering
\includegraphics[width=0.8\linewidth]{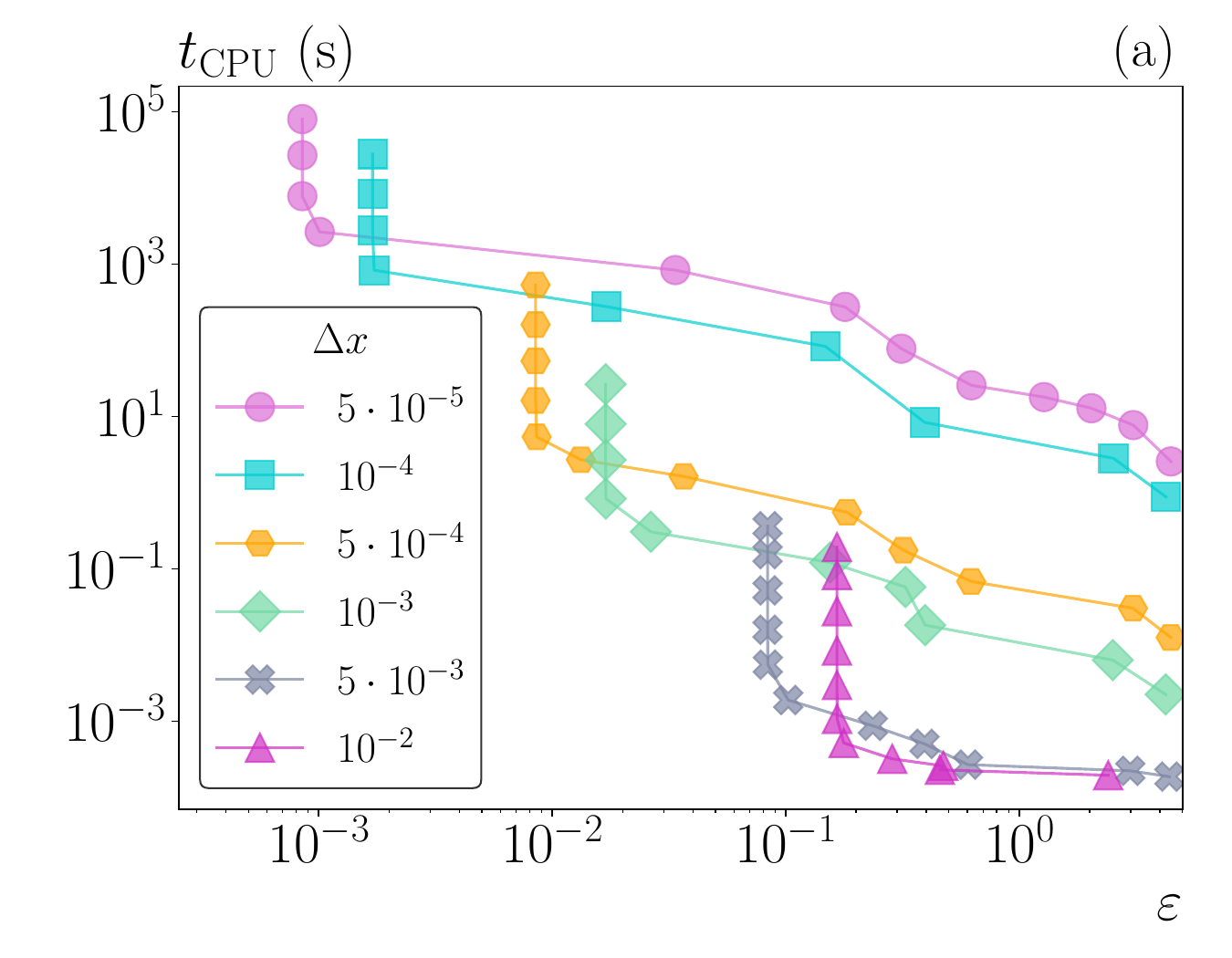}
\includegraphics[width=0.8\linewidth]{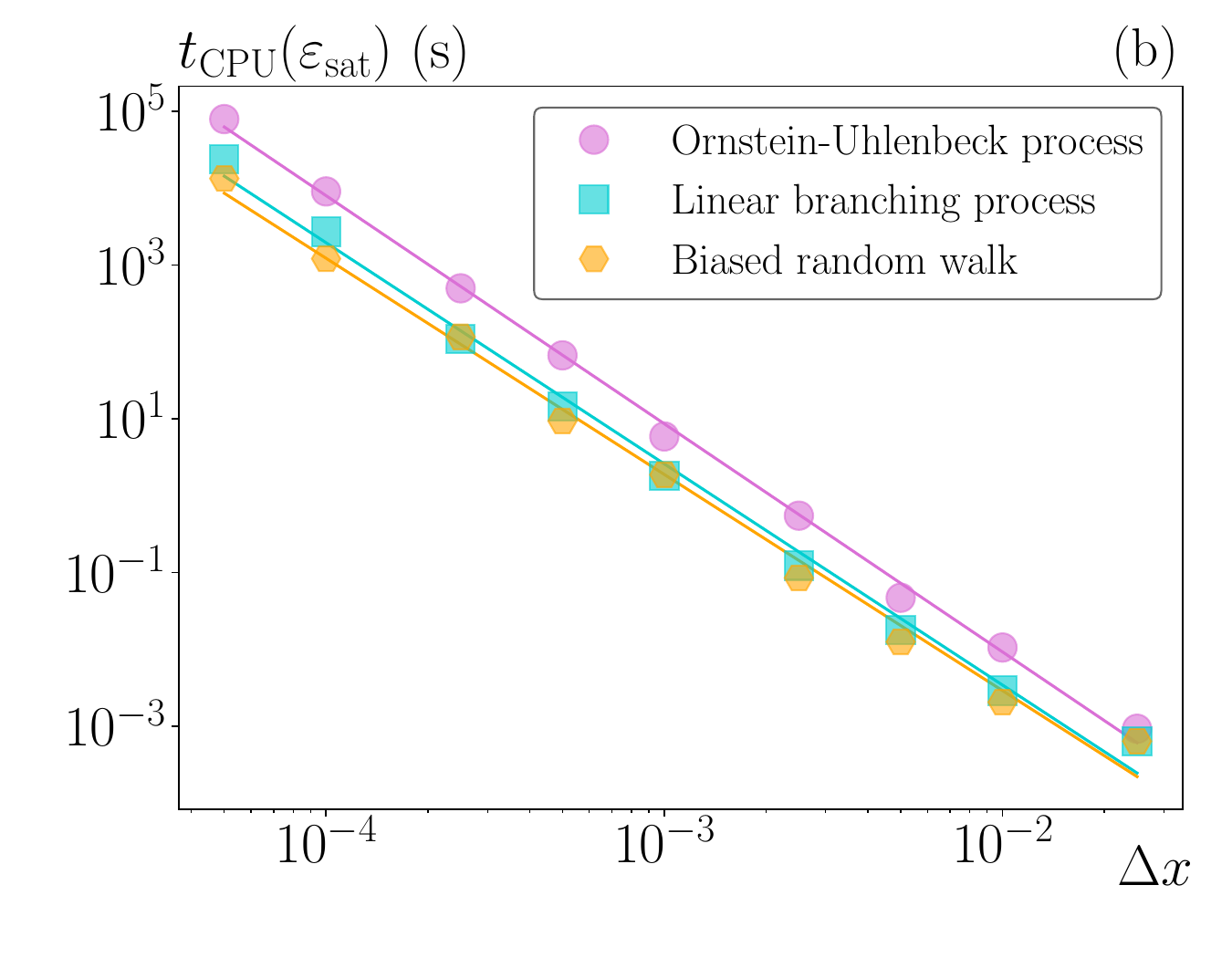}
\caption{\textbf{Impact of $\dx$ on efficiency and accuracy in the iterative algorithm.} (a) Computation time (measured in seconds) as a function of the total error in the numerical estimate of the quasi-stationary distribution, given by Eq.~\eqref{eq:TotalErrorContinuous}. Data points are connected by lines for visual clarity. The results correspond to the continuous linear branching process described in Eq.~\eqref{eq:Branching_DriftDiffusion} with $D = 0.3$ and $f = 5$, for which we have introduced a reflecting upper bound at $L = 2$. (b) Computation time (measured in seconds) required to reach the saturation error as a function of the spatial step $\dx$. Symbols correspond to the data obtained for the Ornstein--Uhlenbeck process in Eq.~\eqref{eq:Ornstein_DriftDiffusion} with $D = 0.3$, $f = 3.5$ and a reflecting boundary at $L = 3$ (pink circles), the linear branching process with same parameters as in (a) (blue squares), and the biased random walk in Eq.~\eqref{eq:BRW_DriftDiffusion} with $D = 0.4$, $f = 2$ and $L = 1$ (orange hexagons). Lines represent the fit $\tcpu (\esat) = b \cdot \dx^a$, yielding values $a \sim -3$ and $b \sim 10^{-8}$ in all cases.}
\label{fig:Iterative_dx_effect}
\end{figure}

In Fig.~\ref{fig:Iterative_dx_effect}(a), we show the computation time, $\tcpu$, required to achieve a target error $\varepsilon$ in the numerical estimate of the quasi-stationary distribution for several choices of the discretisation parameter $\dx$. For any given value of $\dx$, the error eventually saturates even as more computation time is allowed, indicating that the algorithm has reached its maximum possible accuracy. Although we only present results for the continuous linear branching process, all processes examined in this work exhibit similar behaviour.

Next, we analyse how the computation time required to reach the saturation error depends on $\dx$. We define $\tcpu(\esat)$ as the computation time required for $\norm{\bm{Q}^{(k+1)}-\bm{Q}^{(k)}}$ to become less than $10^{-13}$. In Fig.~\ref{fig:Iterative_dx_effect}(b) we show that, for all continuous processes considered in this paper, $\tcpu(\esat)$ follows a power-law relationship with the spatial step $\dx$. These results highlight the importance of choosing a spatial step $\dx$ that strikes a good balance between computational efficiency and accuracy. We have found $\dx = 5 \cdot 10^{-4}$ to be a suitable choice for the models we have studied.

\subsubsection{Impact of $\dx$ and $\dt$ in the Monte Carlo method}\label{sec:MonteCarlo_dxdt}

We now study how the step sizes $\dx$ and $\dt$ influence the performance of the Monte Carlo resetting method.

\begin{figure}[h!]
	\centering
\includegraphics[width=0.8\linewidth]{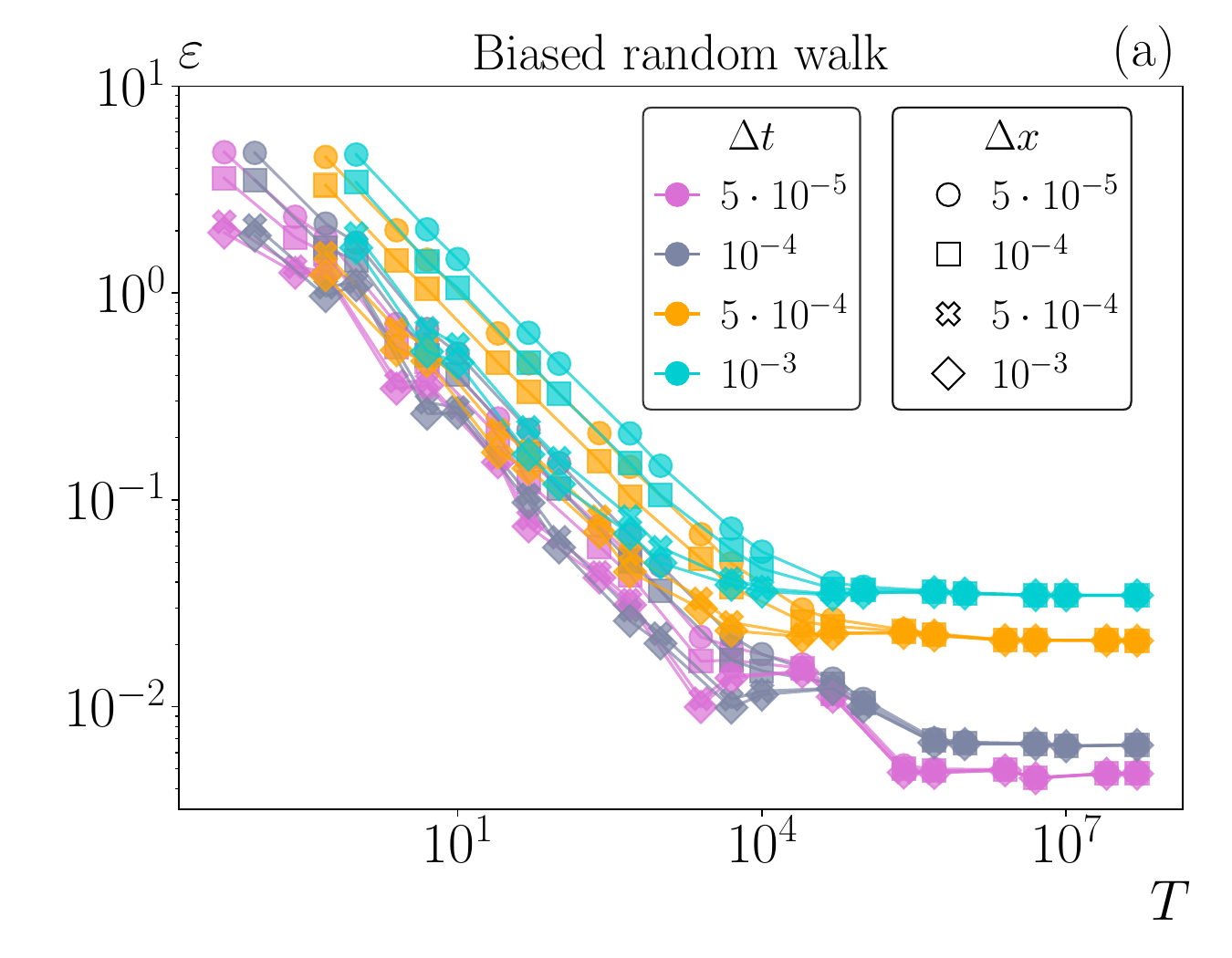}
\includegraphics[width=0.8\linewidth]{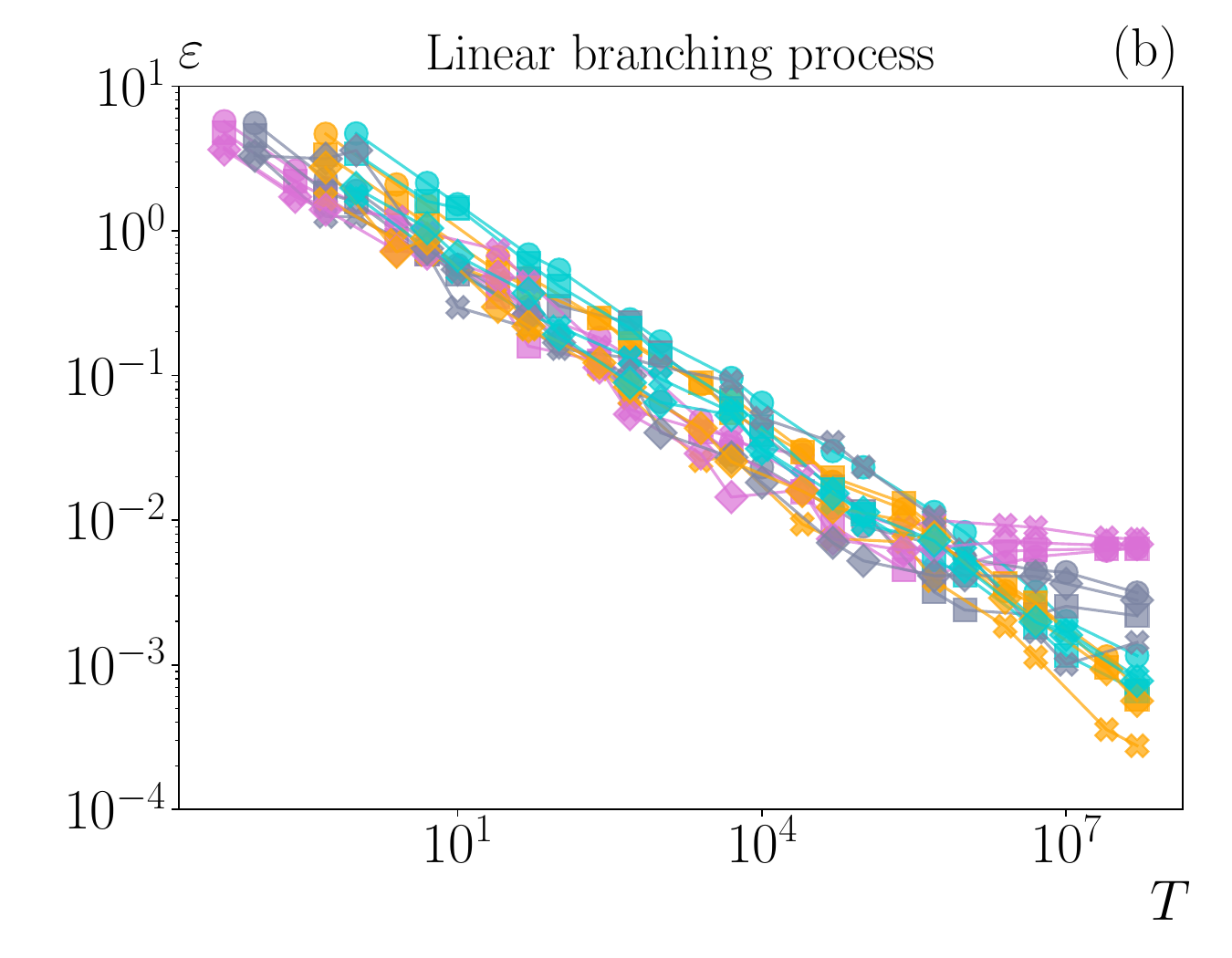}
	\caption{\textbf{Impact of $\dx$ and $\dt$ on error in the Monte Carlo method.} In the Monte Carlo resetting method, total error in the numerical estimate of the quasi-stationary distribution, given by Eq.~\eqref{eq:TotalErrorContinuous}, as a function of the simulation end time, $T$, for several values of the spatial and time discretisation steps, $\dx$ and $\dt$, respectively. Data points have been connected with lines for better readability. Results correspond to: (a) the biased random walk described in Eq.~\eqref{eq:BRW_DriftDiffusion} for $D = 0.4$, $f = 2$ and $L = 1$, and (b) the linear branching process introduced in Eq.~\eqref{eq:Branching_DriftDiffusion} for $D = 0.3$ and $f = 5$.}
\label{fig:MC_errorvsdxdt}
\end{figure}

In Fig.~\ref{fig:MC_errorvsdxdt}, we show the total error in the numerical estimate of the quasi-stationary distribution, $\varepsilon$, as a function of the simulation end time $T$. We present results for the biased random walk and the linear branching process, and for several combinations of discretisation parameters. In all cases, the error decreases with $T$ until it eventually reaches a saturation value. In the saturation regime, the total error is mostly determined by $\dt$, with the influence of $\dx$ becoming almost negligible. We note that the dependence of the saturation error on the discretisation parameter $\dt$ is not systematic, as smaller time steps do not necessarily lead to lower saturation errors. For example, the choice of $\dt = 5 \cdot 10^{-5}$ (the smallest value we considered in the analysis) yields the lowest saturation error in the biased random walk [panel (a)], but the highest in the linear branching process [panel (b)]. Based on these findings, it is not possible to recommend a single optimal value of $\dt$. Nevertheless, we suggest choosing $\dt = 10^{-4}$, as it consistently results in low to moderate errors across all processes we looked at.

\begin{figure}[h!]
	\centering
\includegraphics[width=0.8\linewidth]{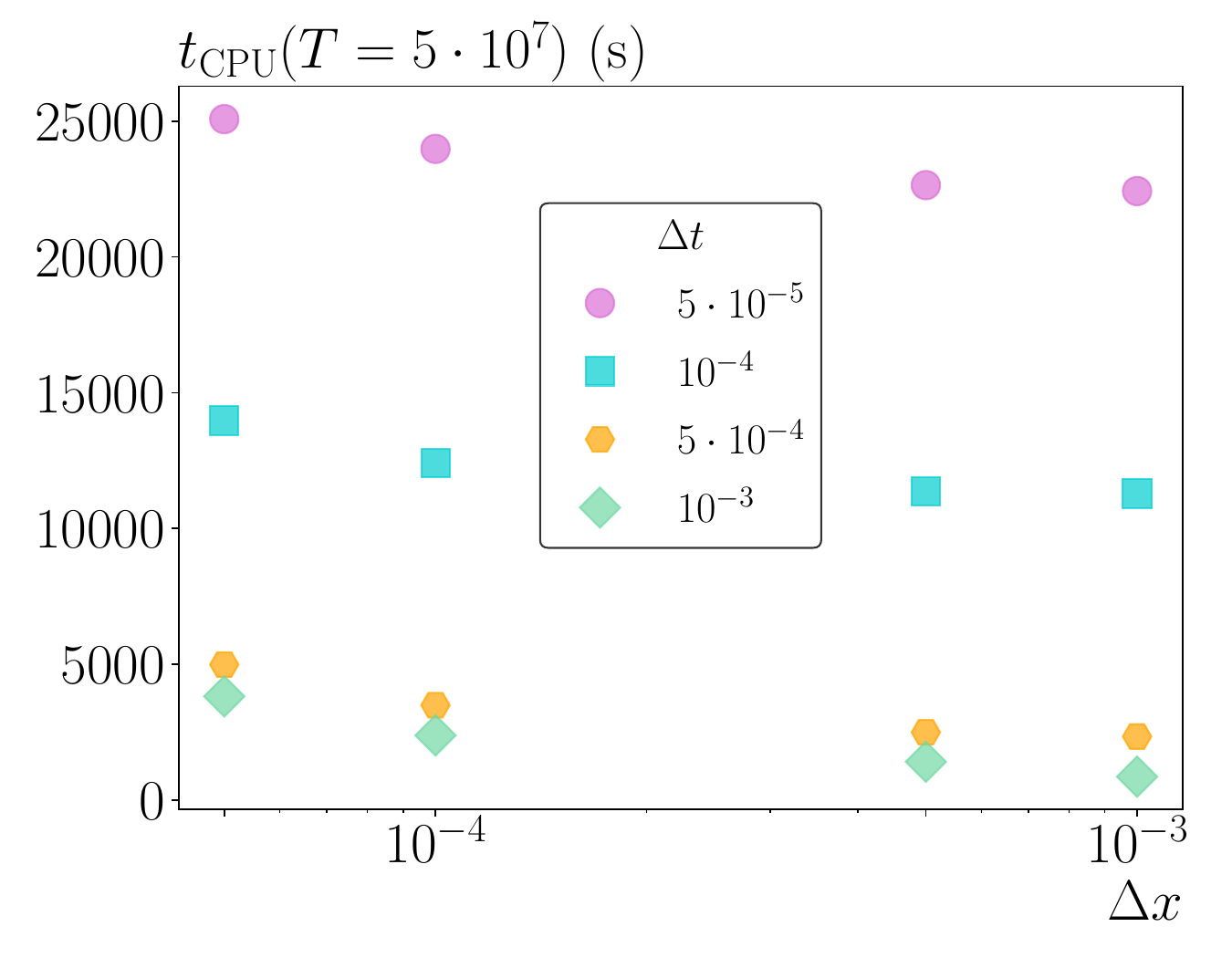}
	\caption{\textbf{Impact of $\dx$ and $\dt$ on computation time in the Monte Carlo method.} In the Monte Carlo method, computation time (measured in seconds) required to reach an end time $T = 5 \cdot 10^{7}$, shown as a function of $\dx$ for different values of $\dt$. The results correspond to the branching process described in Eq.~\eqref{eq:Branching_DriftDiffusion} with same parameters as in Fig.~\ref{fig:MC_errorvsdxdt}.}
\label{fig:MC_tcpuvsdxdt_Branching}
\end{figure}

In Fig.~\ref{fig:MC_tcpuvsdxdt_Branching}, we present the computation time required to reach a simulation end time $T = 5 \cdot 10^7$ as a function of $\dx$, and for several values of $\dt$. We only show the results for the linear branching process, since it reflects the general behaviour in all the processes we studied. A reduction in $\dt$ leads to a substantial increase in computation time, as more simulation steps $K$ are needed to reach the final time $T$. Although its effect is less pronounced, decreasing $\dx$ also increases computation time. Since the number of components in the numerical distribution $\bm{Q}^{(k)}$ increases, more computation time is needed to sample the interval to which the process is reset after absorption.

Since reducing $\dx$ has a negligible effect on the saturation error but increases computation time, we have set $\dx = 10^{-3}$ in the numerical simulations.

\section{Conclusions}\label{sec:Conclusions}

In this work, we have revisited two well-known numerical methods for computing quasi-stationary distributions: an iterative algorithm to solve the non-linear equation that defines the quasi-stationary distribution, and a Monte Carlo simulation method with resetting. Building on prior literature, we have extended the iterative algorithm to handle general Markov processes, with both discrete and continuous state spaces, and without restricting the number of absorbing states. For the Monte Carlo method, we have introduced an approach that simulates a single trajectory and uses this trajectory's history to reset the process after absorption. We have applied both numerical methods to a variety of stochastic models with absorbing states. 

We find that the iterative algorithm generally outperforms the Monte Carlo approach by achieving higher accuracy in significantly shorter computation times. Moreover, the iterative algorithm can compute the probabilities of extremely unlikely states and, consequently, estimate the mean time to extinction in cases where absorption is a rare event. We have also shown that the Monte Carlo method may exhibit convergence issues in certain systems due to the bias error inherent to the method. We have not observed this limitation in the iterative algorithm.

Among the models we have studied, the continuous linear branching process in Eq.~\eqref{eq:Branching_DriftDiffusion} is the only model for which the Monte Carlo method has proved more efficient than the iterative algorithm.  Additionally, the Monte Carlo approach appears to be the better choice for problems with multiple degrees of freedom and complex boundaries.

Finally, we have examined how the free parameters of each numerical method influence their performance. In the iterative algorithm, convergence is significantly accelerated by introducing a small amount of over-relaxation (e.g., $s = 0.1$) and using a Kronecker delta distribution centred near a non-absorbing, likely state of the unconditioned dynamics as initial condition. In the continuous setting, choosing an intermediate spatial step size (e.g., $\dx = 5 \cdot 10^{-4}$) is essential to balance accuracy and computational cost. In the Monte Carlo method, in contrast, the choice of $\dx$ has a negligible effect on accuracy, provided it is chosen reasonably. We therefore recommend setting a larger spatial step (e.g., $\dx = 10^{-3}$) to minimise computation time. Although reducing the time step $\dt$ does not necessarily improve accuracy and no single optimal $\dt$ exists, we find $\dt = 10^{-4}$ to be an effective choice.

The computer codes used in this work are publicly available in the GitHub repository~\cite{NumericalMethodsQSD}, and ready to be applied or adapted to a range of stochastic models with absorbing states.

\begin{acknowledgments}
Partial financial support has been received from Grants PID2021-122256NB-C21/C22 and PID2024-157493NB-C21/C22 funded by MICIU/AEI/10.13039/501100011033 and by “ERDF/EU”, and the María de Maeztu Program for units of Excellence in R\&D, grant CEX2021-001164-M. S.O-B. acknowledges support from the Spanish Ministry of Education and Professional Training under the grant FPU21/04997.
\end{acknowledgments}

\appendix
\makeatletter
\renewcommand{\thefigure}{\Alph{section}\arabic{figure}}
\renewcommand{\thetable}{A\arabic{table}} 
\@addtoreset{figure}{section}
\@addtoreset{table}{section}
\makeatother
\newpage

\begin{widetext}
\section{Summary of the stochastic processes considered in our work}\label{Appendix:SummaryProcesses}

In the following table, we summarise the main characteristics of the stochastic processes used throughout our study:

\begin{table}[ht]
\centering
\renewcommand{\arraystretch}{1.3}
\begin{tabularx}{\textwidth}{c c c c c c c c}
\hline
Process & Eq. & Type & Dim. & Absorbing states & Upper limit & Exact solution \\
\hline
Linear branching & \eqref{eq:Rates_Branching} & Discrete & 1 & $n=0$ & --- & Eq.~\eqref{eq:QSD_Branching} \\
Immigration-death & \eqref{eq:Rates_Mix} & Discrete & 1 & $n=0$ & --- & Eq.~\eqref{eq:QSD_Mix_app} \\
Branching-annihilation-decay & \eqref{eq:Rates_BAD} & Discrete & 1 & $n=0$ & --- & --- \\
Biased voter model & \eqref{eq:Rates_VM} & Discrete & 1 & $n=0$, $n=N$ & $n = N$ (absorbing) & --- \\
Biased random walk & \eqref{eq:Rates_BRW} & Discrete & 1 & $n=0$ & --- & Eq.~\eqref{eq:QSD_BRW} \\
SIS & \eqref{eq:Rates_SIS} & Discrete & 1 & $n=0$ & $n=N$ & --- \\
\hline
Biased random walk & \eqref{eq:BRW_DriftDiffusion} & Continuous & 1 & $x=0$ (artificial) & $x=L$ (reflecting) &  Sec.~\ref{Appendix:QSD_Continuous_BRW} \\
Ornstein--Uhlenbeck & \eqref{eq:Ornstein_DriftDiffusion} & Continuous & 1 & $x=0$ (artificial) & --- & Eq.~\eqref{eq:Ornstein_Analytical} \\
Linear branching & \eqref{eq:Branching_DriftDiffusion} & Continuous & 1 & $x=0$ (natural) & --- & Eq.~\eqref{eq:Branching_Analytical} \\
Biased voter model & \eqref{eq:VM_DriftDiffusion} & Continuous & 1 & $x=0$, $x=1$ (natural) & $x = 1$ (absorbing) & --- \\
\hline
SIRS & \eqref{eq:Rates_SIRS} & Discrete & 2 & $\{(m,0) : 0 \leq m \le N\}$ & $\{(m,n) : m + n = N\}$ & --- \\
Diffusion in $(x,y)$ with circular symmetry & \eqref{eq:2d_evolutionrandomwalk} & Continuous & 2 & $r=r_0$ (artificial) & $r=r_1>r_0$ (reflecting) & Eq.~\eqref{eq:QSD_Brownian2D} \\
\hline
\end{tabularx}
\caption{Overview of the stochastic processes considered in our study. 
For each process, we specify the governing equation, the type (discrete or continuous), 
the system dimension, the absorbing states and upper limit of the state space, and whether an exact solution for the quasi-stationary distribution is available.}
\end{table}

\end{widetext}

\section{Analytical quasi-stationary distributions}\label{Appendix:QSD_Analytical}

In this section, we present analytical expressions for the quasi-stationary distributions of the stochastic processes discussed in the main text. When solutions are already available in the literature, we include the relevant references and provide the main equations. When we could not find known solutions, we present detailed derivations. Depending on the process, the calculations are carried out either directly from the equation for the quasi-stationary distribution or, alternatively, using eigenfunction methods~\cite{meleard2012quasi}.

\subsection{Discrete linear branching process}\label{Appendix:QSD_Discrete_Branching}

The quasi-stationary distribution of the discrete linear branching process introduced in Eq.~\eqref{eq:Rates_Branching}, supported in the regime $R < 1$, is given by~\cite{collet2013quasi}
\begin{equation}\label{eq:QSD_Branching}
 Q(n) = (1 - R) R^{n - 1}.
\end{equation}
Given that $Q(1) = 1 - R$, and taking into account that the per capita
death rate is set to one, the flux into the absorbing state is $1-R$ [see Eq.~\eqref{eq:flux_absorbing}]. According to Eq.~\eqref{eq:MTE}, the mean time to extinction is thus found as
$\tau = (1 - R)^{-1}$.

\subsection{Discrete immigration-death process}\label{Appendix:QSD_Discrete_Mix}

The process is presented in Eq.~\eqref{eq:Rates_Mix}. As we have not been able to find the solution of this model in the literature, we provide a detailed derivation.

Using the known evolution matrix $\mathbb{L}$ for a one-step process~\cite{VanKam,toral2014stochastic} and that the current into the absorbing state $n=0$ is $J_\mathcal{S}=Q(1)$, we find that the quasi-stationary distribution for $n=1,2,\dots$ satisfies 
\begin{equation}\label{eq:QSD_eq_mixed}
 (E^{-1}-1)[R Q(n)]+(E^{1}-1)[n Q(n)]=-Q(1) Q(n),
 \end{equation}
where $E$ is the step operator, defined as \makebox{$E^k [f(n)]=f(n+k)$} for an arbitrary function $f(n)$. Eq.~\eqref{eq:QSD_eq_mixed} is complemented with the condition $Q(0)=0$. To solve this equation for $Q(n)$ we make use of the generating function,
\begin{equation}\label{eq:GF}
 G(s)=\sum_{n=1}^{\infty} s^n Q(n).
\end{equation}

Using Eq.~\eqref{eq:QSD_eq_mixed}, we find that $G(s)$ satisfies the differential equation
\begin{equation}\label{eq:ODE_generating_function_mixed}
 (s-1)\left[R\, G(s) -G'(s)\right] +Q(1)\left[G(s)-1\right]=0,
\end{equation}
and the initial condition $G(0)=0$. Here the prime denotes the derivative of the function.

By solving the homogeneous part of Eq.~\eqref{eq:ODE_generating_function_mixed} and finding a particular solution of the non-homogeneous equation with the method of variations of constants, we find the solution
\begin{equation}
 G(s)=(1-s)^{Q(1)}e^{Rs}\,Q(1)\,\int_0^sdx\,
\frac{e^{-Rx}
}{(1-x)^{Q(1)+1}}.
\end{equation}
Using the integral representation of Kummer's hypergeometric function~\cite{gradshteyn2014table},
\begin{equation}
M[a,b,z]=\frac{\Gamma(b)}{\Gamma(a)\Gamma(b-a)}\int_0^1dt\,e^{t z}t^{a-1}(1-t)^{b-a-1},
\end{equation}
with $a=-Q(1)$ and $b=1-Q(1)$, together with the functional relation $e^{-z}M[a,b,z]=M[b-a,b,-z]$, one arrives after some algebra to 
\begin{align}
G(s)=&M\left[1,1-Q(1),-R(1-s)\right] - \nonumber \\
&-e^{Rs}(1-s)^{Q(1)}M[1,1-Q(1),-R].
\end{align}
Besides the initial condition, this solution fulfills the normalisation condition \makebox{$G(s=1)=1$}, and the self-consistent condition $G'(0)=Q(1)$ for all values of $Q(1)$. As the term containing $(1-s)^{Q(1)}$ is non-analytic at $s=1$, we set it to zero. This requirement fixes the value $Q(1)$ through the equation
\begin{equation}\label{eq:Mix_eqforQ1}
M[1,1-Q(1),-R]=0,
\end{equation}
and gives the final solution
\begin{equation}
G(s)={M}[1,1-Q(1),-R(1-s)].
\end{equation}
We now expand this solution around $s=0$,
\begin{align}
G(s)&=\Gamma[1-Q(1)]\sum_{k=0}^\infty\frac{[-R(1-s)]^k}{\Gamma[1-Q(1)+k]}= \nonumber \\
&=\Gamma[1-Q(1)]\sum_{k=0}^\infty\sum_{n=0}^k\frac{(-R)^k}{\Gamma[1-Q(1)+k]}{k\choose n}(-s)^n.
\end{align}
Exchanging the order of the sums, we get
\begin{equation}
G(s)=\Gamma[1-Q(1)]\sum_{n=0}^\infty\left[\sum_{k=n}^\infty\frac{(-1)^n(-R)^k}{\Gamma[1-Q(1)+k]}{k\choose n}\right]s^n,
\end{equation}
and, after performing the $k$-sum, we obtain the following expression for the quasi-stationary distribution
\begin{equation}\label{eq:QSD_Mix_app}
Q(n)=\Gamma[1-Q(1)]\frac{ M[n+1;1-Q(1)+n;-R]}{\Gamma[1-Q(1)+n]}R^n,
\end{equation}
in which we have to replace $Q(1)$ by the value obtained by solving numerically Eq.~\eqref{eq:Mix_eqforQ1}. According to Eqs.~\eqref{eq:flux_absorbing} and~\eqref{eq:MTE}, the mean time to extinction can then be calculated as $\tau = 1 / Q(1)$.

\subsection{Discrete biased random walk}\label{Appendix:QSD_Discrete_BRW}

The process is defined by Eq.~\eqref{eq:Rates_BRW}. The quasi-stationary distribution of the discrete biased random walk with an absorbing barrier at $n=0$, supported in the regime $R < 1$, reads~\cite{collet2013quasi}:
\begin{equation}\label{eq:QSD_BRW}
 Q(n) = n \left(1 - \sqrt{R}\right)^2 \left(\sqrt{R}\right)^{n - 1}.
\end{equation}
As the current into the absorbing state is \makebox{$J_\mathcal{S}=Q(1)$}, Eq.~\eqref{eq:MTE} yields the mean extinction time \makebox{$\tau = \left(1 - \sqrt{R}\right)^{-2}$}.

\subsection{Continuous biased random walk}\label{Appendix:QSD_Continuous_BRW}

We now derive the quasi-stationary distribution for the continuous biased random walk bounded by an absorbing point in $x=0$ and a reflecting boundary in $x=L$.

The drift and diffusion functions of this process are given by Eq.~\eqref{eq:BRW_DriftDiffusion}, while its Fokker--Planck equation reads
\begin{equation}
 \begin{cases}
 \partial_t P(x,t) = -f\, \partial_x \,P(x,t)+ D\, \partial^2_x \,P(x,t), \\
 P\left(x=0,t\right)=0,\quad\forall t,\\
 J(L)=f\,P(L,t)-D\, \partial_x \,P(x,t)|_{x=L}.
\end{cases}
\end{equation}

As shown in~\cite{meleard2012quasi,collet2013quasi}, to find the quasi-stationary distribution one can first solve the eigenvalue problem of the Fokker--Planck operator $\mathbb{L}$, namely 
\begin{equation}\label{Eigenproblem}
 \begin{cases}
 -f\, \partial_x \,R_\lambda(x)+D\, \partial^2_x \,R_\lambda(x) =-\lambda \,R_\lambda(x), \\
 R_\lambda(0)=0,\quad\forall t,\\
 f\,R_\lambda(L)-D\, \partial_x \,R_\lambda(x)|_{x=L}=0.
 \end{cases}
\end{equation}
Since the eigenvalues of the Fokker--Planck operator are non-negative~\cite{Risken1991}, we order the possible eigenvalues as $0=\lambda_0<\lambda_1<\lambda_2\cdots$. The quasi-stationary distribution $Q(x)$ is then the, properly normalised, eigenfunction $R_{\lambda_1}(x)$ with the smallest non-null eigenvalue.

The solution of the second-order differential equation in Eq.~\eqref{Eigenproblem} with the boundary condition $R_\lambda(0)=0$ depends on the value of $\lambda$:
\begin{equation}\label{eq:general_solution_brownian}
R_\lambda(x) =
\begin{cases}
 e^{v x}\sinh(w_\lambda x),&\lambda\in [0,f^2/(4D)), \text{ case I},\\
e^{v x}\sin(w_\lambda x),&\lambda > f^2/(4D), \text{ case II},
 \end{cases}
\end{equation}
with $v=f/(2 D)$ and $w_\lambda=\sqrt{|f^2-4D\lambda|}/(2D)$. We only need to consider $w_\lambda>0$, since a negative value of $w_\lambda$ leads, after normalisation, to the same quasi-stationary distribution than the opposite positive value, whereas a value $w_\lambda=0$ leads to $R_\lambda(x)=0$, which is not acceptable.

We now analyse each case separately.

\begin{itemize}
 \item \textbf{Case I}: The reflecting boundary condition leads to the following equation for the eigenvalue $\lambda$:
\begin{equation}\label{eq:eigenvalues_CaseI}
\tanh(w_\lambda L) = \frac{w_\lambda}{v}.
\end{equation}
 If $v>1/L$, the above equation admits a unique positive solution $w_{\lambda_1}$ such that $0<w_{\lambda_1}<v$ and a corresponding eigenvalue $\lambda_1$, while there are no valid solutions for $v<1/L$ (this includes $v\le 0$). The value $w_{\lambda_1}$ must be found using numerical methods. As $0<\lambda_1<f^2/(4D)$, the eigenvalue is always smaller than those corresponding to case II and it provides, after normalisation, the quasi-stationary solution
\begin{align}\label{SolQ1}
 v&>1/L,\nonumber\\
 Q(x)&=\frac{\lambda_1}{D w_{\lambda_1}}e^{v x}\sinh(w_{\lambda_1}x),\\
 \lambda_1&=\frac{f^2-4D^2w_{\lambda_1}^2}{4D}\nonumber.
\end{align}
If $v< 1/L$, the quasi-stationary solution is provided instead by case II. 

 \item \textbf{Case II}: Similarly, the reflecting boundary condition leads to the transcendental equation
\begin{equation}\label{eq:eigenvalues_Case2}
\tan(w_\lambda L) = \frac{w_\lambda}{v}.
 \end{equation}
In this case, there is an infinite countable set of positive solutions $w_\lambda$ for any value of $v$ and corresponding eigenvalues $\lambda>f^2/(4D)$, all larger than the ones found in case I. Therefore it is only for $v<1/L$ ---where there was no solution in case I--- that we need to consider the smallest solution for case II, $w_{\lambda_1}$. This is such that $w_{\lambda_1}L\in(0,\pi/2)$ if $v>0$, or $w_{\lambda_1}L\in(\pi/2,\pi]$ if $v\le 0$. As before, the value of $w_{\lambda_1}$ needs to be found numerically.
After normalisation, the quasi-stationary solution reads
 \begin{align}\label{SolQ2}
 v&< 1/L,\nonumber\\
 Q(x)&=\frac{\lambda_1}{Dw_{\lambda_1}}e^{v x}\sin(w_{\lambda_1}x),\\
 \lambda_1&=\frac{f^2+4D^2w_{\lambda_1}^2}{4D}\nonumber.
\end{align}

Both solutions, Eqs.~(\ref{SolQ1},~\ref{SolQ2}), merge continuously for $v=1/L$, where the eigenfunction is found in the limit $w_{\lambda_1}\to 0$ as $Q(x)=L^{-2}xe^{x/L}$.

\end{itemize}
According to Eq.~\eqref{eq:flux_continuousx1}, the flux into the absorbing state $x = 0$ is $ J_\mathcal{S}=\left.-fQ(x)+D\partial_xQ(x)\right|_{x=0}=\lambda_1$, in all cases. From Eq.~\eqref{eq:MTE}, the average time it takes the process to reach the absorbing state is $\tau =\lambda_1^{-1}$.

For completeness, we discuss two special cases in which the eigenvalue equation can be solved analytically. The first is the unbiased case, obtained by setting $v=f=0$. According to Eq.~\eqref{eq:general_solution_brownian}, the quasi-stationary distribution in this case will be a sine (case II). Moreover, the equation for the eigenvalues reduces to $\cos\left(w_\lambda L\right) = 0$, 
with solutions $w_{\lambda_n} = (2n-1)\dfrac{\pi}{2\,L},\ n=1,2,\dots$. As $\lambda_n=Dw_{\lambda_n}^2$, the smallest eigenvalue corresponds to $w_{\lambda_1} = \dfrac{\pi}{2\,L}$, leading to
\begin{equation}\label{qsin}
 Q(x)=\frac{\pi}{2L}\sin\left(\frac{\pi x}{2\,L}\right).
\end{equation}
Accordingly, the mean time to extinction is $\tau = \dfrac{4L^2}{D\pi^2}$.

The second is the unbounded case, obtained in the limit $L\to\infty$. Here we only consider $v<0$ (i.e., case II), since the exponential term in Eq.~\eqref{eq:general_solution_brownian} diverges otherwise. The limit $L\to \infty$ simplifies Eq.~\eqref{eq:eigenvalues_Case2} as $\lim_{L\to\infty}\tan\left(w_\lambda \,L\right) = 0,$ with solution \makebox{$w_n = \lim_{L\to\infty} \frac{\pi}{L}n$}, with $n=1,2,\dots$. Thus, under this limit, the spectrum of eigenvalues becomes continuous in the interval $\lambda\in [D v^2,+\infty]$, where the minimum eigenvalue, $\lambda=D v^2$, comes from inserting $w_1=0$ in Eq.~\eqref{SolQ1}. Therefore, the quasi-stationary distribution reads
\begin{equation}
 Q(x) = \frac{1}{v^2}\,x\,e^{v x},
\end{equation}
and the extinction time is $\tau = \dfrac{1}{D v^2}.$

\subsection{Ornstein--Uhlenbeck process}\label{Appendix:QSD_Continuous_Uhlenbeck}

The quasi-stationary distribution of the Ornstein--Uhlenbeck process in Eq.~\eqref{eq:Ornstein_DriftDiffusion} reads~\cite{Aalen2004}
\begin{equation}\label{eq:Ornstein_Analytical}
 Q(x) = \frac{f}{D}x\, e^{-\frac{f}{2D}x^2}.
\end{equation}

From the knowledge of the quasi-stationary distribution, we can determine the mean absorption time by combining Eqs.~\eqref{eq:MTE} and~\eqref{eq:flux_continuousx1} as $\tau =1/f.$

\subsection{Continuous branching process}\label{Appendix:QSD_Continuous_Branching}

The quasi-stationary distribution of the continuous branching process introduced in Eq.~\eqref{eq:Branching_DriftDiffusion}, which is supported for $f > 0$, is given by~\cite{Lambert2007}
\begin{equation}\label{eq:Branching_Analytical}
 Q(x) = \frac{f}{D} e^{-\frac{f}{D}x}.
\end{equation}

By combining Eq.~\eqref{eq:Branching_Analytical} with Eqs.~\eqref{eq:MTE} and~\eqref{eq:flux_continuousx1}, the mean time to extinction can be expressed as $ \tau = 1/f$.

\subsection{Two-dimensional random with with circular symmetry}\label{Appendix:QSD_2D_Brownian}

The problem is defined by Eq.~\eqref{eq:2d_evolutionrandomwalk}, where the radial variable is confined to the interval $r\in[r_0,r_1]$, absorbing at $r_0$ and reflecting at $r_1$. This problem exhibits circular symmetry, which becomes evident upon changing variables from Cartesian to polar coordinates, $(x, y) \to (r, \theta)$ with $\theta = \arctan(y/x)$, and $r^2=x^2+y^2$. In~\cite{gardiner1985handbook,magalang2024optimal}, this change of variables is performed in detail, showing that the angular diffusion is decoupled from the radial one,
\begin{equation}
 \dot{r} = \sqrt{2D} \xi_r(t).
\end{equation}
As a result, since the boundary conditions are also independent of the angular variable, the quasi-stationary distribution of the process becomes effectively one-dimensional. In particular, the dynamics reduces to diffusion in the radial coordinate, for which the quasi-stationary distribution $Q(r)$ is given by a straightforward modification of Eq.~\eqref{qsin},
\begin{equation}\label{eq:QSD_Brownian2D_1D}
 Q(r) = w \sin \left[ w (r - r_0) \right],
\end{equation}
with $w = \pi / [2 (r_1 - r_0)]$.

The quasi-stationary distribution can be expressed in Cartesian coordinates, $Q(x, y)$, by a change of variables,
\begin{equation}\label{eq:QSD_Brownian2D}
 Q(x, y) = \frac{w}{2 \pi \sqrt{x^2 + y^2}} \sin \left[ w (\sqrt{x^2 + y^2} - r_0) \right].
\end{equation}

\section{Efficiency comparison of the single-trajectory and multiple-trajectory approaches}\label{Appendix:MC_MSComparison}
We now compare the efficiency of the single-trajectory and multiple-trajectory Monte Carlo approaches across various examples. Specifically, we evaluate the computation time required by each method to achieve a specified target error in the numerical estimate of the quasi-stationary distribution. To perform the analysis, we consider all processes introduced in the main text (see Sec.~\ref{sec:Examples}) with known analytical solution, for which the total error can be computed.

In the single-trajectory method, we obtain different error values in the estimation of the quasi-stationary distribution by performing computations at various final steps~$K$, which is the only parameter controlling the errors. In contrast, the multiple-trajectory method has two main parameters: the number of trajectories~$M$ and the simulation time~$T$. To explore different error values in this method, we perform simulations for several values of $M$ while keeping the total simulation time $T$ fixed. This choice is motivated by the fact that $T$ governs convergence, while finite $M$ introduces both statistical and bias errors (the latter arising from the finite sampling of the empirical distribution used for resets).

\begin{figure}[h!]
	\centering
\includegraphics[width=0.8\linewidth]{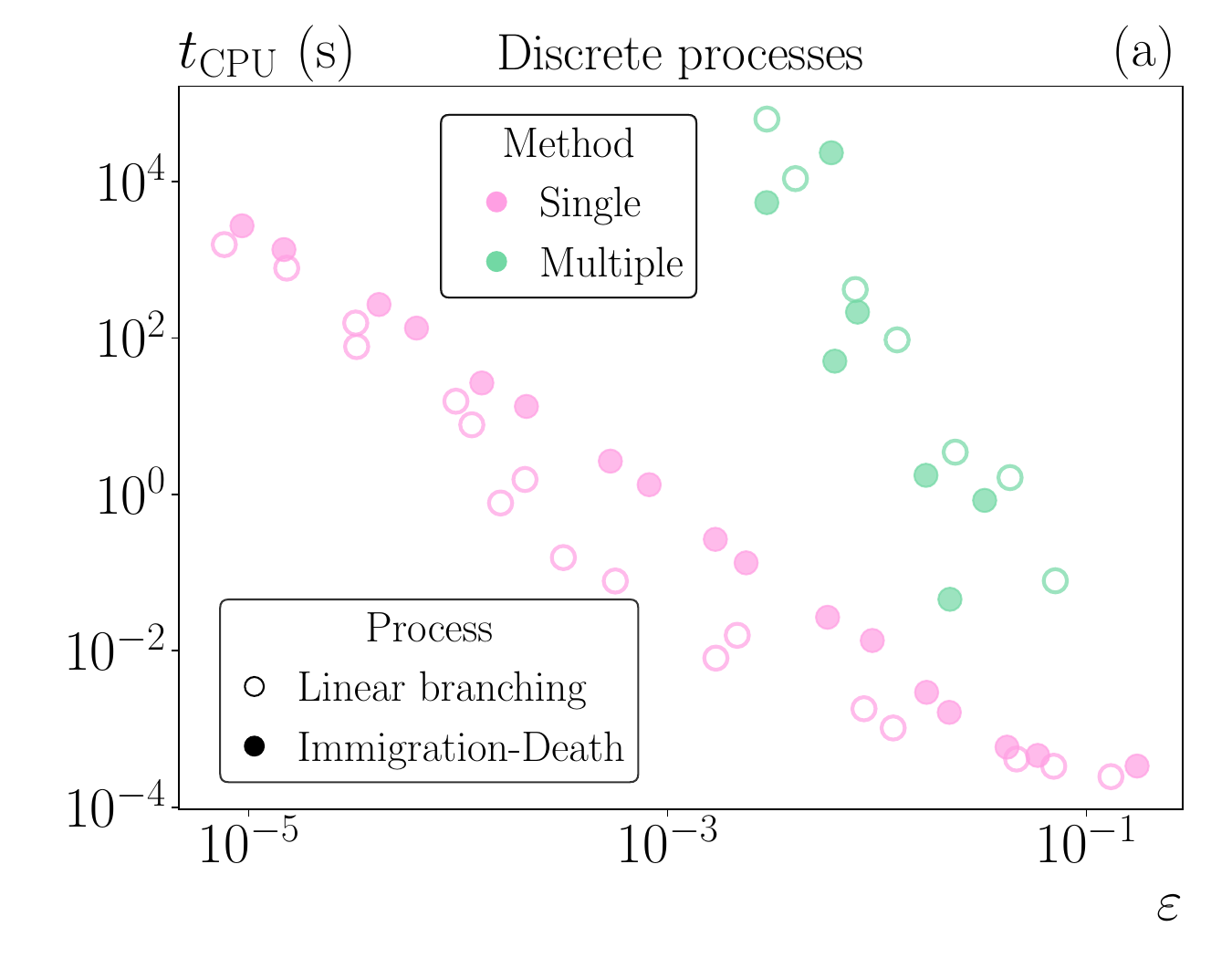}
\includegraphics[width=0.8\linewidth]{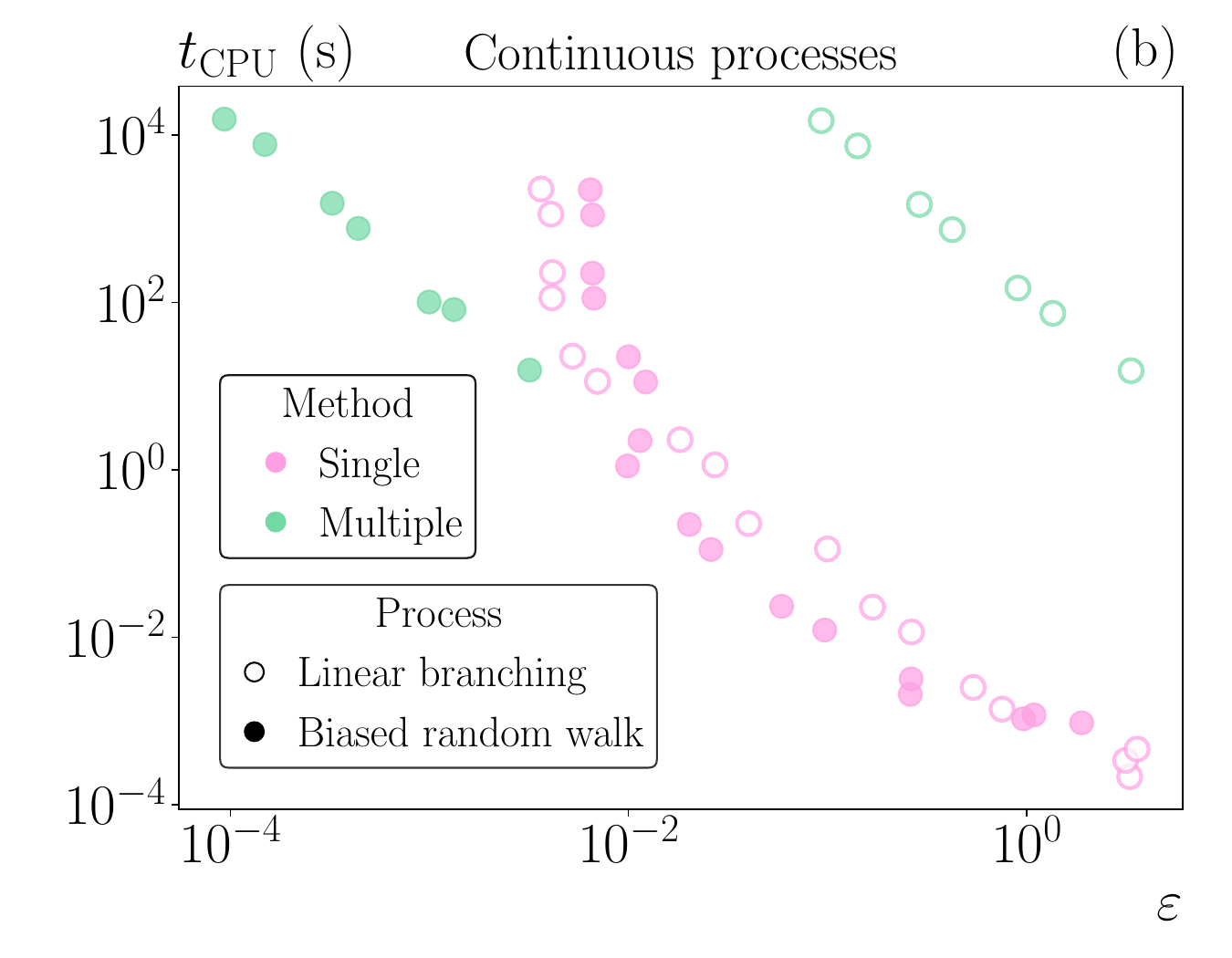}
\caption{\textbf{Comparison of efficiency and accuracy of the single-trajectory and multiple-trajectory Monte Carlo approaches.} Computation time (measured in seconds) as a function of the total error in the numerical estimate of the quasi-stationary distribution. Pink symbols come from the single-trajectory method, where we have performed $K = 10^{11}$ simulation steps and used the same initial condition $n_0$ (or $x_0$, if the process is continuous) as in Sec.~\ref{sec:Examples} (see Figs.~\ref{fig:QSD_Discrete} and~\ref{fig:QSD_Continuous}). Green symbols come from the multiple-trajectory method, where we have set a total simulation time $T = 10^3$ and combined the results from simulations with $M = 10^2, 5 \times 10^2, 10^3, 5 \times 10^3, 10^4, 5 \times 10^4$ and $10^5$ trajectories. In all simulations, the initial condition for each trajectory $i = 1, \dots, M$ has been randomly selected from the set $[1, 2, \dots, N]$ in the discrete case, or from the interval $(0, L]$ in the continuous case. (a) Results corresponding to the discrete
linear branching process (empty symbols) and the immigration-death process (filled symbols) introduced in Sec.~\ref{sec:ExamplesDiscrete}. In both models, we have set $R = 0.6$ and computed the errors using Eq.~\eqref{eq:TotalErrorDiscrete}. (b) Results obtained for the continuous linear branching process (empty symbols) for $D = 0.3$ and $f = 5$, and the continuous biased random walk (filled symbols) for $D = 0.4, f = 2$ and $L = 1$ (see Sec.~\ref{sec:ExamplesContinuous}). Errors have been calculated using Eq.~\eqref{eq:TotalErrorContinuous}. In both numerical methods, we have set the time step to $\dt = 10^{-4}$ and the spatial step size to $\dx = 10^{-4}$.}
\label{fig:MS_tcpuvserror}
\end{figure}

Fig.~\ref{fig:MS_tcpuvserror} shows the computation time required to reach a given target error. In the single-trajectory method, we have performed up to $K = 10^{11}$ simulation steps. In the multiple-trajectory approach, we have set the total simulation time to $T = 10^3$ and combined the results from seven values of $M$. We have verified that further increasing $T$ does not lead to a significant improvement in accuracy. 

In Fig.~\ref{fig:MS_tcpuvserror}(a), we present the results for two stochastic models with discrete state spaces. In both cases, the single-trajectory approach yields more efficiency and accuracy than the multiple-trajectory method. 

In Fig.~\ref{fig:MS_tcpuvserror}(b), we show the results for the continuous biased random walk and the continuous linear branching process. For the linear branching process, the single-trajectory method outperforms the multiple-trajectory approach, leading to smaller errors in shorter computation times. This behaviour is also observed for the Ornstein--Uhlenbeck process, although not shown here. In contrast, for the biased random walk, the multiple-trajectory method proves more efficient. Thus, among the three processes with continuous states examined, the biased random walk is the only case where the multiple-trajectory approach outperforms the single-trajectory one.

\section{Implementation of the iterative algorithm to the SIRS model}\label{Appendix:Iterative_SIRS}

In this section, we detail the implementation of the iterative algorithm (see Sec.~\ref{sec:IterativeDiscrete}) to the SIRS model described in Sec.~\ref{sec:Examples2D}, which involves two discrete stochastic variables. 

We recall that the transition rates for the SIRS model are 
\begin{align}\label{eq:SIRS_rates}
W[(m, n) \to (m-1, n+1)] &\equiv W_\text{I}(m, n) = \beta \frac{n m}{N}, \nonumber \\
W[(m, n) \to (m, n-1)] &\equiv W_\text{R}(m, n) = \gamma n, \nonumber \\
W[(m, n) \to (m+1, n)] &\equiv W_\text{S}(m, n) = \mu (N - n - m).
\end{align}
The probability flux into the set of absorbing states, corresponding to $\mathcal{S} = \{(m,0) : 0 \leq m \le N\}$, is given by 
\begin{equation}\label{eq:flux_SIRS}
J_\mathcal{S} = \gamma \, Q(n=1),
\end{equation}
where $Q(n=1)$ represents the marginal probability of states with one infected individual, that is, \makebox{$Q(n=1) = \sum_{m = 0}^{N-1} Q(m, 1)$}. Note that in this sum the variable $m$ does not take the value $N$ because the state $(N, 1)$ is out of the state space.

By introducing the transition rates in Eq.~\eqref{eq:SIRS_rates} and the probability flux in Eq.~\eqref{eq:flux_SIRS} into the two-dimensional form of Eq.~\eqref{eq:QSD} and rearranging terms, we get the following relation 
\begin{widetext}
\begin{equation}
Q(m,n) = \frac{W_\text{I}(m+1,n-1) \, Q(m+1,n-1) + W_\text{R}(m,n+1) \, Q(m,n+1) + W_\text{S}(m-1,n) \, Q(m-1,n)}{W_\text{I}(m,n) + W_\text{R}(m,n) + W_\text{S}(m,n) - \gamma \, Q(n=1)} \equiv f(m, n, \bm{Q}).
\end{equation}
\end{widetext}
The iterative algorithm with over-relaxation is analogous to the one used for processes with a single stochastic variable [see Eq.~\eqref{eq:Iterative_Scheme}]. Starting from an initial guess $\bm{Q}^{(0)}$, we iterate the following relation for $k = 0, 1, 2, \dots$:
\begin{equation}
Q(m,n)^{(k+1)} = s \, Q(m,n)^{(k)} + (1 - s) \, f\left(m, n,\bm{Q}^{(k)}\right)
\end{equation}
for all non-absorbing states, i.e., for all states $(m, n) \in \mathcal{R}$. Specifically, the set of non-absorbing states is given by \makebox{$\mathcal{R} = \{(m,n) : m = 0, 1, \dots, N - 1, \, n = 1, 2, \dots, N - m\}$}. The probability distribution must be normalised after each iteration step.

The iteration proceeds until the norm of the difference $\norm{\bm{Q}^{(k+1)}-\bm{Q}^{(k)}}$ stabilises at a constant value, where the norm here is defined as $\norm{\bm{a}}^2=\sum_{(m,n)\in \mathcal{R}} a(m,n)^2$.\\[10pt]

\section{Oscillatory behaviour in the iterative algorithm without over-relaxation}\label{Appendix:IterativeQSDOscillations}

In this section, we examine the oscillatory behaviour exhibited in the iterative algorithm introduced in Sec.~\ref{sec:Iterative} when no over-relaxation is applied ($s = 0$) and the iteration is initialised with a Kronecker delta distribution. This phenomenon arises both in one-step discrete stochastic processes and in continuous processes.

We begin with the discrete case. For one-step processes, the iteration scheme defined by Eqs.~\eqref{eq:Q_iterative0} and~\eqref{eq:Iterative_Scheme} reduces, for $s = 0$, to the form
\begin{widetext}
\begin{equation}\label{eq:IterativeSchemeAppendix}
Q(n)^{(k)}= \frac{W(n+1 \to n) Q(n+1)^{(k-1)} + W(n-1 \to n) Q(n-1)^{(k-1)}}{W(n \to n+1) + W(n \to n-1) - J_\mathcal{S}^{(k-1)}},
\end{equation}
\end{widetext}
where, for simplicity, we omit the explicit expression of the flux $J_\mathcal{S}$ into the set of absorbing states.

Let us assume that the algorithm is initialised with the distribution $Q(n)^{(0)} = \delta_{n, n_0}$ for any $n_0 \in \mathcal{R}$. From Eq.~\eqref{eq:IterativeSchemeAppendix}, it follows that at step 1, $Q(n)^{(1)} > 0$ only for the states $n_0 \pm 1$; at step 2, $Q(n)^{(2)} > 0$ only for the states $n_0$ and $n_0 \pm 2$; at step 3, $Q(n)^{(3)} > 0$ only for the states $n_0 \pm 1$ and $n_0 \pm 3$; and so on. Thus, the choice of this initial condition leads, after a transient regime during which the initial condition propagates throughout the state space, to a period-2 oscillatory cycle. In this regime, the numerical distribution alternates its support between odd and even states at each iteration step: at step $k$, $Q(n)^{(k)} > 0$ only for odd values of $n$, while at step $k + 1$, $Q(n)^{(k+1)} > 0$ only for even values. Even introducing an infinitesimal amount of over-relaxation ($s > 0$) is enough to prevent this phenomenon. 

More generally, this oscillatory behaviour arises in any process with discrete states such that all of its allowed transitions $n \to n + \ell$ involve steps $\ell$ of the same parity.

A completely analogous argument applies to continuous processes. The use of the iterative algorithm defined in Eqs.~\eqref{eq:Qcontinuous_iterative1} and~\eqref{eq:Iterative_Scheme_Continuous} with $s = 0$ and initial condition $Q_i^{(0)} =\delta_{x_i, x_0}$ for any $x_0$ also leads to a period-2 oscillatory regime.

\begin{figure}[h]
	\centering
\includegraphics[width=0.8\linewidth]{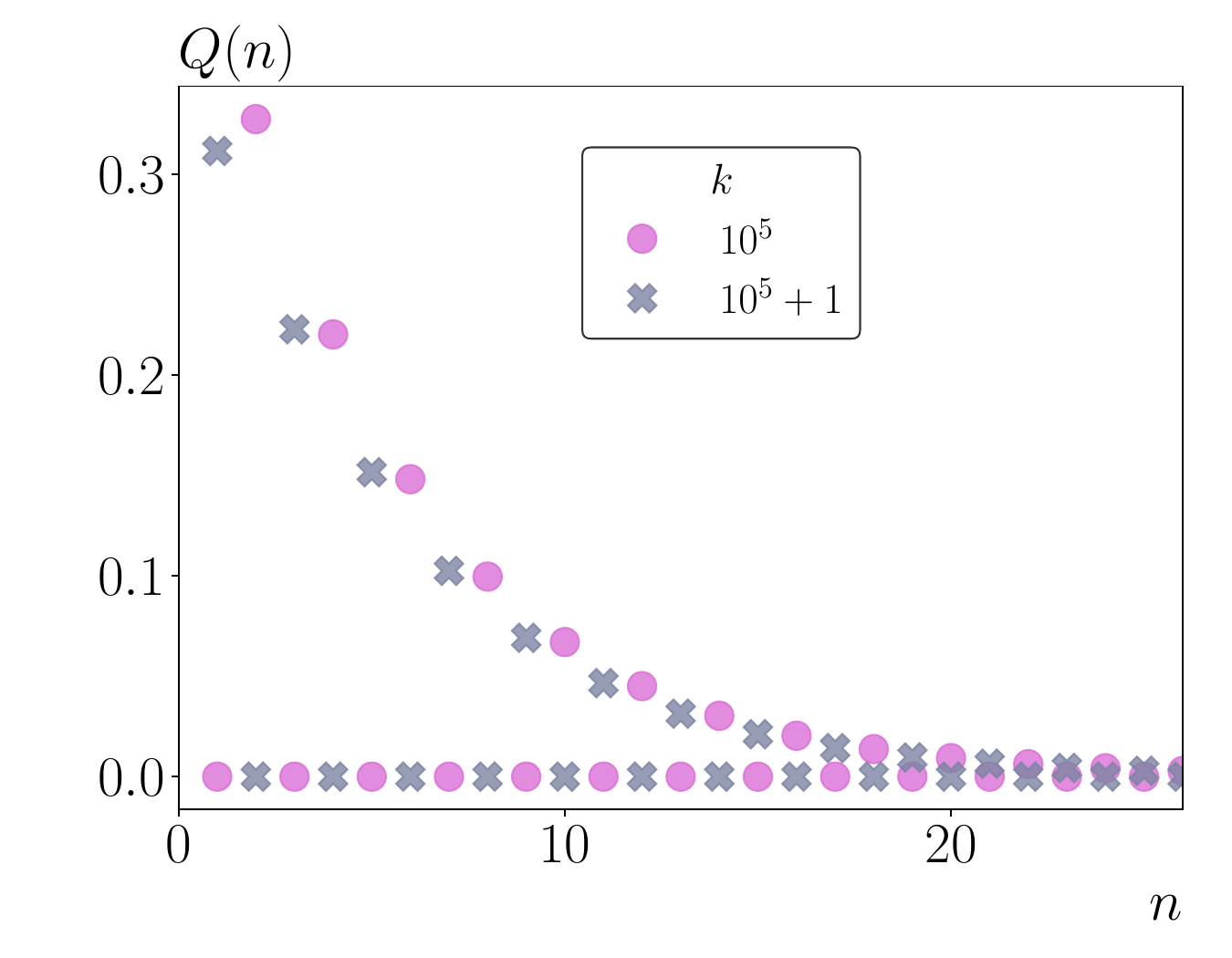}
\caption{\textbf{Pathological oscillations in the iterative algorithm without over-relaxation.} Quasi-stationary distribution of the discrete linear branching process in Eq.~\eqref{eq:Rates_Branching} obtained with the iterative algorithm in the absence of over-relaxation ($s = 0$) and with a Kronecker delta distribution centred at $n_0 = 10$ as initial condition. The pink dots (resp. grey crosses) correspond to a number of iteration steps $k = 10^5$ (resp. $k = 10^5 + 1$). We have set the reproduction ratio to $R = 0.8$ and introduced an upper bound at $N = 150$.}
\label{fig:Iterative_QSDOscillations_Branching}
\end{figure}

In Fig.~\ref{fig:Iterative_QSDOscillations_Branching}, we show the quasi-stationary distribution of the discrete linear branching process obtained with $s = 0$ and a Kronecker delta distribution as initial condition. We present the results of two consecutive iteration steps, illustrating that the numerical quasi-stationary distribution oscillates between two distinct, physically meaningless distributions.

\section{Effect of system size on computation time and memory consumption in the SIS model}\label{Appendix:SIScriticality}

We now consider the susceptible-infected-susceptible (SIS) model~\cite{keeling} to study how system size affects computation time and memory consumption. The transition rates of the model are given by
\begin{align}\label{eq:Rates_SIS}
W(n\to n + 1) &= R \frac{n (N - n)}{N}, \nonumber \\
W(n\to n - 1) &= n,
\end{align}
where $N$ is the population size, $R$ is the basic reproduction ratio, and $0 \le n \le N$ denotes the number of infected individuals. The model has a single absorbing state at $n = 0$, which corresponds to the fading of the epidemic. The critical value $R = 1$ separates the subcritical regime ($R < 1$), in which the epidemic rapidly dies out, from the supercritical regime ($R > 1$), in which the system reaches an endemic state.

\begin{figure}[h]
	\centering
\includegraphics[width=0.7\linewidth]{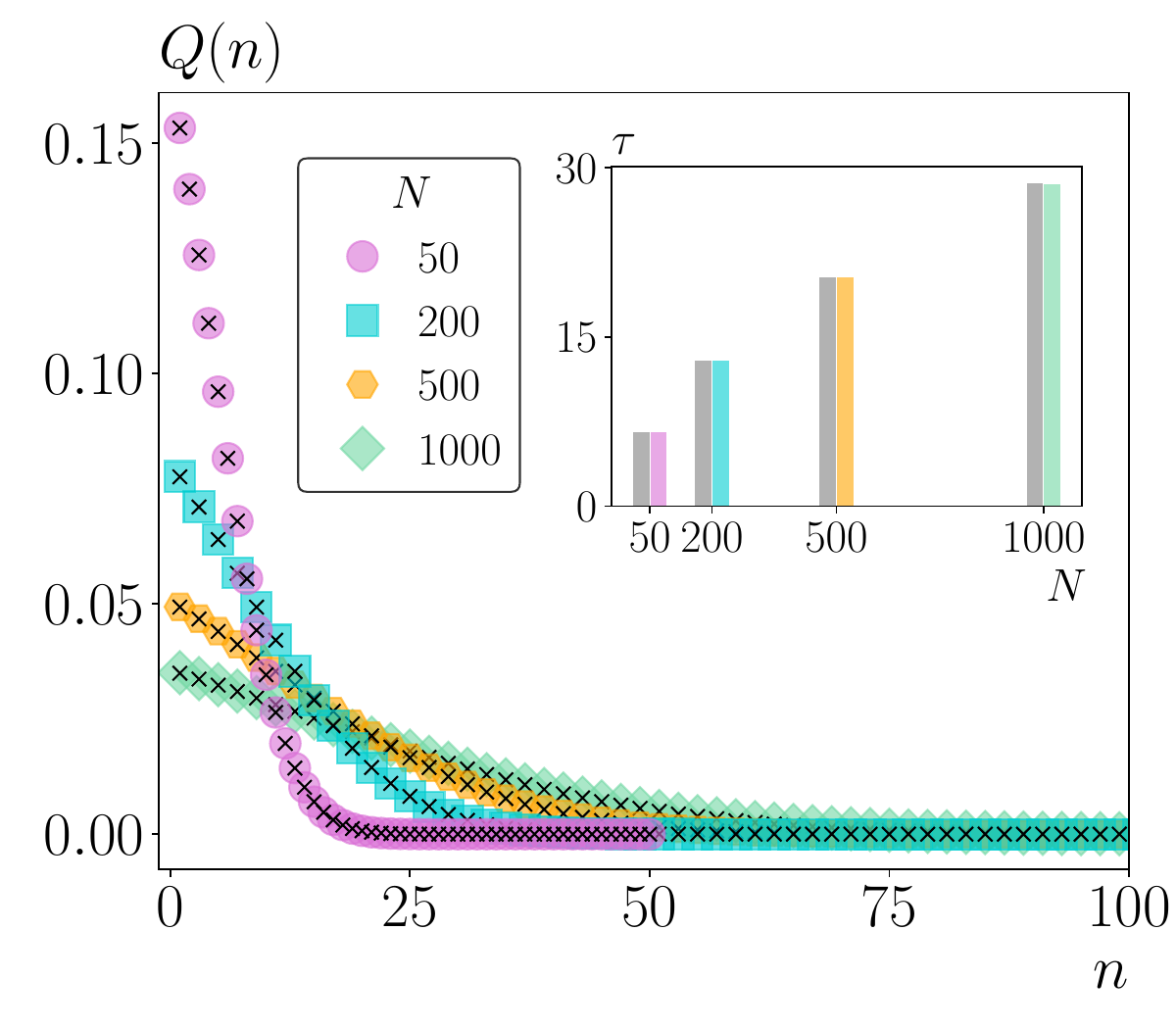}
\caption{\textbf{Quasi-stationary distribution and mean time to extinction of the SIS model in Eq.~\eqref{eq:Rates_SIS}}. We have set the reproduction ratio to $R = 1$, and considered several system sizes $N$. Black crosses correspond to the iterative algorithm with a relaxation factor $s = 0.1$ (see Sec.~\ref{sec:IterativeDiscrete}), while coloured symbols come from the Monte Carlo method with resetting (see Sec.~\ref{sec:MonteCarloDiscrete}). For better visualisation, for \makebox{$N = 200, 500, 1000$} we only plot the distributions for odd values of $n$. The initial condition has been set to $n_0 = 10$ in both numerical methods.  Inset: Mean time to extinction, $\tau$, as a function of $N$ obtained with the iterative algorithm (grey bars) and with the Monte Carlo method (coloured bars).}
\label{fig:QSD_Discrete_SIS}
\end{figure}

In Fig.~\ref{fig:QSD_Discrete_SIS}, we show the quasi-stationary distribution and the mean time to extinction of the SIS model at criticality for several system sizes $N$. The results obtained using the iterative algorithm and the Monte Carlo method are in good agreement with each other.

The SIS model is not included in the main text because it does not introduce qualitatively new features with respect to the models already discussed. However, it provides a convenient framework for studying size-dependent computational effects. We first analyse the dependence of computation time on system size at criticality, where the relaxation time diverges~\cite{hohenberg1977critical} and simulations become increasingly demanding. We then examine how the memory requirements of the numerical methods scale with system size.

\subsection{Computation time}

First, we analyse the effect of system size on the computation time of the iterative algorithm and the Monte Carlo method. Since the quasi-stationary distribution of the SIS model is not known analytically, we cannot evaluate the error of the numerical estimates using Eq.~\eqref{eq:TotalErrorDiscrete}, which prevents a direct comparison of efficiency and accuracy such as that performed in Sec.~\ref{sec:EfficiencyAccuracy} of the main text.

\begin{figure}[h]
	\centering
\includegraphics[width=0.95\linewidth]{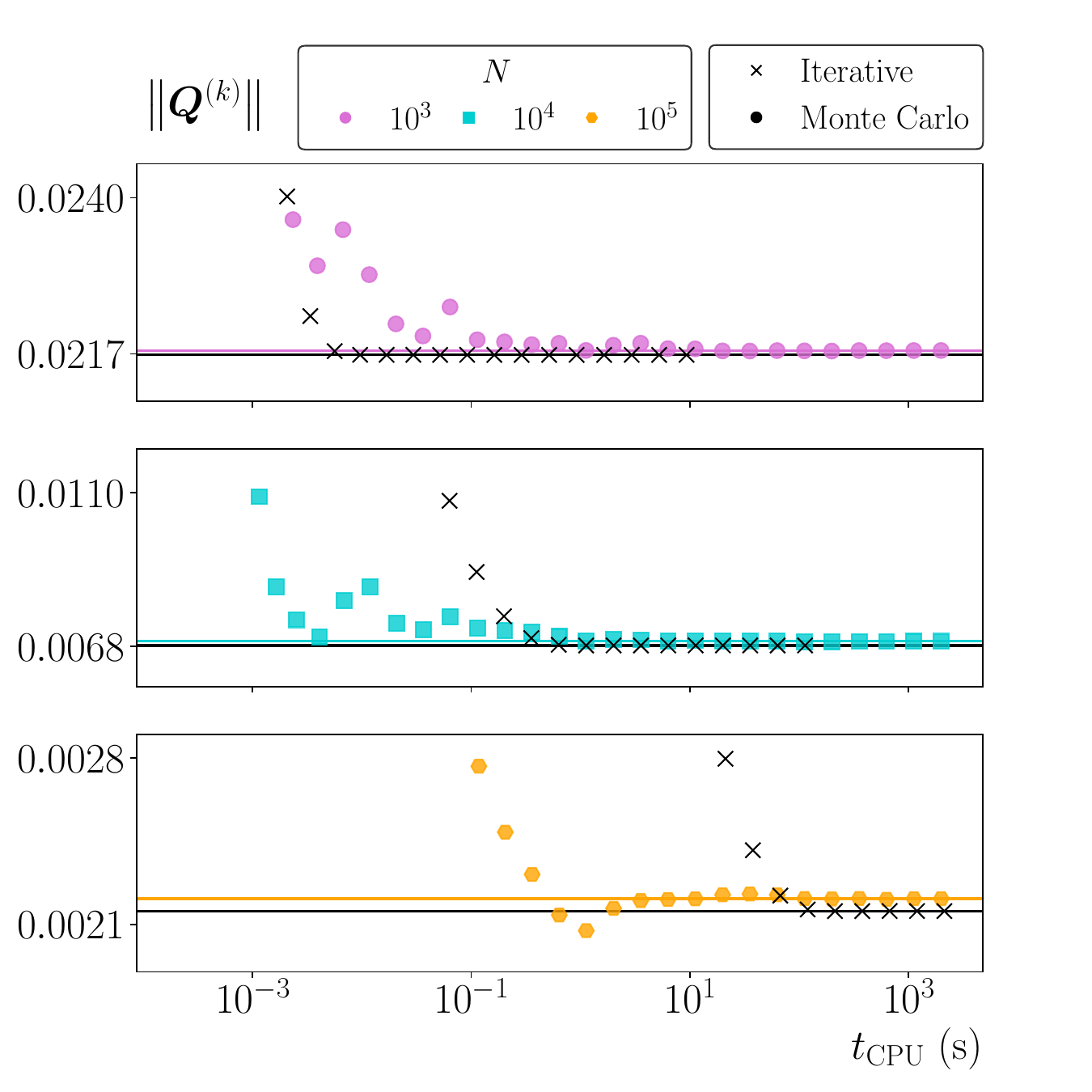}
\caption{\textbf{Convergence behaviour at criticality for different system sizes.} Convergence measure $\|\bm{Q}^{(k)}\|$ as a function of computation time (measured in seconds) for the SIS model at the critical point $R = 1$, and for several system sizes $N$. Black crosses and coloured symbols correspond to the iterative algorithm and to the Monte Carlo method, respectively. The black horizontal lines (resp. coloured horizontal lines) indicate the final value of the convergence measure reached by the iterative algorithm (resp. Monte Carlo simulations).}
\label{fig:SIScriticality_convergence}
\end{figure}

In Fig.~\ref{fig:SIScriticality_convergence}, we show the evolution of the norm $\|\bm{Q}^{(k)}\|$ as a function of computation time for the SIS model at criticality, and for three system sizes: $N = 10^3, 10^4$, and $10^5$. The convergence of the numerical methods, indicated by the stabilisation of this norm, requires longer computation times for larger systems, as expected. We note that the stabilised values of the norm $\norm{\bm{Q}^{(k)}}$ obtained with the two approaches exhibit slight discrepancies. In the absence of an analytical solution, it is not possible to determine which estimate is more accurate. However, based on the analyses presented in Sec.~\ref{sec:EfficiencyAccuracy} for problems with known analytical solution, the results obtained with the iterative algorithm appear to be more reliable. The deviations observed in the Monte Carlo estimates can be attributed to the stochastic nature of the method and to its inherent bias error.

We next compare the convergence times of the two methods. From a qualitative inspection of Fig.~\ref{fig:SIScriticality_convergence} one could conclude that the iterative algorithm converges faster than the Monte Carlo method for small systems, while the Monte Carlo approach becomes more efficient as the system size increases. However, a quantitative analysis is required to properly evaluate this behaviour. To this end,  we introduce a precise definition of the saturation time and determine its scaling with system size for both approaches. Specifically, we define the saturation time as the time required for the norm \makebox{$\Delta^{(k,\bar{k})}=\norm{\bm{Q}^{(k)}-\bm{Q}^{(k-\bar{k})}}$} to fall below a prescribed threshold $\vartheta$. We set $\vartheta = 10^{-8}$ to allow for a meaningful comparison between the two approaches. For the Monte Carlo method,  we compute the convergence measure using distributions separated by $\bar{k} = N$ steps so that, on average, each component of the histogram is updated once over that interval. For the iterative algorithm, in contrast, we use $\bar{k} = 1$, since all components of the quasi-stationary distribution are updated at each iteration.

\begin{figure}[h]
	\centering
\includegraphics[width=0.8\linewidth]{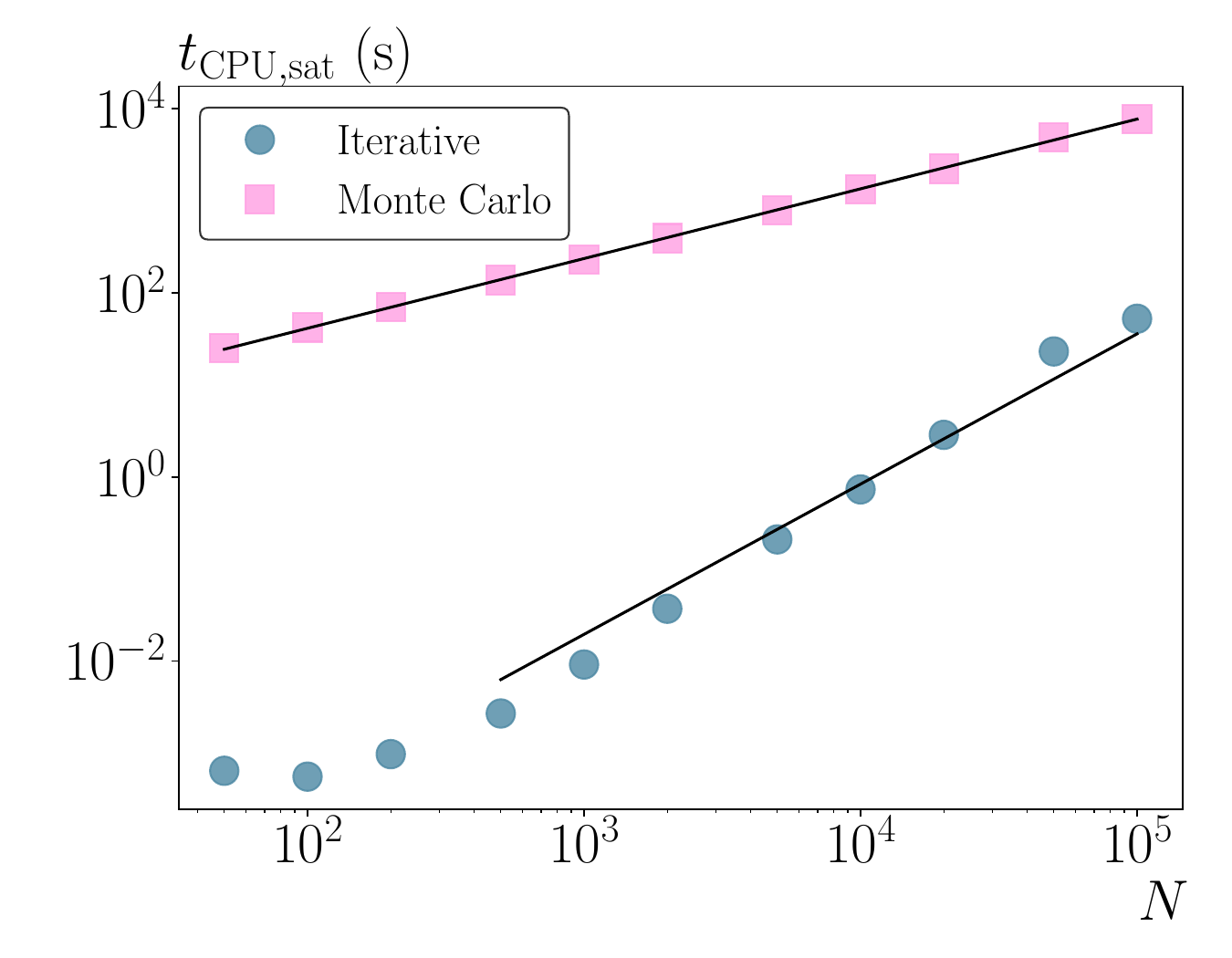}
\caption{\textbf{Effect of system size on the convergence time at criticality.} For the SIS model at the critical point \makebox{$R = 1$}, saturation time (measured in seconds) as a function of system size for the iterative algorithm (blue circles) and the Monte Carlo method (pink squares). The saturation time is defined as the time required for the norm $\Delta^{(k,\bar{k})}=\norm{\bm{Q}^{(k)}-\bm{Q}^{(k-\bar{k})}}$ to fall below the threshold $\vartheta = 10^{-8}$, with $\bar{k} = 1$ for the iterative algorithm and $\bar{k} = N$ for the Monte Carlo method. Solid lines correspond to power-law fits of the form \makebox{$t_\mathrm{CPU,sat} \propto N^b$}, yielding exponents $b \simeq 1.6$ and \makebox{$b \simeq 0.75$} for the iterative algorithm and the Monte Carlo method, respectively.}
\label{fig:SIScriticality_tcpusat}
\end{figure}

\begin{figure}[h!]
	\centering
\includegraphics[width=0.8\linewidth]{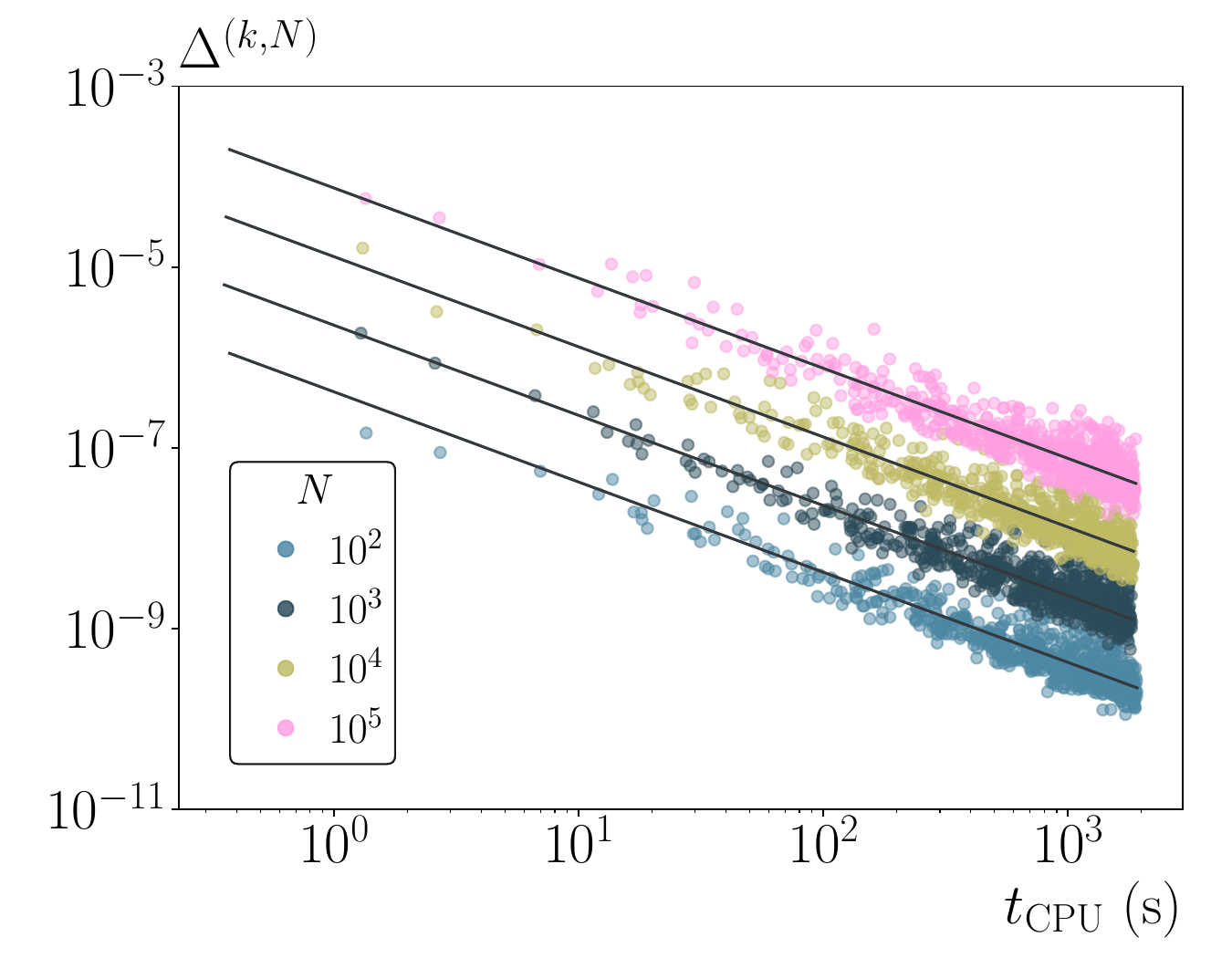}
\caption{\textbf{Estimation of the saturation time in the Monte Carlo method.} In the Monte Carlo method, norm of the difference $\Delta^{(k,N)}= \norm{\bm{Q}^{(k)}-\bm{Q}^{(k-N)}}$ as a function of computation time (measured in seconds) for the SIS model at criticality, and for several system sizes. Data points come from numerical simulations, and solid lines represent power-law fits of the form $\Delta^{(k,N)} \propto t_\mathrm{CPU}^b$, with $b \simeq -1$ in all cases.}
\label{fig:SIScriticality_diffversustcpu_MonteCarlo}
\end{figure}

In the iterative algorithm, the quantity $\Delta^{(k,1)}$ decreases monotonically with computation time, making it straightforward to determine the saturation time as the point where $\Delta^{(k,1)}$ falls below the threshold $\vartheta$. According to this criterion, the saturation time of the iterative algorithm, shown in Fig.~\ref{fig:SIScriticality_tcpusat}, exhibits a power-law scaling with system size, $t_\mathrm{CPU,sat} \propto N^{1.6}$. In the Monte Carlo method, in contrast, the quantity $\Delta^{(k,N)}$ fluctuates indefinitely due to the stochastic nature of the simulation. As a result, the saturation time is extracted from a power-law fit of $\Delta^{(k,N)}$ as a function of computation time for each system size (see Fig.~\ref{fig:SIScriticality_diffversustcpu_MonteCarlo}). Using this procedure, the resulting saturation times follow a power-law scaling of the form $t_\mathrm{CPU,sat} \propto N^{0.75}$ (see Fig.~\ref{fig:SIScriticality_tcpusat}).

The results in Fig.~\ref{fig:SIScriticality_tcpusat} indicate that, although the growth of the computation time with system size is more pronounced for the iterative algorithm, the convergence times of the iterative approach remain shorter than those of the Monte Carlo method for all system sizes considered. Only in the limit of very large systems does the Monte Carlo approach appear to outperform the iterative algorithm.

\subsection{Memory consumption}

\begin{figure}[h]
	\centering
\includegraphics[width=0.8\linewidth]{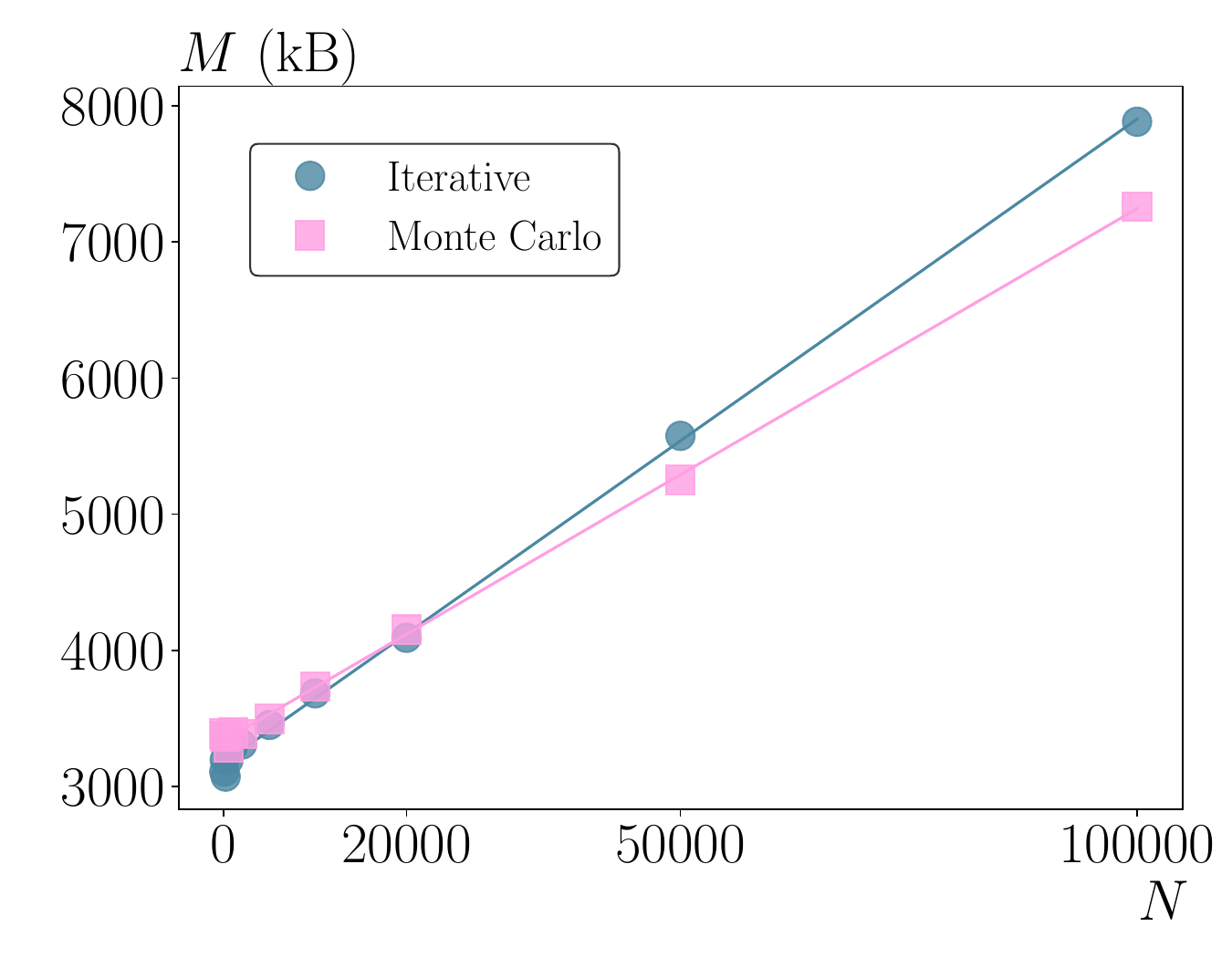}
\caption{\textbf{Dependence of memory usage on system size.} Peak memory usage $M$ (measured in kilobytes) as a function of system size $N$ in the SIS model for $R = 1$. Symbols represent the data obtained with the iterative algorithm (blue circles) and with the Monte Carlo method (pink squares). Solid lines correspond to linear fits, yielding an intercept of approximately 3000 and a slope of $\simeq 0.4$ for both methods.}
\label{fig:SIScriticality_memory}
\end{figure}

We now analyse the effect of system size on the peak memory usage $M$ recorded during execution. Memory consumption is not expected to depend on the reproduction ratio $R$, hence we analyse its scaling with system size at $R=1$ for consistency with the computation time analysis. In Fig.~\ref{fig:SIScriticality_memory}, we show $M$ as a function of system size. The memory consumption of both the iterative algorithm and the Monte Carlo method grows approximately linearly with system size, and is of comparable magnitude for both approaches.

\clearpage
\bibliography{pqst}

@inbook{traulsen2009stochastic,
author = {Traulsen, Arne and Hauert, Christoph},
publisher = {John Wiley \& Sons, Ltd},
isbn = {9783527628001},
title = {Stochastic Evolutionary Game Dynamics},
booktitle = {Reviews of Nonlinear Dynamics and Complexity},
chapter = {2},
volume={2},
pages = {25-61},
doi = {https://doi.org/10.1002/9783527628001.ch2},
url = {https://onlinelibrary.wiley.com/doi/abs/10.1002/9783527628001.ch2},
year = {2009},
}

@article{hohenberg1977critical,
  title = {Theory of dynamic critical phenomena},
  author = {Hohenberg, P. C. and Halperin, B. I.},
  journal = {Rev. Mod. Phys.},
  volume = {49},
  issue = {3},
  pages = {435--479},
  numpages = {0},
  year = {1977},
  month = {Jul},
  publisher = {American Physical Society},
  doi = {10.1103/RevModPhys.49.435},
  url = {https://link.aps.org/doi/10.1103/RevModPhys.49.435}
}

@article{aguilar2022sampling,
  title={Sampling rare trajectories using stochastic bridges},
  author={Aguilar, Javier and Baron, Joseph W and Galla, Tobias and Toral, Ra{\'u}l},
  journal={Physical Review E},
  volume={105},
  number={6},
  pages={064138},
  year={2022},
  publisher={APS},
  url={https://journals.aps.org/pre/abstract/10.1103/PhysRevE.105.064138}
}

@article{assaf2017wkb,
  title={{WKB} theory of large deviations in stochastic populations},
  author={Assaf, Michael and Meerson, Baruch},
  journal={Journal of Physics A: Mathematical and Theoretical},
  volume={50},
  number={26},
  pages={263001},
  year={2017},
  publisher={IOP Publishing},
  url={https://iopscience.iop.org/article/10.1088/1751-8121/aa669a}
}

@article{assaf2010extinction,
  title={Extinction of metastable stochastic populations},
  author={Assaf, Michael and Meerson, Baruch},
  journal={Physical Review E},
  volume={81},
  number={2},
  pages={021116},
  year={2010},
  publisher={APS},
  url={https://journals.aps.org/pre/abstract/10.1103/PhysRevE.81.021116}
}

@article{azaele2016statistical,
  title={Statistical mechanics of ecological systems: Neutral theory and beyond},
  author={Azaele, Sandro and Suweis, Samir and Grilli, Jacopo and Volkov, Igor and Banavar, Jayanth R and Maritan, Amos},
  journal={Reviews of Modern Physics},
  volume={88},
  number={3},
  pages={035003},
  year={2016},
  publisher={APS},
  url={https://journals.aps.org/rmp/abstract/10.1103/RevModPhys.88.035003}
}

@book{collet2013quasi,
  title={Quasi-stationary distributions: Markov chains, diffusions and dynamical systems},
  author={Collet, Pierre and Mart{\'\i}nez, Servet and San Mart{\'\i}n, Jaime},
  year={2013},
  publisher={Springer},
  series={Collection Probability and its Applications},
  place={Heidelberg},
  url={https://link.springer.com/book/10.1007/978-3-642-33131-2}
}

@book{marro2005nonequilibrium, 
    place={Cambridge}, 
    series={Collection Alea-Saclay: Monographs and Texts in Statistical Physics}, 
    title={Nonequilibrium Phase Transitions in Lattice Models}, 
    publisher={Cambridge University Press}, 
    author={Marro, Joaquin and Dickman, Ronald}, 
    year={1999}, 
    collection={Collection Alea-Saclay: Monographs and Texts in Statistical Physics},
    url={https://www.cambridge.org/core/books/nonequilibrium-phase-transitions-in-lattice-models/7D601962354E3A23F77A8D834AE17A5C}
}

@article{de2005simulate,
  title={How to simulate the quasistationary state},
  author={de Oliveira, Marcelo Martins and Dickman, Ronald},
  journal={Physical Review E},
  volume={71},
  number={1},
  pages={016129},
  year={2005},
  publisher={APS},
  url={https://journals.aps.org/pre/abstract/10.1103/PhysRevE.71.016129}
}

@article{dickman2002numerical,
  title={Numerical analysis of the master equation},
  author={Dickman, Ronald},
  journal={Physical Review E},
  volume={65},
  number={4},
  pages={047701},
  year={2002},
  publisher={APS},
  url={https://journals.aps.org/pre/abstract/10.1103/PhysRevE.65.047701}
}

@book{gardiner1985handbook,
  title={Handbook of Stochastic Methods},
  author={Gardiner, Crispin W},
  edition={3},
  year={2004},
  publisher={Springer Berlin}
}

@book{gradshteyn2014table,
  title={Table of integrals, series, and products},
  author={Gradshteyn, Izrail Solomonovich and Ryzhik, Iosif Moiseevich},
  year={2014},
  publisher={Academic press},
  doi={10.1016/C2010-0-64839-5}
}

@article{hinrichsen2000non,
  title={Non-equilibrium critical phenomena and phase transitions into absorbing states},
  author={Hinrichsen, Haye},
  journal={Advances in Physics},
  volume={49},
  number={7},
  pages={815--958},
  year={2000},
  publisher={Taylor \& Francis},
  url={https://www.tandfonline.com/doi/pdf/10.1080/00018730050198152}
}

@book{jacobs2010stochastic,
  title={Stochastic Processes for Physicists: Understanding Noisy Systems},
  author={Jacobs, Kurt},
  year={2010},
  publisher={Cambridge University Press},
  url={https://www.cambridge.org/core/books/stochastic-processes-for-physicists/A4DA560863F148D920B6A3165996D5D7}
}

@article{Lambert2007,
   author = {Amaury Lambert},
   doi = {10.1214/EJP.v12-402},
   issn = {10836489},
   journal = {Electronic Journal of Probability},
   month = {1},
   pages = {420-446},
   title = {Quasi-stationary distributions and the continuous-state branching process conditioned to be never extinct},
   volume = {12},
   year = {2007},
}

@article{Lipowski2001,
   author = {Adam Lipowski and Michel Droz},
   doi = {10.1103/PhysRevE.64.031107},
   issn = {1063-651X},
   issue = {3},
   journal = {Physical Review E},
   month = {8},
   pages = {031107},
   title = {Criticality of natural absorbing states},
   volume = {64},
   url = {https://link.aps.org/doi/10.1103/PhysRevE.64.031107},
   year = {2001},
}

@article{magalang2024optimal,
  title = {Optimal switching strategies in multidrug therapies for chronic diseases},
  author = {Magalang, Juan and Aguilar, Javier and Esguerra, Jose Perico and Rold\'an, \'Edgar and Sanchez-Taltavull, Daniel},
  journal = {Physical Review E},
  volume = {112},
  issue = {3},
  pages = {034408},
  numpages = {16},
  year = {2025},
  month = {Sep},
  publisher = {American Physical Society},
  doi = {10.1103/htck-mcby},
  url = {https://link.aps.org/doi/10.1103/htck-mcby}
}

@article{gillespie1976general,
  title={A general method for numerically simulating the stochastic time evolution of coupled chemical reactions},
  author={Gillespie, Daniel T},
  journal={Journal of Computational Physics},
  volume={22},
  number={4},
  pages={403--434},
  year={1976},
  publisher={Elsevier},
  url={https://www.sciencedirect.com/science/article/abs/pii/0021999176900413}
}

@article{gillespie1977exact,
  title={Exact stochastic simulation of coupled chemical reactions},
  author={Gillespie, Daniel T},
  journal={The Journal of Physical Chemistry},
  volume={81},
  number={25},
  pages={2340--2361},
  year={1977},
  publisher={ACS Publications},
  url={https://pubs.acs.org/doi/10.1021/j100540a008}
}

@misc{NumericalMethodsQSD,
  title        = {Public {GitHub} repository: \url{https://github.com/saraoliver-bonafoux/NumericalMethodsQSD}}
}

@article{meleard2012quasi,
author = {Meleard, Sylvie and Villemonais, Denis},
year = {2012},
month = {12},
pages = {340-410},
title = {Quasi-stationary distributions and population processes},
volume = {9},
journal = {Probability Surveys},
doi = {10.1214/11-PS191}
}

@article{czaplicka2022biased,
  title={Biased-voter model: How persuasive a small group can be?},
  author={Czaplicka, Agnieszka and Charalambous, Christos and Toral, Raul and San Miguel, Maxi},
  journal={Chaos, Solitons \& Fractals},
  volume={161},
  pages={112363},
  year={2022},
  publisher={Elsevier},
  url={https://www.sciencedirect.com/science/article/pii/S0960077922005732}
}

@article{Nasell1996,
	author = {N{\aa}sell, Ingemar},
	doi = {https://doi.org/10.2307/1428186},
	journal = {Advances in Applied Probability},
	number = {3},
	pages = {895--932},
	title = {{The Quasi-Stationary Distribution of the Closed Endemic SIS Model}},
	volume = {28},
	year = {1996}}

@article{Aalen2004,
   author = {Odd O. Aalen and HAkon K. Gjessing},
   doi = {10.1007/s10985-004-4775-9},
   issn = {1380-7870},
   issue = {4},
   journal = {Lifetime Data Analysis},
   month = {12},
   pages = {407-423},
   title = {Survival Models Based on the Ornstein-Uhlenbeck Process},
   volume = {10},
   url = {http://link.springer.com/10.1007/s10985-004-4775-9},
   year = {2004},
}

@book{naasell2011extinction,
  title={Extinction and Quasi-Stationarity in the Stochastic Logistic SIS Model},
  author={N{\aa}sell, Ingemar},
  year={2011},
  publisher={Springer},
  url={https://link.springer.com/book/10.1007/978-3-642-20530-9}
}

@article{volkov2003neutral,
  title={Neutral theory and relative species abundance in ecology},
  author={Volkov, Igor and Banavar, Jayanth R and Hubbell, Stephen P and Maritan, Amos},
  journal={Nature},
  volume={424},
  number={6952},
  pages={1035--1037},
  year={2003},
  publisher={Nature Publishing Group UK London},
  url={https://www.nature.com/articles/nature01883}
}

@article{pastor2001epidemic,
  title={Epidemic dynamics and endemic states in complex networks},
  author={Pastor-Satorras, Romualdo and Vespignani, Alessandro},
  journal={Physical Review E},
  volume={63},
  number={6},
  pages={066117},
  year={2001},
  publisher={APS},
  url={https://journals.aps.org/pre/abstract/10.1103/PhysRevE.63.066117}
}

@article{Risken1991,
   author = {H. Risken and T. K. Caugheyz},
   doi = {10.1115/1.2897281},
   issn = {0021-8936},
   issue = {3},
   journal = {Journal of Applied Mechanics},
   pages = {860-860},
   title = {{The Fokker-Planck Equation: Methods of Solution and Application, 2nd ed.}},
   volume = {58},
   year = {1991},
}

@book{keeling,
 ISBN = {9780691116174},
 URL = {http://www.jstor.org/stable/j.ctvcm4gk0},
 author = {Matt J. Keeling and Pejman Rohani},
 publisher = {Princeton University Press},
 title = {Modeling Infectious Diseases in Humans and Animals},
 urldate = {2023-08-28},
 year = {2008}
}

@article{cavender1978quasi,
  title={Quasi-stationary distributions of birth-and-death processes},
  author={Cavender, James A},
  journal={Advances in Applied Probability},
  volume={10},
  number={3},
  pages={570--586},
  year={1978},
  publisher={Cambridge University Press},
  url={https://www.cambridge.org/core/journals/advances-in-applied-probability/article/abs/quasi-stationary-distributions-of-birth-and-death-processes/D41EEA3B8438EAB7F7E827674DC1AC9E}
}

@article{Magalang2023,
  title = {Analytic and {M}onte {C}arlo approximations to the distribution of the first-passage time of drifted diffusion with stochastic resetting and mixed boundary conditions},
  author = {Magalang, Juan and Turin, Riccardo and Aguilar, Javier and Colombani, Laetitia and Sanchez-Taltavull, Daniel and Gatto, Riccardo},
  journal = {Phys. Rev. E},
  volume = {111},
  issue = {5},
  pages = {054117},
  numpages = {19},
  year = {2025},
  month = {May},
  publisher = {American Physical Society},
  url = {https://link.aps.org/doi/10.1103/PhysRevE.111.054117}
}

@article{lelievre2016partial,
  title={Partial differential equations and stochastic methods in molecular dynamics},
  author={Lelievre, Tony and Stoltz, Gabriel},
  journal={Acta Numerica},
  volume={25},
  pages={681--880},
  year={2016},
  publisher={Cambridge University Press},
  url={https://www.cambridge.org/core/journals/acta-numerica/article/partial-differential-equations-and-stochastic-methods-in-molecular-dynamics/60F8398275D5150AA54DD98F745A9285}
}

@book{krapivsky2010kinetic,
  title={A Kinetic View of Statistical Physics},
  author={Krapivsky, Pavel L and Redner, Sidney and Ben-Naim, Eli},
  year={2010},
  publisher={Cambridge University Press},
  url={https://www.cambridge.org/core/books/kinetic-view-of-statistical-physics/773F488A893B060A5A5FA287158AB229}
}

@book{redner2001guide,
  title={A Guide to First-Passage Processes},
  author={Redner, Sidney},
  year={2001},
  publisher={Cambridge University Press},
  url={https://www.cambridge.org/core/books/guide-to-firstpassage-processes/59066FD9754B42D22B028E33726D1F07}
}

@article{pollock2020quasi,
  title={{Quasi-stationary Monte Carlo and the ScaLE algorithm}},
  author={Pollock, Murray and Fearnhead, Paul and Johansen, Adam M and Roberts, Gareth O},
  journal={Journal of the Royal Statistical Society Series B: Statistical Methodology},
  volume={82},
  number={5},
  pages={1167--1221},
  year={2020},
  publisher={Oxford University Press},
  url={https://academic.oup.com/jrsssb/article/82/5/1167/7056133}
}

@article{Meleard,
author = {Sylvie M{\'e}l{\'e}ard and Denis Villemonais},
title = {{Quasi-stationary distributions and population processes}},
volume = {9},
journal = {Probability Surveys},
number = {none},
publisher = {Institute of Mathematical Statistics and Bernoulli Society},
pages = {340 -- 410},
year = {2012},
doi = {10.1214/11-PS191},
URL = {https://doi.org/10.1214/11-PS191}
}

@article{castellano2009statistical,
  title={Statistical physics of social dynamics},
  author={Castellano, Claudio and Fortunato, Santo and Loreto, Vittorio},
  journal={Reviews of Modern Physics},
  volume={81},
  number={2},
  pages={591},
  year={2009},
  publisher={APS},
  url={https://journals.aps.org/rmp/abstract/10.1103/RevModPhys.81.591}
}

@article{axelrod1997dissemination,
  title={{The Dissemination of Culture: A Model with Local Convergence and Global Polarization}},
  author={Axelrod, Robert},
  journal={Journal of Conflict Resolution},
  volume={41},
  number={2},
  pages={203--226},
  year={1997},
  publisher={Sage Periodicals Press 2455 Teller Road, Thousand Oaks, CA 91320},
  url={https://www.jstor.org/stable/174371}
}

@article{dickman2002quasi,
  title={Quasi-stationary distributions for stochastic processes with an absorbing state},
  author={Dickman, Ronald and Vidigal, Ronaldo},
  journal={Journal of Physics A: Mathematical and General},
  volume={35},
  number={5},
  pages={1147},
  year={2002},
  publisher={IOP Publishing},
  url={https://iopscience.iop.org/article/10.1088/0305-4470/35/5/303}
}

@book{toral2014stochastic,
  title={Stochastic numerical methods: an introduction for students and scientists},
  author={Toral, Ra{\'u}l and Colet, Pere},
  year={2014},
  publisher={John Wiley \& Sons},
  url={https://www.wiley.com/en-us/Stochastic+Numerical+Methods%3A+An+Introduction+for+Students+and+Scientists-p-9783527683123}
}

@incollection{blanchet2013empirical,
  title={Empirical analysis of a stochastic approximation approach for computing quasi-stationary distributions},
  author={Blanchet, Jose and Glynn, Peter and Zheng, Shuheng},
  booktitle={EVOLVE-A Bridge between Probability, Set Oriented Numerics, and Evolutionary Computation II},
  pages={19--37},
  year={2013},
  publisher={Springer}
}

@article{van2013quasi,
  title={Quasi-stationary distributions for discrete-state models},
  author={van Doorn, Erik A and Pollett, Philip K},
  journal={European Journal of Operational Research},
  volume={230},
  number={1},
  pages={1--14},
  year={2013},
  publisher={Elsevier},
  url={https://www.sciencedirect.com/science/article/pii/S0377221713000799}
}

@book{karlin2014first,
  title={A First Course in Stochastic Processes},
  author={Karlin, Samuel},
  year={2014},
  publisher={Academic press},
  url={https://www.sciencedirect.com/book/9780080570419/a-first-course-in-stochastic-processes}
}

@article{dornic2005integration,
  title={{Integration of Langevin Equations with Multiplicative Noise and the Viability of Field Theories for Absorbing Phase Transitions}},
  author={Dornic, Ivan and Chat{\'e}, Hugues and Munoz, Miguel A},
  journal={Physical Review Letters},
  volume={94},
  number={10},
  pages={100601},
  year={2005},
  publisher={APS},
  url={https://journals.aps.org/prl/abstract/10.1103/PhysRevLett.94.100601}
}

@article{moro2004numerical,
  title={Numerical schemes for continuum models of reaction-diffusion systems subject to internal noise},
  author={Moro, Esteban},
  journal={Physical Review E},
  volume={70},
  number={4},
  pages={045102},
  year={2004},
  publisher={APS},
  url={https://journals.aps.org/pre/abstract/10.1103/PhysRevE.70.045102}
}

@book{VanKam,
   author = {N. G. van Kampen},
   doi = {10.1016/B978-0-444-52965-7.X5000-4},
   isbn = {9780444529657},
   publisher = {Elsevier},
   title = {Stochastic Processes in Physics and Chemistry},
   url = {https://linkinghub.elsevier.com/retrieve/pii/B9780444529657X50004},
   year = {2007},
}

@article{Vezzani2023,
	author = {Vezzani, Alessandro and Mu{\~{n}}oz, Miguel A. and Burioni, Raffaella},
	doi = {10.1103/PhysRevE.107.014105},
	issn = {24700053},
	journal = {Physical Review E},
	month = {jan},
	number = {1},
	pages = {014105},
	pmid = {36797930},
	publisher = {American Physical Society},
	title = {{Anomalous finite-size scaling in higher-order processes with absorbing states}},
	url = {https://link.aps.org/doi/10.1103/PhysRevE.107.014105},
	volume = {107},
	year = {2023}}

@article{NasellEpidemics2D,
 ISSN = {13697412, 14679868},
 URL = {http://www.jstor.org/stable/2680643},
 author = {Ingemar Nasell},
 journal = {Journal of the Royal Statistical Society. Series B (Statistical Methodology)},
 number = {2},
 pages = {309--330},
 publisher = {[Royal Statistical Society, Oxford University Press]},
 title = {On the Time to Extinction in Recurrent Epidemics},
 urldate = {2024-11-06},
 volume = {61},
 year = {1999}
}

\end{document}